\documentclass[11pt]{article}
\pdfoutput=1
\usepackage{jheppub}
\usepackage[utf8]{inputenc}

\usepackage{tikz-cd}
\usepackage{cancel}
\usepackage{amsmath}
\usepackage{amsfonts}
\usepackage{mathtools}
\usepackage[mathscr]{eucal}
\usepackage{tensor}
\usepackage{bigints}
\usepackage{calc}
\usepackage{makecell}
\usepackage{bbm}
\usepackage{amsmath}
\usepackage{bm}
\usepackage{verbatim}

\usepackage[labelfont={small, bf, sf}, textfont={small,sf}]{caption}
\usepackage{subcaption}
\usepackage{tikz}
\usepackage{pgfplots}
\pgfplotsset{compat=1.15}

\usepackage{accents}

\newcommand{\be}{\begin{equation}}
\newcommand{\ee}{\end{equation}}
\newcommand{\bea}{\begin{eqnarray}}
\newcommand{\eea}{\end{eqnarray}}

\newcommand{\R}{\mathbb{R}}

\newcommand{\rd}{\mathrm{d}}
\newcommand{\Lie}{\mathcal L}

\newcommand{\Sl}{\mathfrak{sl}}

\newcommand{\slr}{\Sl(2,\R)}
\newcommand{\SL}{\mathrm{SL}}
\newcommand{\SO}{\mathrm{SO}}
\newcommand\Diff{\text{Diff}}
\newcommand\SDiff{\text{SDiff}}

\def\N{N}

\def\tP{\widetilde{P}}
\def\w{\boldsymbol{w}}
\def\tn{\widetilde{n}}

\def\n{\nu}

\def\sg{\mathfrak{c}}

\def\P{P}
\def\Ad{\mathrm{Ad}}

\newcommand{\si}{\sigma}
\newcommand{\pa}{\partial}
\newcommand{\upi}{\nu}

\newcommand{\pb}[2]{\{ #1 , #2 \}_{\nu}}

\DeclareMathOperator{\diff}{\mathfrak{diff}}
\DeclareMathOperator{\sdiff}{\mathfrak{sdiff}}

\DeclareMathOperator{\tr}{Tr}

\def\Seq{\stackrel{S}=}

\newcommand{\lp}{\ell}
\newcommand{\lpup}[1]{\ell^{#1}}
\newcommand{\lpdn}[1]{\ell_{#1}}
\newcommand{\lm}{\bar{\ell}}
\newcommand{\lmup}[1]{\lm^{#1}}
\newcommand{\lmdn}[1]{\lm_{#1}}
\newcommand{\Bp}{B}
\newcommand{\Bm}{\bar{B}} 
\newcommand{\Ap}{A}
\newcommand{\Am}{\bar{A}} 
\newcommand{\km}{\bar{\kappa}}

\newcommand{\Hajicek}{H\'aj\'{i}\v{c}ek }

\newcommand{\mfk}[1]{\mathfrak{#1}}
\newcommand{\mbs}[1]{\boldsymbol{#1}}
\newcommand{\mbb}[1]{\mathbb{#1}}

\newcommand{\mcal}[1]{\mathcal{#1}}

\newcommand{\tenofo}[1]{\text{\normalfont #1}}
\newcommand{\newl}{\medskip\noindent}

\newcommand{\wt}[1]{\widetilde{#1}}
\newcommand{\gsl}{\mfk{g}_{\slr}(S)}

\newcommand{\cog}{G_{\SL(2,\mbb{R})}(S)}
\newcommand{\hyg}{G_{\mbb{R}}(S)}
\newcommand{\hya}{\mfk{g}_{\mbb{R}}(S)}

\newcommand{\coa}{\mfk{g}_{\slr}(S)}

\usepackage{changepage}
\usepackage{fontawesome}

\newcommand{\subsubsubsection}[1]{\paragraph{\it #1}\mbox{}\\[.3cm]}
\allowdisplaybreaks

\newcommand{\SU}[1]{\text{SU}(#1)}

\def\AA{\mathcal{A}}

\newcommand{\beq}{\begin{equation}}
\newcommand{\eeq}{\end{equation}}

\title{\centering Gravitational Edge Modes, Coadjoint Orbits, \\ and Hydrodynamics
}
\author[1]{William Donnelly,}
\author[1]{Laurent Freidel,}
\author[1,2]{Seyed Faroogh Moosavian,}
\author[1]{Antony J. Speranza}
\affiliation[1]{Perimeter Institute for Theoretical Physics, 31 Caroline St. N.,  Waterloo ON, N2L 2Y5, Canada}
\affiliation[2]{Department of Physics, McGill University, Ernest Rutherford Physics Building, 3600 Rue University, Montr\'eal, QC H3A 2T8, Canada}
\emailAdd{wdonnelly@pitp.ca\,\!}
\emailAdd{lfreidel@pitp.ca\,\!}
\emailAdd{sfmoosavian@physics.mcgill.ca\,\!}
\emailAdd{asperanz@gmail.com}

\abstract{
The phase space of general relativity in a finite subregion is characterized by edge modes localized at the codimension-2 boundary, transforming under an infinite-dimensional group of symmetries.  The quantization of this symmetry algebra is conjectured to be an important aspect of quantum gravity. As a step towards quantization, we derive a complete classification of the positive-area coadjoint orbits of this group for boundaries that are topologically a 2-sphere.  
This classification parallels Wigner's famous classification of representations of the Poincar\'e group since both groups have the structure of a semidirect product.
We find that the total area is a Casimir of the algebra, analogous to mass in the Poincar\'e group.
A further infinite family of Casimirs can be constructed from the 
curvature of the normal bundle of the boundary surface. These arise as invariants of the little group, which is the group of area-preserving diffeomorphisms, and are the analogues of spin. 
Additionally, we show that the symmetry group of hydrodynamics appears as 
a reduction of the corner symmetries of general relativity.  Coadjoint orbits of both groups are 
classified by the same set of invariants, and, in the case of the hydrodynamical group, 
the invariants are interpreted as the generalized enstrophies of the fluid.}

\begin{document}
\maketitle

\section{Introduction and summary of results}

Symmetries provide a fundamental organizational tool in physics.
One of the primary lessons of quantum mechanics, culminating in Wigner's Theorem, is that quantization of a classical system with a physical symmetry group $G$ furnishes a unitary representation of the group \cite{Wigner1931}.
Thus, the problem of quantizing gravity would benefit from the existence 
of a large physical symmetry group associated with gravitational subsystems,
whose representation theory would control the quantum gravitational Hilbert space.
Realizing this idea is a central theme of the present work.

General relativity contains gauge redundancies encoded in the infinite-dimensional group of  diffeomorphisms of the spacetime manifold.
Gauge symmetries are not physical symmetries in the sense of Wigner; rather, they are redundancies of the physical description.
These redundancies are represented trivially on the space of physical states, and cannot serve as a useful organizational principle.
The situation drastically changes when boundaries are introduced (be they asymptotic or finite) to decompose  spacetime into a collection of subregions. 
The presence of boundaries on Cauchy slices, called \emph{corners}, transmutes gauge redundancies into physical symmetries \cite{Regge:1974zd,Carlip:1994gy,Balachandran:1995qa}.
The Noether charges associated to these physical symmetries, which we call \emph{corner symmetries}, are nonvanishing and can be written as local integrals over the codimension-2 corner.
Colloquially, one can understand these corner charges as ``handles'' which can be used to couple to a gravitational system from the outside \cite{Rovelli:2013fga,Rovelli:2020mpk}. The idea that corner degrees of freedom play a central role in the quantum mechanical description of black holes was recognised very early on by \cite{Carlip:1994gy, Balachandran:1995qa,Smolin:1995vq, Ashtekar:1997yu, Strominger:1997eq}.
The importance of corner symmetries for the quantization of gravity was first formulated in \cite{Freidel:2015gpa, Donnelly:2016auv}.

The mechanism by which physical symmetries arise from gauge
is sometimes described as a breaking of gauge symmetry by the boundaries, 
and is associated with 
the appearance of new physical degrees of freedom.
In \cite{Donnelly:2016auv}, it was shown how to 
realize the new degrees of freedom locally on the boundary
using a corner version of the Stueckelberg mechanism.  
This procedure maintains formal diffeomorphism invariance by 
introducing a new field, representing the \emph{edge modes}, which transforms
nontrivially under diffeomorphisms.  
The ``broken gauge symmetries'' are then recognized as a physical symmetry acting
solely on the edge modes. 
Following Ref.~\cite{Freidel:2020xyx, Freidel:2020svx, Freidel:2020ayo} we refer to these as \emph{corner symmetries}.\footnote{The term ``surface symmetry'' was used in Ref.~\cite{Donnelly:2016auv}, but here we use the term ``corner symmetry'' to emphasize that it acts at a codimension-2 surface.}

For general relativity in metric variables, it was shown in \cite{Donnelly:2016auv} that the relevant \emph{corner symmetry group} of a finite region bounded by 
a codimension-2 corner $S$ is given by 
\begin{equation}\label{eq:the corner symmetry-group I}
    \cog=\Diff(S)\ltimes \SL(2,\mbb{R})^S.
\end{equation}
Here, $\Diff(S)$ is the group of diffeomorphisms of $S$, and $\SL(2,\mbb{R})^S$ is the space of $\SL(2,\mbb{R})$-valued maps on $S$. 
The group $\SL(2,\R)^S$ acts via linear transformations on the two-dimensional plane normal to $S$.
It is a generalization of a loop group in which
the underlying space is a sphere, rather than a circle, 
and we refer to it as a \emph{sphere group}, following \cite{loop,Frappat:1989gn, Dowker:1990ss,Neeb}. 
The group $\cog$ is the automorphism group of the  normal bundle 
of a codimension-2 sphere embedded in spacetime (see appendix 
\ref{app:fiber} for details on the fiber bundle description
of this group).
The Lie algebra of the corner symmetry group $\cog$, which we denote by $\coa$, is
\begin{equation}\label{eq:the corner symmetry algebra}
    \coa=\diff(S) \oplus_{\Lie} \mathfrak{sl}(2,\mathbb{R})^S.
\end{equation}
The $\diff(S)$ generators are realized as vector fields on the sphere, and the $\slr^S$ generators are realized as $\slr$-valued functions on the sphere; the subscript $\Lie$ on the semidirect sum indicates that the infinitesimal diffeomorphisms of $S$ act on the local $\slr^S$ generators in a natural way by the Lie derivative of scalar functions.

We note that different formulations of gravity, in particular tetrad gravity, have additional gauge symmetries which can lead to an enlarged surface symmetry group 
\cite{Geiller:2017whh,Freidel:2020xyx,Freidel:2020svx,Freidel:2020ayo}. 
For more general diffeomorphism-invariant theories, including 
higher curvature theories and couplings to non-metric fields, 
it was shown in Ref.~\cite{Speranza:2017gxd} that the group of symmetries 
can be reduced to (\ref{eq:the corner symmetry-group I}), and a generalized
expression for the associated charges was derived using the Iyer-Wald
formalism \cite{Iyer1994a}.  
In this work, we will focus on metric general relativity and the group \eqref{eq:the corner symmetry-group I}.
In fact, we will make two further assumptions.
First, we only consider  four-dimensional spacetimes, and as such, the corner $S$ is a two-dimensional surface. 
In this case, one of the nice features of \eqref{eq:the corner symmetry-group I} is that the edge modes live in two dimensions and the relevant subgroups of $\tenofo{Diff}(S)$ are well understood.
Second, we specialize to the case that $S$ is a 2-sphere which further simplifies the analysis. We expect that relaxing these assumptions to work
with higher genus surfaces and  
higher spacetime dimensions to be straightforward, and we outline these generalizations in Section \ref{subsec:generalizations}.

At the classical level, the symmetry \eqref{eq:the corner symmetry algebra} is implemented by the Poisson bracket on the phase space of the gravitational theory \cite{Donnelly:2016auv}.
In the quantum theory, we expect the Hilbert space to carry a unitary representation of this symmetry. A powerful method to study representations of $G$ at the semiclassical level is Kirillov's orbit method \cite{Kirillov196202,Kirillov1976,Kirillov199908, kirillov2004lectures}. 
In this formalism, one first studies the coadjoint orbits of $G$. 
Each coadjoint orbit of $G$ is a symplectic manifold and 
can be quantized using the available tools from the geometric quantization, i.e.\ one can associate an irreducible unitary representation to each coadjoint orbit of $G$ that satisfies an integrality condition. 
Coadjoint orbits of semidirect product groups such as \eqref{eq:the corner symmetry-group I} can be constructed by the method of symplectic induction, starting from coadjoint orbits of certain subgroups of $G$.
This can be viewed as a classical analog of Mackey's machinery of induced representations \cite{Mackey195201,Mackey195309, Mackey:1978za}.
In particular, irreducible unitary representations obtained using these two methods should 
agree, though there are various subtleties involved \cite{DuvalElhadadGotaySniatyckiTuynman199102}. 

The present work is the first in a series of papers in which we study various aspects of corner symmetry-group \eqref{eq:the corner symmetry-group I},
and is dedicated to the study of the coadjoint orbits of 
$\cog$ as a prequel to quantization. 
The coadjoint orbits of a number of other symmetry groups 
relevant to special and general relativity have been studied previously,
including  Poincar\'e  \cite{KimNoz1986, Hudon2009}, 
Virasoro \cite{Witten198803, BalogFeherPalla199801},
$\text{BMS}_3$ \cite{BarnichOblak201408, BarnichOblak201403, BarnichOblak201502},
and loop groups \cite{Frenkel1984}.  
In each example involving a semidirect product of groups, the normal 
subgroup is abelian; by contrast, the corner symmetry group \eqref{eq:the corner symmetry-group I} has a nonabelian normal factor, and hence 
the analysis of its orbits is more involved.  

The general classification of coadjoint orbits of semidirect product groups $G=H\ltimes N$ with an abelian normal subgroup $N$ has been studied in Refs.~\cite{Rawnsley197501,Baguis199705}.
Our normal subgroup is the nonabelian sphere group $\slr^S$, so to study it we first reduce it to a semidirect product with an abelian normal subgroup by diagonalizing the local $\slr$ generator.
This reduces the problem to classification of orbits of the \emph{hydrodynamical group}
\begin{equation}\label{eq:the hydroynamical group I}
    G_{\mbb{R}}(S):=\tenofo{Diff}(S)\ltimes \mathbb{R}^S.
\end{equation}
This group appears in compressible hydrodynamics, where the generators of $\Diff(S)$ and $\R^S$ are the momentum density and mass density respectively \cite{MarsdenRatiuWeinstein1984a,MarsdenRatiuWeinstein1984b,HolmMarsdenRatiu1998,Holm200103,khesin2020geometric}. It is also closely related to
the so-called \emph{generalized} BMS group of asymptotic symmetries 
of flat space,\footnote{Note that in the extended BMS group, the factor $\R^S$ represents densities of weight $1/2$, while in hydrodynamics it represents  densities of weight $0$. The density weight affects the action of $\Diff(S)$ on the abelian factor.} introduced by Campiglia and Laddha \cite{Campiglia:2014yka} and further developed in the canonical setting by Comp\`ere et al. \cite{Compere:2018ylh, Ruzziconi:2020cjt} (see also \cite{Campiglia:2020qvc}). 
It also appears in the recent investigations of the  near-horizon symmetry group \cite{Chandrasekaran:2018aop,Donnay:2019jiz}.
Importantly, it enters in  the study of soft theorems as  Ward identities 
for the S-matrix  
\cite{Campiglia:2015yka, Donnay:2020guq}. Penna was the first one to emphasize the analogy between the gravitational and hydrodynamical symmetry groups \cite{Penna201703}.
Finally, let us mention that 
the canonical duality  between the local area and boost symmetry parameter was noticed early on by Hayward, 
Carlip, and Teitelboim \cite{Hayward:1993my, Carlip:1993sa}.

The hydrodynamical group \eqref{eq:the hydroynamical group I} has an abelian normal subgroup and so can be studied using the general framework developed in \cite{Rawnsley197501,Baguis199705}.
For a semidirect product of the form $H \ltimes N$ one fixes a generator of $N$ and studies the \emph{little group} which is the subgroup of $H$ that fixes it.
In the hydrodynamical group, the normal subgroup generator is a positive density $|\tn|$ and the little group is the group of area-preserving diffeomorphisms:
\begin{equation}
    \tenofo{SDiff}(S)_{|\wt n|}:=\{g\in\tenofo{Diff}(S)|\,g^*(|\wt n|)=|\wt n|\},
\end{equation}
where diffeomorphisms act by pullback. 
The coadjoint orbits of the hydrodynamical group are then classified by
\begin{enumerate}
    \item The total mass of the fluid, 
    \begin{equation} \label{eq:fluid mass}
        M = \bigintsss_S d^2 \sigma \; |\wt n|,
    \end{equation}
    \item A coadjoint orbit of the group of area-preserving diffeomorphisms.
\end{enumerate}
These invariants are analogous to the mass and spin in the classification of orbits of the Poincar\'e group.

Coadjoint orbits of the subgroup of area-preserving diffeomorphisms can be expressed in terms of the vorticity
\begin{equation}
    \mbs w = \rd p,
\end{equation}
where $p$ is the fluid momentum one-form.
In two dimensions the vorticity 2-form can be expressed entirely in terms of the scalar vorticity $w$ such that $\mbs w = w \, |\tn| \rd^2\sigma$, and the coadjoint orbits of $\SDiff$ are in one-to-one correspondence with orbits of $w$ under area-preserving diffeomorphisms \cite{izosimov2016coadjoint}.
The complete classification involves the \emph{measured Reeb graph} which will be described in section \ref{subsec:area-preserving diffeomorphisms}, but one can construct an infinite sequence of invariants
\begin{equation}\label{eq:the invariants of G_R coadjoint orbits}
    C_k:=     \bigintsss_S\rd^2\si\,|\wt{n}|\,w^k, \qquad k=2,3,\cdots,
\end{equation}
known as \emph{generalized enstrophies}.

The classification of coadjoint orbits of the surface symmetry group $\cog$ then follows from the classification of coadjoint orbits of the hydrodynamical group.
The generator of local $\slr$ transformations $n = n^a\tau_a$ transforms in the adjoint representation and breaks the $\slr^S$ symmetry to an $\R^S$ subgroup.
The new feature is that to describe invariants we have to construct a vorticity function which is invariant under local $\slr$ transformations.
Given a momentum 1-form $p$ and $\slr$ generator $n^a$, we define the \emph{dressed vorticity}
\begin{equation}\label{eq:the definition of dressed 2-form vorticity in terms of velocity}
    \bar{\mbs{w}}:= \rd p - \frac{1}{2}\varepsilon_{abc}n^a\rd n^b\wedge \rd n^c. 
\end{equation}
The construction of the dressed vorticity closely parallels the construction of the electromagnetic field strength in the $\SU{2}$ Georgi-Glashow model: in that context the adjoint Higgs breaks $\SU{2}$ down to $\mathrm{U}(1)$ leading to an expression similar to \eqref{eq:the definition of dressed 2-form vorticity in terms of velocity} for the electromagnetic field strength \cite{Corrigan, Shnir2005}.
Equipped with the dressed scalar vorticity $\bar w$ such that $\bar {\mbs w} = \bar w \, |\tn| \rd^2\sigma$, the invariants are constructed just as for the hydrodynamical group and consist of the total mass $M$ and the  measured Reeb graph associated to $\bar w$.
In particular, the Casimirs can be constructed by simply replacing the vorticity $w$ with its dressed version:
\begin{equation}\label{eq:the invariants of G_{SL(2,R)} coadjoint orbits}
\bar{C}_k = \bigintsss_S d^2 \sigma |\wt n| \bar w^k, \qquad k = 2,3,\ldots.
\end{equation}
This provides the complete classification of 
coadjoint orbits of $\cog$ which possesses a continuous and strictly positive area element. 

Having completed the classification of coadjoint orbits of the surface symmetry group, we turn to the realization of this symmetry on the gravitational phase space.
The surface symmetry group acts via diffeomorphisms 
in a neighbourhood of S embedded in spacetime, and its generators are related to the normal geometry of $S$.
In particular, the $\slr$ generators are constructed from the normal metric,
and the hydrodynamical group \eqref{eq:the hydroynamical group I} corresponds to the subgroup which preserves the normal metric.
The generator of this subgroup is the area form on the surface, so the Casimir corresponding to the fluid mass \eqref{eq:fluid mass} is the total area of the surface.

To complete the geometric description of the coadjoint orbits, we have to study the little group, which consists of diffeomorphisms which fix the area form on the surface.
The generators of these area-preserving diffeomorphism generators are related to a geometric invariant of the surface known as the outer curvature \cite{Carter1992, Speranza:2019hkr}.
The spacetime covariant derivative defines a connection on the normal bundle of $S$, given by $D_A := \pa_A + M_A$, whose curvature is known as the outer curvature tensor:
\beq
W\indices{^i_j_A_B}= 
\partial_A M\indices{_B^i_j} 
-\partial_B M\indices{_A^i_j}+[M_A, M_B]^i{}_j.
\eeq
Upon lowering an index with the normal metric, $W_{ijAB}$ is antisymmetric in both its tangent indices $AB$ and normal indices $ij$, so it can be captured by a single scalar function $W$.
It is shown in Section \ref{sec:vorticity}
smeared functions of the the outer curvature $W[\phi]$ satisfy an 
algebra of area-preserving diffeomorphisms of $S$.

Under the moment map sending the gravitational phase space to the 
space of coadjoint orbits, 
we find that $W$ maps precisely to the dressed vorticity $\bar w$. 
This correspondence therefore elucidates the geometric 
origin of the dressed vorticity as an orbit invariant, 
since it arises from a curvature invariant of the 
embedded surface $S$.  Conversely, using the abstract classification
of the invariants of the corner symmetry coadjoint orbits and the universality
of the moment map \cite{kirillov2004lectures}, we find that 
the complete set of invariants of the gravitational corner phase space
is given by the total surface area, together with the outer curvature scalar, up to diffeomorphism.  This classification theorem is a major achievement of the present work.

The paper is organized as follows. In Section \ref{sec:generalities of coadjoint orbits}, we describe the general features of coadjoint orbits, including 
the canonical symplectic form they possess. 
As an illustration of the general method, we give a brief overview of the coadjoint orbits of the $4$-dimensional Poincar\'e group in Section \ref{subsec:coadjoint orbits of the poincare group}. 
Following this, in Section \ref{sec:coadjoint orbits of hydrodynamical group} we explore the coadjoint orbits of the hydrodynamical group
as a preliminary step toward analyzing the full corner symmetry group.
We review the reduction of the  orbits to the little group of area-preserving diffeomorphisms and then provide a classification of its orbit invariants, which
lift to invariants on the full hydrodynamical group.
In Section \ref{sec:corner_symmetry_group}, we then show how this classification can be extended to coadjoint orbits of the corner symmetry group. Sections \ref{sec:algebra} and \ref{sec:poisson} are dedicated to the realization of the corner symmetry group on the phase space of general relativity in a finite region. 
Section \ref{sec:algebra} relates the generators of the corner symmetry group to the normal geometry of the surface $S$, and we explain how the generators are sent
to corresponding objects in the coadjoint orbit via the moment map. Finally, we show how the natural symplectic structure on the coadjoint orbits is realized in terms of Poisson brackets on the gravitational phase space in Section \ref{sec:poisson}. 
Section \ref{sec:discussion} gives an extended discussion of the implications 
of the present work for classical and quantum gravity and describes several avenues for
future work.
Appendices \ref{app:fiber} and \ref{app:frame fields for the normal bundle} 
provide additional interpretational details that supplement the discussions in the 
main text.  In particular, the description of the symmetry group as the automorphism
group of a principle fiber bundle is described in Appendix \ref{app:fiber},
which in addition describes how the coadjoint orbits can be naturally constructed 
in terms of objects defined directly on the bundle.  
Appendix \ref{app:frame fields for the normal bundle} gives an alternative 
description of the normal bundle geometry in terms of frame fields, 
which allows additional objects such as the dressed momentum $\bar p$ defined 
in section \ref{lifting} to be interpreted on the gravitational phase space.  
The derivations of some key identities are presented in Appendix \ref{app:additional identities}.

\subsection{Conventions}
Here we summarize the conventions used throughout the paper. 
\begin{itemize}
    \item[$-$] Small Greek letters $\mu,\nu,\ldots$ are  used for spacetime indices.
    \item[$-$] Small letters $i,j,k,\ldots$ are indices for directions normal to $S$.
    \item[$-$] Capital letters $A,B,C,\ldots,$ are indices for directions tangential to $S$. 
    \item[$-$] Small letters $a,b,c,\ldots=1,2,3$ are $\slr$ Lie algebra indices. 
\item[$-$] We use a notation that distinguishes 2-form, its density and the corresponding scalar function that depends on a choice of measure on the sphere. Given a 2-form $\w$ on $S$ with components $w_{AB}$, we denote the corresponding density by $\wt{w}$ and the corresponding scalar by  $w$.
 This gives the following correspondence for an arbitrary 2-form $\w \in \Omega^2(S)$ 
\be
\w= \frac12 w_{AB}\,\rd \sigma^A \wedge \rd \sigma^B = \wt{w}\,\rd^2 \si
= w\,\n.
\ee
where 
\be 
\n := \sqrt{q} \rd^2 \sigma= \wt n \rd^2 \si,
\ee
It is important to keep in mind that the scalar $w$ depends on a choice of measure, but does not require a metric on $S$.
In the following, and with a slight abuse of notation, we will often use the density notation to denote the integral over $S$:
\be 
\bigintsss_S \widetilde{w}:=\bigintsss_S \w.
\ee 

\item[$-$] We use the following definition of the commutator of Lie-algebra-valued forms on $S$ \cite[Ch.\ IV, Sec.\ A.6]{Choquet2004}
 \begin{equation}
     [\mbs{\alpha},\mbs{\beta}]=
     \mbs \alpha^a\wedge \mbs \beta^b \, [\tau_a, \tau_b]^c \tau_c,
 \end{equation}
 where $\mbs{\alpha} = \mbs\alpha^a\tau_a$ and $\mbs{\beta}= \mbs\beta^b \tau_b$, and 
 $\tau_a$ are a basis for the Lie algebra. This bracket satisfies
 \begin{equation}
     [\mbs{\alpha},\mbs{\beta}]=(-1)^{pq+1}[\mbs{\beta},\mbs{\alpha}].
 \end{equation}
 
\end{itemize}

\section{Generalities of coadjoint orbits}\label{sec:generalities of coadjoint orbits}

We begin with a review of some relevant aspects of the method of coadjoint orbits,
focusing particularly on properties of orbits for semidirect products.
After describing the general theory, we discuss a key example: the Poincar\'e group. 
This example allows us to develop concepts and tools that exist for arbitrary semidirect products, including the infinite-dimensional ones we will consider in what follows. The analogy between coadjoint orbits of the Poincar\'e group is presented in Table 
\ref{figure:analogy}, and will be expanded upon in subsequent sections.

\begin{center}
\begin{table}[t!]
    \hskip-.95 cm
    \setlength{\arrayrulewidth}{1pt}
\begin{tabular}{|c|c|c|c|}
\cline{2-4}
    \multicolumn{1}{c|}{} &  Poincar\'e group &
    Hydrodynamical group &
    \makecell{Corner symmetry \\ group} \\
\hline
Physical system & Relativistic particle & Compressible fluid & Spacetime region \\
\hline
Group & $\SO(1,3) \ltimes \R^{1,3}$ &
$\Diff(S) \ltimes \R^S$ & 
$\Diff(S) \ltimes \SL(2,\R)^S$ \\
\hline 
Normal subgroup & $\R^{1,3}$ & $\R^S$ & $\SL(2,\R)^S$ \\
\hline 
Quotient group & $\SO(1,3)$ & $\Diff(S)$ & $\Diff(S)$ \\
\hline 
Normal character  & momentum $P_\mu$ & mass density $\tn$ & area form $\sqrt{q}$ \\
 \hline 
\makecell{Quotient group \\ generator} & \makecell{angular momentum \\ $J_{\mu \nu}$} & \makecell{momentum density \\ $\tilde{p}_A$} & \makecell{ twist density \\ $\tilde P_A$ }\\
\hline 
\makecell{Homogeneous \\ orbit label} & \makecell{ mass \\ $m^2 = -P^2$} & \makecell{total mass \\ $M = \int_S \tilde n$} & \makecell{total area \\ $\mathcal{A} = \int_S \sqrt{q} $} \\
\hline 
Little group & $\SO(3)$ & SDiff($S$) & SDiff($S$) \\
\hline 
\makecell{Little group \\ generator } & \makecell{Pauli-Lubanski \\ pseudovector \\
$W^\mu = \tfrac12 \epsilon^{\mu \nu \rho \sigma} P_\nu J_{\rho \sigma}$
} & \makecell{vorticity  \\ $\mbs{w} = \rd p$} &
\makecell{outer curvature \\    $W:= \rd P - \frac{1}{2}\varepsilon_{abc}N^a\rd N^b\wedge \rd N^c$ 
 } \\
\hline \makecell{Little group \\ invariant} & 
\makecell{ total spin \\ $W^\mu W_\mu = m^2 s(s+1)$ } & 
\makecell{generalized \\ enstrophies \\ $C_k = \int_S \tn w^k$} &
\makecell{outer curvature \\ moments \\ $\bar C_k = \int_S \sqrt{q}\,W^k$} \\
\hline 
\end{tabular}
\caption{
The analogy between classification of coadjoint orbits for the Poincar\'e group (Section \ref{subsec:coadjoint orbits of the poincare group}), the symmetry group $\Diff(S) \ltimes \R^S$ of compressible hydrodynamics (Section \ref{sec:coadjoint orbits of hydrodynamical group}) and our corner symmetry group (Sections \ref{sec:corner_symmetry_group} and \ref{sec:algebra}).
In each case we have a physical system which naturally realizes the symmetry, and functions on the coadjoint orbit are identified with physical quantities.
In the case of the corner symmetry group, we have used the notation associated with the gravitational phase space, which are related to the coadjoint orbit by the moment map described in section \ref{sec:moment_map}.
Fleshing out this analogy is a major result of this work and will occupy a large portion of sections \ref{sec:generalities of coadjoint orbits}-\ref{sec:algebra}.
} 
\label{figure:analogy}
\end{table}
\end{center}

\vspace*{-1.2cm}

\subsection{Coadjoint actions and coadjoint orbits}\label{subsec:coadjoint action and coadjoint orbits}
Given a group $H$, we denote its Lie algebra $\mfk{h}$, with elements $\xi \in \mfk{h}$, its dual Lie algebra by $\mfk{h}^*$, with element $p\in \mfk{h}^*$.
The adjoint action of $H$ on $\mfk{h}$ is defined by 
$\Ad_h(X)= h X h^{-1}$, for $h\in H$ and $X\in \mfk{h}$.
The coadjoint action of $H$ on $\mfk{h}^*$ is defined by the pairing\footnote{The convention 
is chosen so that $\Ad_h^*$ defines a left action,  $\Ad^*_h\Ad^*_{h'}=\Ad^*_{hh'}$.}
\be \label{addef}
\langle \Ad^*_h(p), X \rangle =
\langle p, \Ad_{h^{-1}}(X) \rangle,\quad p\in \mfk{h}^*, X \in \mfk{h},  h\in H.
\ee

A central aspect of quantum representation theory is the fact that it is possible to draw a correspondence between a  subset of coadjoint orbits for a  group and the set of irreducible unitary representation of the same group.
The subset is selected by imposition of some  integrality conditions on the coadjoint orbit.
This  powerful correspondence has been established rigorously in a  large number of cases, 
including for nilpotent groups \cite{Kirillov196202}, compact and noncompact semisimple groups
 \cite{Rossman}, and even some infinite dimensional groups \cite{Frenkel1984}.
In our work we are interested in using this correspondence to  study the set of coadjoint orbits associated to the corner symmetry group.

Given an element $p_0 \in \mfk{h}^*$ its coadjoint orbit $\mcal{O}^H_{p_0}$ is given by 
\be 
\mcal{O}^H_{p_0} =\{ p_0= \Ad^*_h(p_0)\in \mfk{h}^*, \forall h \in H\}.
\ee 
The coadjoint orbit is isomorphic to 
$H/H_{p_0}$ where $H_{p_0}$ is the isotropy group
\be
H_{p_0}:=\{ k \in H|\, \Ad^*_k(p_0)=p \}.
\ee

\subsection{The canonical symplectic form on coadjoint orbits}\label{subsec:the canonical symplectic form on coadjoint orbits}

A central result of Kirillov, Kostant, and Souriau is that coadjoint orbits are symplectic manifolds \cite{Souriau1970,Kostant1965,Kostant1970,kirillov2004lectures,Marsden-Ratiu,GuilleminSternberg1977,GuilleminSternberg1980}. The symplectic structure on $\mcal{O}^H_{p_0}$ is given by
\be\label{sympdef}
\omega_{p_0} = \frac12 \langle p_0, [h_p^{-1}\rd h_p , h_p^{-1}\rd h_p ]\rangle, \quad\mathrm{with}\quad\Ad_{h_p}(p_0) =p,
\ee
which can be written equivalently as
\be\label{sympdef2}
\omega_{p} = \frac12 \langle p, [\rd h_p h_p^{-1} , \rd h_p h_p^{-1} ]\rangle, \quad\mathrm{with}\quad\Ad_{h_p}(p_0) =p.
\ee
The right-hand side of \eqref{sympdef}  is invariant under the transformation $h\to h k$ with $k \in H_p$ and therefore depends only on the orbit element $p$ and not on 
 on the choice of orbit representative
$h_p$.
 This means that $\omega_{p_0} \in \Omega^2(\mcal{O}^H_{p_0})$. The fact that this form is closed follows by a direct use of Jacobi identity. The fact that it is invertible is also direct to establish: Given $X,Y\in \mfk{h}$ we have 
\be 
\omega_{p_0}(X,Y):= \frac12 \langle p_0 ,[X,Y]\rangle = 
-\frac12 \langle \mathrm{ad}^*_X(p_0),Y\rangle.
\ee 
Demanding that this vanish for all $Y$ means that $X \in \mfk{h}_{p_0}$, hence 
$[X]$ vanishes as a tangent vector to $\mcal{O}^H_{p_0}$.

\subsection{Coadjoint orbits of the Poincar\'e  group}\label{subsec:coadjoint orbits of the poincare group}
The Poincar\'e group is a semidirect product
$G= H \ltimes N$ with homogeneous subgroup 
$H=\SO(1,3)$ the Lorentz group and normal subgroup $N = \mathbb{R}^{1,3}$, the translation group.
The main classification theorem states that the nondegenerate coadjoint orbits of the Poincar\'e group are labelled by the mass and spin, $(m,s)$.
The coadjoint orbit itself represents the phase space of a relativistic spinning particle.
The mass represents the choice of a Lorentz orbit inside $\mathbb{R}^{1,3}$, the mass-shell orbit,
while the spin represents the choice of rotation subgroup orbit.
The rotation group  $\SO(3)$ appears naturally as the \emph{little group}, i.e., the subgroup of the Lorentz group that fixes a given element of the mass-shell orbit.

Let us now delve further into the details of the construction.
The Lie algebra dual elements are given by a pair of momentum and angular momentum $(J_{\mu \nu}, P_\nu)$.
The angular momentum labels an element of the Lorentz dual, $J\in \mfk{h}^*$, while the momentum labels an element in the dual of the translation algebra.
The coadjoint action of $(h,x)\in \SO(1,3)\ltimes \R^{1,3} $ on a given element $(J_0,P_0)$ is explicitly given by 
\be\label{coadjoint-orbit-P}
J_{\mu \nu} =   h_{\mu}{}^{\rho}  h_{\nu}{}^{\sigma} J_{0 \rho \sigma}  +  P_{[\mu} x_{\nu]}, \qquad 
P_\mu= h_\mu{}^{\rho} P_{0\rho} .
\ee
The first step in the classification of the coadjoint orbit is to identify the possible homogeneous orbits, $\mcal{O}^{\SO(1,3)}_{P_0}:=\{P =  h\cdot  P_{0}\}$, i.e., the orbits of $P_0$ under the Lorentz group. 
These are the different mass shells. 
There are 4 different classes of orbits depending on the value of the mass $m^2=-P^2$: the massive orbits $m^2>0$, the tachyonic orbits with $m^2<0$, the massless orbits $m^2=0$, and the trivial orbit $P_0=0$.

We are interested in the description of the massive orbit. 
In this case, we can write $P= m n$, where $n$ is an element of the unit hyperboloid.
Any element in the hyperboloid can be obtained from the Lorentz action on a fixed representative vector $n_0$. We take $n_0=(1,0,0,0)$, which allows us to represent the  unit hyperboloid as the orbit $\mcal{O}^{\SO(1,3)}_{n_0}$.
Given an element $n$ in the hyperboloid, we define its little group, denoted $\SO(3)_{n}$, to be the subgroup  of Lorentz transformations fixing $n$.
The different little groups associated to different points on the orbits are conjugate to each other 
$\SO(3)_{h\cdot n_0} = h\SO(3)_{ n_0}h^{-1}$,
and we denote the subgroup fixing $n_0$ simply by $\SO(3)$.
This means that the massive orbit is a homogeneous space,
\be
\mcal{O}^{\SO(1,3)}_{n_0} = \SO(1,3)/\SO(3).
\ee

Given a point $n$ on the unit mass shell, we can consider its isotropy subgroup in Poincar\'e. 
This is the subgroup
\be
G_{n}= \SO(3)_n\ltimes \R^{1,3},
\ee
which contains both the little group and the translation group.
The goal is now to construct the invariant associated with the action of this subgroup on the angular momentum.
First, we deal with the action of translations, which are controlled by the orbital angular momentum
\be 
[L_P(x)]_{\mu \nu} := P_{[\mu} x_{\nu]}.
\ee 
In order to construct an orbit invariant, one needs to construct from $J$ an operator which is invariant under translation. 
This map, denoted $S_P$, is the celebrated Pauli-Lubanski spin observable. It is given by\footnote{The notation distinguishes the spin map $S_P:\mfk{so}(3,1)^* \to \mfk{so}(3)_n^*$ from the Pauli-Lubanski spin itself
$W:(\mfk{so}(3,1)\ltimes \mathbb{R}^{1,3})^* \to \mfk{so}(3)_n^*$} 
\be 
W^\mu(P,J) = [S_P (J)]^\mu =  \frac12\epsilon^{\mu \nu \rho \sigma} P_\nu J_{\rho \sigma}.
\ee
The Pauli-Lubanski spin map  possesses two essential properties that guarantee the success of the construction. 
First, it is a translation invariant, as follows from the fact that 
\be 
\mathrm{Ker}(S_P)=\mathrm{Im}(L_P),
\ee 
which can be directly checked. 
This  means that the orbital angular momentum does not contribute to the spin. 
Moreover, the Pauli-Lubanski spin map is covariant under the action of the little group: $S_P(k J k^{-1})= h\cdot S_P(J)$ for $k \in \SO(3)_{n}$, and the Pauli-Lubanski spin $W^\mu$ transforms as a  pseudovector under  Lorentz transformations:
$h\cdot W(P,J)= W(h \cdot P, hJh^{-1})$ for $h\in \SO(1,3)$.

These two properties mean that the component of $J$ which is translationally and rotationally invariant --- that is, invariant under the isotropy subgroup $G_n$ --- is given by the spin $s$, where 
$m s= |W|$.
More formally, the spin $s$ really labels a little group orbit $\mcal{O}^{\SO(3)}_{W_0}$,
where we denote the little group representative by $W_0:=S_{P_0}(J_0)$. 
The little group orbit is a homogeneous space 
\be 
\mcal{O}^{\SO(3)}_{W_0} = \SO(3)/\SO(2).
\ee
This is a sphere of radius $s$, which represents the classical phase space associated with the spin.

Returning to the description of the orbit, we can use the action of translation to chose a representative where $J_0^{\mu \nu} n_{0\nu}=0$ and the action of rotation to choose a representative  $W_0$ which points in a fixed direction.
Overall, this means that the massive spinning orbit representative can be taken to be \be
J_{0\mu \nu}= s \delta_{[\mu}^1\delta_{\nu]}^2,\qquad P_{0\mu} = m \delta_\mu^0, \qquad W_0^\mu = ms \delta^\mu_3.
\ee
The isotropy group of the Poincar\'e orbit is the subgroup that fixes $P_0$ and $J_0$. It is the group
\be 
G_{(J_0,P_0)}= \SO(2)\times \R,
\ee 
where $\SO(2)$ is the subgroup of rotation that fixes $W_0$ and $\R$ is the subgroup of translation that fixes the reference orbital momenta $L_{P_0}$. 
This is the set of translations along $P_0$ or time translation.
This shows that the Poincar\'e orbit is the homogeneous space $G/G_{(J_0,P_0)}$.

The last element of the construction is the description of the symplectic structure for the spinning relativistic particle.  This symplectic structure is the sum of two terms
\be\label{symp1}
\omega_{(m,s)}
= m\, \rd x^\mu \wedge  \rd n_\mu +
 \frac{s}2  [h^{-1} \rd h,h^{-1} \rd h]_{12},
\ee
with $n= h \cdot n_0 $ and $[\cdot,\cdot]_{12}$ denotes the commutator's matrix element  along $J_0$.
The first factor is the canonical structure on $T^*\mcal{O}^{\SO(1,3)}_{n_0}$, which descends from the canonical structure on the cotangent bundle $T^*\R^{1,3}$.
The second factor reduces, when $h \in \SO(3)$, to the canonical symplectic structure on the unit sphere multiplied by the spin.
\eqref{symp1} shows that the relativistic particle phase space structure can be obtained by symplectic induction from the symplectic structure of the little group orbit (the sphere). 
This structure is a semiclassical analog of the construction of irreducible representations of a semidirect product by induction  \cite{Mackey:1978za,Piard197702}.

\section{Coadjoint orbits of the hydrodynamical group}\label{sec:coadjoint orbits of hydrodynamical group}

In the next section we will show that the coadjoint orbits of corner symmetry group can be reduced to those of the so-called hydrodynamical group by a symmetry breaking.
As such, we first discuss the coadjoint orbits of hydrodynamical group in this section, before turning to the orbits for the full corner symmetry in the following section.

\subsection{Hydrodynamical group}\label{subsec:hydrodynamical group}

The hydrodynamical group is defined to be the semidirect product
\be\label{gh}
G_\R(S) = \Diff(S)\ltimes \R^S,
\ee 
where $\mathbb{R}^S = C^\infty(S)\equiv C(S)$  denotes the space of real functions on the sphere, and the diffeomorphism action on it  is by pullback 
$g^*\phi =  \phi\circ g $ for $g\in \Diff(S)$ and $\phi \in C(S)$.
This group arises naturally as a subgroup of the corner symmetry group, but it is simpler because its normal subgroup is abelian.
It therefore constitutes an essential example for us. Such a group has also appeared as a symmetry group for gravity in the study of null surfaces \cite{Chandrasekaran:2018aop}.
A similar group (where the abelian factor represents half densities) appears as the extended BMS group, which is the symmetry group of asymptotically flat gravity \cite{Campiglia:2014yka, Penna201703, Compere:2018ylh}.  

The group \eqref{gh} has an important physical application: it is the symmetry group  of an ideal barotropic\footnote{A barotropic fluid is a compressible fluid whose pressure is a function of the density only, while for a general compressible fluid the pressure also depends on the entropy.} fluid \cite{MarsdenRatiuWeinstein1984a,Khesin-barotropic, arnold1999topological,Morrison}.
In the fluid dynamical context, the generator of $\Diff(S)$ is the fluid momentum, while the generator of $\R^S$ is the fluid density.
This symmetry group belongs to a larger class of symmetry group called Euler-Poincar\'e hydrodynamical groups, which are symmetry  groups of compressible perfect fluids \cite{Arnold,HolmMarsdenRatiu1998,Holm200103,khesin2020geometric}. 

The coadjoint orbit of $G_\R(S)$ is the phase space for a barotropic fluid in much the same way as a coadjoint orbit of the the Poincar\'e group is the phase space of a relativistic particle. 
We find that the coadjoint orbits of the hydrodynamical groups are labelled by the total mass of the fluid and by the fluid vorticity $w$.
The set of homogeneous orbits $\mathcal{O}^{\Diff(S)}_\nu$, where $\nu$ is a volume form on $S$, are labelled by the total mass of the fluid; these orbits are the analogs of the mass shells.
The little group preserving each volume form $\nu$ is the group of area-preserving diffeomorphisms (this group also arises naturally in \emph{in}compressible hydrodynamics as the symmetry group of the Euler equations \cite{Arnold,arnold1999topological}).

Let us now delve into the detailed construction.
The group law of $G_\R(S)$ is given by\footnote{To 
be consistent with the action on functions via pullbacks, the diffeomorphism group
multiplication must be defined by $h h' = (h' \circ h)$, which is opposite
to the usual definition.  This ensures that the Lie algebra is given by the Lie
bracket of vector fields as in (\ref{commutator1}).
The standard group composition law for the diffeomorphism group leads to a Lie 
bracket which is minus the vector field bracket \cite{Milnor1984}.} 
\be
(h,x)(h',x')= (hh', x + h^*x'),
\ee
with $h,h'\in \Diff(S)$ and $x,x' \in C(S)$.
In the following we will refer to $C(S)$ as the subgroup of translations. 
The Lie algebra $\mfk{g}_\R(S)$ consists of pairs $(\xi,\alpha)$ where $\xi = \xi^A(\sigma) \partial_A$ is a vector field on $S$ and $\alpha \in C(S)$ is a function on $S$.
The commutator is given by 
\begin{equation} \label{commutator1} 
[(\xi,\alpha), (\eta,\beta)] = \left([\xi,\eta]_\text{Lie}, \Lie_\xi\beta - \Lie_\beta \alpha \right).
\end{equation}
where $[\xi,\eta]_\text{Lie}^B = \Lie_\xi \eta^B = \xi^A \partial_A \eta^B - \eta^A \partial_A \xi^B$ denotes the Lie bracket of vector fields, and the Lie derivative acts on the scalar functions $\alpha$ and $\beta$ as $\Lie_\xi \beta = \xi^A \partial_A \beta$.

The coadjoint representation consists of pairs $(\tilde p, \wt{n})$ where $\tilde p = \tilde p_A(\sigma) \rd \sigma^A$ is a covector density\footnote{ The tilde is here to emphasize the density weight.} on $S$ which represents the fluid momentum, and $\wt{n}$ is a scalar density on $S$ which represents the mass density.
The adjoint and coadjoint representations have a natural pairing
\begin{equation} \label{pairing1}
\big \langle (\tilde p , \wt{n}), (\xi,\alpha) \big \rangle = \int_{S} (\tilde p_A \xi^A + \tn \alpha).
\end{equation}
This pairing allows us to define the action of the coadjoint representation. 
The action of the generator $(\xi,\alpha)$ on $(\tilde p, \wt{n})$ is denoted $(\xi,\alpha) \vartriangleright (\tilde p,\wt{n})$ and is defined by the relation
\begin{equation}
    \big \langle (\xi,\alpha) \vartriangleright (\tilde p,\wt{n}), (\eta,\beta) \big \rangle = - \big \langle (\tilde p, \wt{n}), [(\xi,\alpha), (\eta,\beta)] \big \rangle
\end{equation}
for all adjoint vectors $(\eta,\beta)$.
Using the definitions \eqref{commutator1} and \eqref{pairing1} and integrating by parts, we obtain 
\begin{equation} \label{coadjoint1}
(\xi, \alpha) \vartriangleright (\tilde p, \wt{n}) = ( \Lie_\xi \tilde p +  \wt{n}\rd \alpha, \Lie_\xi \wt{n}).
\end{equation}
In coordinates, the quantities appearing in \eqref{coadjoint1} are as follows: $\Lie_\xi \tilde p_B = \xi^A \partial_A \tilde p_B + (\partial_B \xi^A) \tilde p_A + (\partial_A \xi^A) \tilde p_B$ is the Lie derivative of a covector density, $(\wt{n}\rd \alpha )_A = \wt{n}\partial_A \alpha$ is the product  of the one-form $d \alpha$ with the density $\wt{n}$, $\Lie_\xi \wt{n}= \xi^A \partial_A \wt{n}+ (\partial_A \xi^A) \wt{n}$ is the Lie derivative of the density $\wt{n}$. 

The infinitesimal action \eqref{coadjoint1} can be exponentiated to obtain the coadjoint action of the group $\Diff(S) \ltimes \R^S$:
\be \label{fluid-coadjoint}
(h,x)\triangleright (\tilde{p},\wt{n}) = (h^*\tilde{p} + \wt{n}\rd x , h^*\wt{n})
\ee
where $h \in \Diff(S)$ and $x\in C(S)$ 
This is the analog of the finite coadjoint action \eqref{coadjoint-orbit-P} for the Poincar\'e group; the additional term $\wt{n}\rd x$ is analogous to the orbital angular momentum term in the Poincar\'e group.
Choosing an orbit representative $(\tilde{p}_0,\wt{n}_0)$ any element $(\tilde p, \wt{n})$ of the orbit can be obtained by the group action as $(\tilde{p}, \wt{n}) = (h,x) \triangleright (\tilde{p}_0, \wt{n}_0)$.

The first step in the classification of the coadjoint orbits is to identify the possible homogeneous orbits  $\mcal{O}^{\Diff(S)}_{\wt{n}_0}=\{\wt{n}=  h^* \wt{n}_{0}| h \in \Diff(S)\}$, which in this case are the possible orbits of a density under diffeomorphisms. 
If $\wt{n}$ is a generic density, not necessarily positive, we can split $S$ into three sectors as $S = S_+\cup S_0\cup S_-$ where $S_+$ denotes the set of points where $\wt{n}$ is strictly positive, $S_-$ where $\wt{n}$ is strictly negative and $S_0$ the set where $\wt{n}$ vanishes.
In this general case, orbit invariants will include topological invariants of the sets $S_\pm,S_0$ and the total density in each connected component of $S_\pm$.

We will focus our discussion on the analog of the massive orbits, which are those in which $\wt{n}>0$, 
corresponding to an everywhere-positive fluid density.
This is not only the physically relevant case for fluid dynamics, but will also be the the case of interest for us when
comparing to the corner symmetry group in section \ref{sec:corner_symmetry_group}.
In this case, we can appeal to Moser's theorem \cite{Moser1965}, which states that two scalar densities $\wt{n}$ and $\wt{n}_0$ with the same total mass $M = \int_S \wt{n}=\int_S \wt{n}_0$ can be transformed into one another by a diffeomorphism.
Thus, just as in the Poincar\'e group, the homogeneous orbits are labelled by the total mass of the fluid.
We choose the representative $\wt{n}_0$ to be the constant density on the round sphere of total mass $M$.
In other words, we pick 
\be
\wt{n}_0\rd^2\sigma  = \frac{M}{4\pi} \sin(\theta) \rd \theta \rd \phi.
\ee
In the following, we denote by $\SDiff(S)_{\wt{n}}$ the subgroup of diffeomorphisms preserving $\wt{n}$, and by $\sdiff(S)_{\wt{n}}$ its Lie algebra.

Having chosen the element of the orbit in which $\wt{n}= \wt{n}_0$,  we consider its isotropy subgroup $\SDiff(S)_{\wt{n}} \ltimes \R^S$.
The coadjoint orbit is then determined by the orbit of $\tilde p$ under this isotropy subgroup, with the coadjoint action given by
\begin{equation} \label{apd-coadjoint1}
(\xi, \alpha) \vartriangleright \tilde p = \Lie_{\xi} \tilde p +  \wt{n}\rd \alpha, \qquad  \Lie_{\xi}\wt{n}=0.
\end{equation}
We will see shortly that this orbit corresponds precisely to a coadjoint orbit of $\SDiff(S)_{\wt{n}}$.
The next step in classifying coadjoint orbits of $G_\R(S)$ is therefore to classify coadjoint orbits of $\SDiff(S)_{\wt{n}}$, which is the subject of the next section.

\subsection{Area-preserving diffeomorphisms}\label{subsec:area-preserving diffeomorphisms}

The group of area-preserving diffeomorphisms of the sphere has been extensively considered in fluid dynamics.
This group arises naturally in incompressible hydrodynamics as the symmetry group of the Euler equations \cite{Arnold,arnold1999topological}. 

The algebra $\sdiff(S)$ is given by divergenceless vector fields $\xi$
equipped with the vector-field Lie bracket.
The dual Lie algebra consists of covector densities, with the pairing given by
\begin{equation} \label{sdiff-pairing1}
    \langle \tilde p, \xi \rangle = \int_S \tilde p_A \xi^A.
\end{equation}
The action on $\tilde p$ is via the Lie derivative, $\xi \vartriangleright \tilde p = \Lie_\xi \tilde p$.
This pairing is, however, degenerate: since the divergence of $\xi$ vanishes, we have for any scalar function $\alpha$
\begin{equation}
\langle \wt{n}\rd \alpha, \xi \rangle = 0.
\end{equation} 
The dual space is therefore given by equivalence classes $[\tilde p]$ of densitized one-forms up to the equivalence relation $\tilde p \sim \tilde p +\wt{n}\rd \alpha$. 
This shows that  the coadjoint orbits of $\sdiff(S)$ are precisely the orbits of the action \eqref{apd-coadjoint1}.

To classify the coadjoint orbits of $\sdiff$, it will be useful to give an equivalent description of the Lie algebra in terms of scalar fields.
Any divergence-free vector field can be written in terms of a scalar field $\phi$, called its \emph{stream function}, as
\begin{equation}
\xi_\phi^B = \nu^{AB} \partial_A \phi
\end{equation}
where $\n^{AB} = \frac{\varepsilon^{AB}}{\wt{n}}$ is an antisymmetric tensor defined in terms of the antisymmetric Levi-Civita symbol $\varepsilon^{AB}$ with $\varepsilon^{01} = 1$.
The stream function is defined up to a shift by a constant zero mode $\phi \to \phi + c$, which can be fixed by demanding that $\int_S \wt{n}\phi = 0$.
In this representation, the algebra is defined via the Poisson bracket $\{ \cdot,\cdot \}_{\wt{n}}$ between functions
\begin{equation}
    \{ \phi, \psi \}_{\wt{n}} := \nu^{AB} \partial_A \phi \, \partial_B \psi,
\end{equation}
which reproduces the vector field Lie bracket due to the 
identity $[\xi_\phi, \xi_\psi] = \xi_{ \{\phi, \psi \}_{\wt{n}}}$.

Having expressed the $\sdiff$ generators in terms of the scalar stream functions, 
the coadjoint representation can similarly be defined using these scalars.  
The pairing between the adjoint and coadjoint vectors is then given by
\begin{equation} \label{sdiff-pairing}
\langle \wt p, \xi_\phi \rangle = - \int \wt n (\nu^{AB} \partial_A p_B) \phi = -\int \w\,\phi = -\int \wt n \, w \, \phi.
\end{equation}
Here we have introduced the vorticity 2-form $\w = \rd p$, where $p = \frac{\tilde p}{\tilde n}$ is 
the de-densitized momentum, and the vorticity scalar $w = \tfrac12 \nu^{AB} \w_{AB}$.
Thus the coadjoint representation can be conveniently parametrized by the scalar vorticity $w$ with the pairing given by \eqref{sdiff-pairing}.
The coadjoint action of $\phi$ on $w$ is given by the Poisson bracket
\begin{equation} \label{sdiff-coadjoint-action}
\phi \vartriangleright w = \{\phi, w \}_{\wt{n}}.
\end{equation}

We have now reduced the problem of classifying coadjoint orbits of $\sdiff_{\wt{n}}$ to the problem of classifying orbits of a scalar function $w$ under the action \eqref{sdiff-coadjoint-action}.
From the scalar vorticity, we can immediately write down an infinite set of invariants, the Casimirs 
\begin{equation}
    C_k = \int_S \wt{n}\; w^k, \qquad k = 2,3,\ldots.
\end{equation}
These are known as the \emph{generalized enstrophies}.
Note that $C_1 = \int_S \tilde n w$ vanishes because $w$ arises
as the dual of an exact $2$-form, and hence must integrate 
to zero.  
The generalized enstrophies are not quite a complete set of invariants, so to give a complete classification of coadjoint orbits, a more refined invariant is required.
The problem of classifying coadjoint orbits of $\sdiff$ was solved for a certain class of generic orbits in Ref.~\cite{izosimov2016coadjoint}.
A complete invariant of the function $w$ under area-preserving diffeomorphisms is the \emph{measured Reeb graph} associated to $w$, whose construction we now describe.

Suppose that $w$ is a \emph{simple Morse function}, i.e., that the critical points of $w$ are isolated and all have distinct values.\footnote{Although the simple Morse functions are dense in the space of functions (in a sense defined in Ref.~\cite{izosimov2016coadjoint}), there can be certain degenerate cases not captured by this classification.
For example, the gradient of $w$ could vanish on a set of nonzero measure, or on a line.}
Let $\sim_w$ denote the equivalence relation on $S$ defined by $x \sim_w y$ if $x$ and $y$ lie on the same connected component of a level set of $w$.
The Reeb graph $\Gamma_w$ is defined as the quotient $S/\sim_w$ with the quotient topology,
and we let 
$\pi: S \to \Gamma_w$ denote the quotient map. 
As a topological space, $\Gamma_w$ is a graph, with each point on an edge or vertex corresponding
to a connected component of a level set of $w$.   
The topology of the graph reflects the way in which the level sets of $w$ split and merge as the value of the function is varied.

\begin{figure}[htb]
     \centering
     \begin{subfigure}[b]{0.25\textwidth}
         \centering
         \includegraphics[width=\textwidth]{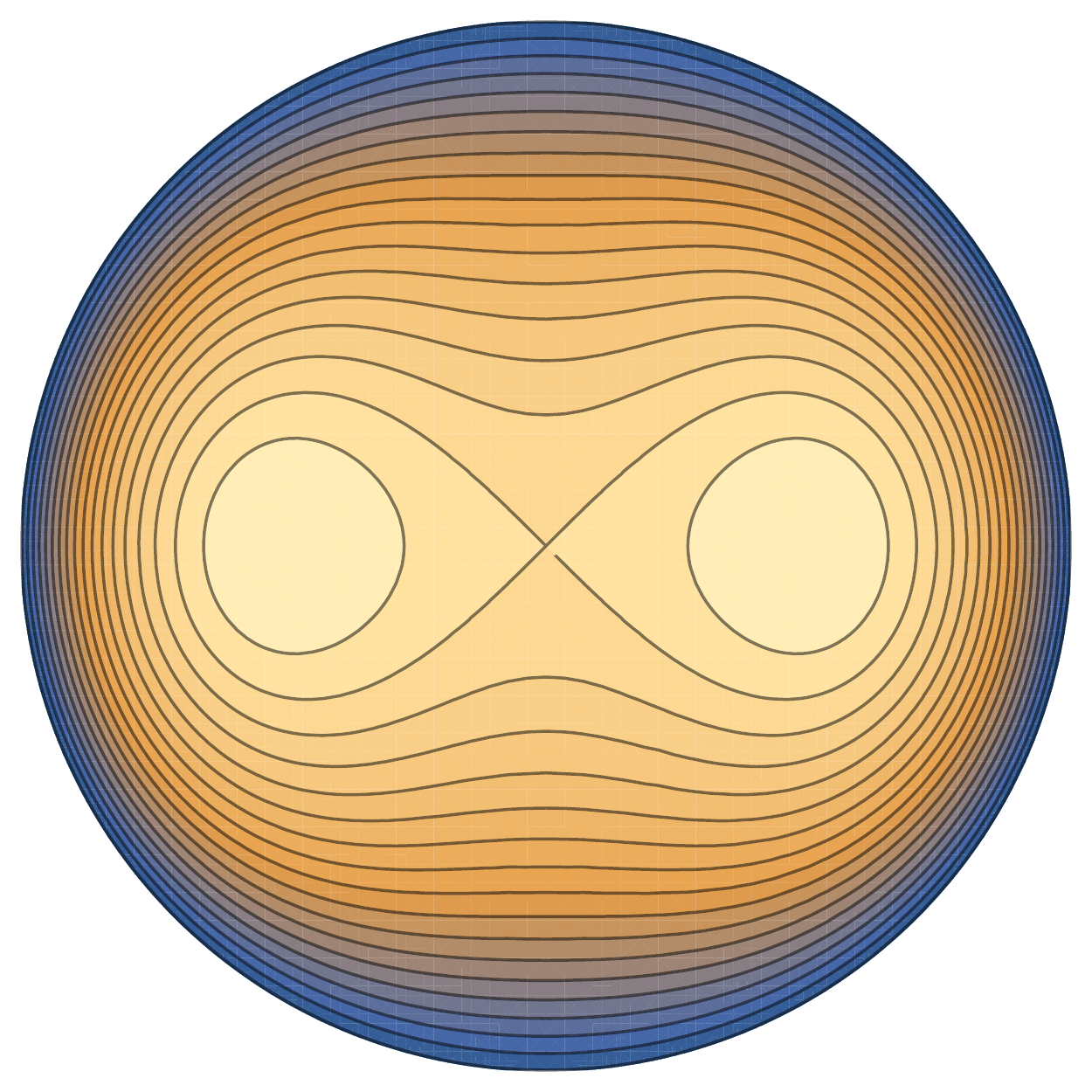}
         \caption{Contour plot of $f(u,v)$.}
     \end{subfigure}
     \hfill
     \begin{subfigure}[b]{0.4\textwidth}
         \centering
         \includegraphics[width=\textwidth]{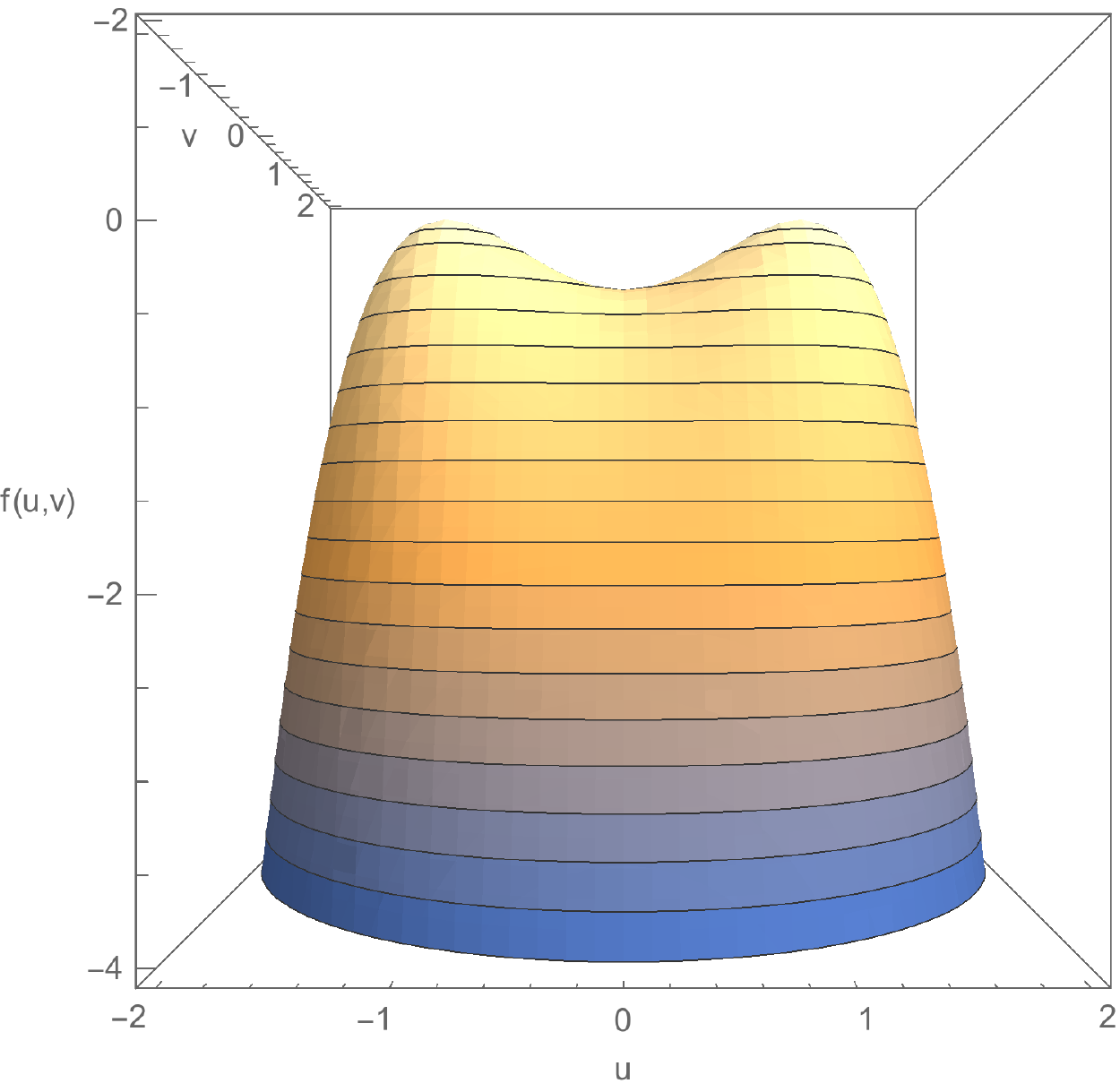}
         \caption{3D plot of $f(u,v)$}
     \end{subfigure}
     \hfill
     \begin{subfigure}[b]{0.25\textwidth}
        \centering
				\begin{tikzpicture}
				\begin{scope}[xshift=-2cm]
				\draw[<->] (0,-4.25) -- (0,1.25);
				\foreach \i in {-15,...,3}
				{
				  \draw (-0.05,\i/4) --  (0.05,\i/4);
				}
				\foreach \i in {-4,...,1}
				{
				  \draw (-0.1,\i) -- node[left]{\i} (0.1,\i);
				}
				\end{scope}
				
				\fill[black] (-0.5,0.5) circle (0.05);
				\fill[black] (0.5,0.5) circle (0.05);
				\fill[black] (0,0) circle (0.05);
				\fill[black] (0,-4) circle (0.05);
				\draw (0,0) -- node[left] {$e_1$} (-0.5,0.5);
				\draw (0,0) -- node[right] {$e_2$} (0.5,0.5);
				\draw (0,0) -- node[right] {$e_3$} (0,-4);
				
				\end{tikzpicture}
         \caption{The Reeb graph $\Gamma_f$.}
         
     \end{subfigure}
    \vspace{0.3cm}

          \begin{subfigure}[b]{\textwidth}
         \centering
         \includegraphics{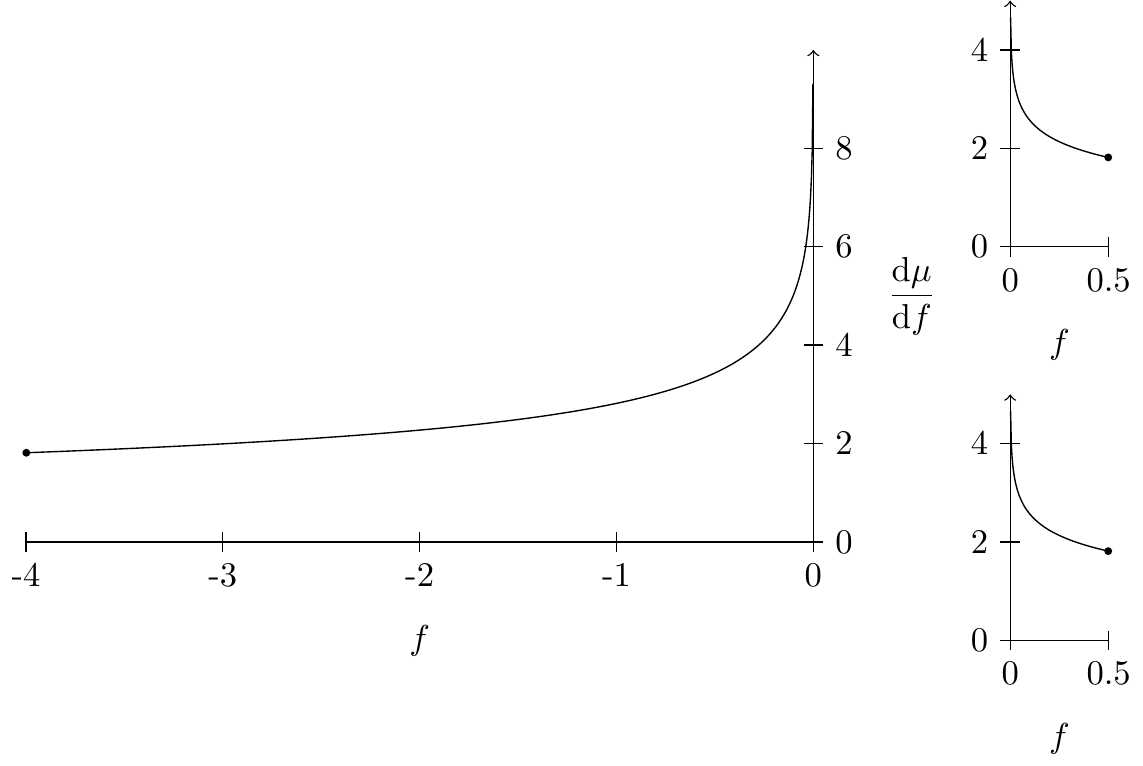}
         \caption{The density $\rd \mu/ \rd f$ of the measure on the Reeb graph: on the left is the measure on edge $e_3$, and on the right are the measures on $e_1$ and $e_2$ (which are equal by symmetry).}
     \end{subfigure}

  \caption{An example function $f$ \eqref{example-function}, its Reeb graph, and the measure on the Reeb graph.
  Figure (a) shows the contours of $f$ in the Lambert azimuthal equal-area coordinates. When plotted as a height function (b) we clearly see the saddle point and two maxima.
  The corresponding Reeb graph (c) shows the splitting of the orbits as the height is varied, going from a single orbit below the critical orbit to a pair of orbits above.
  The measure on the Reeb graph, whose density is displayed in (d), displays clearly the logarithmic singularities at $f = 0$ associated with the saddle point.
  }
  \label{figure:reeb_example}
\end{figure}
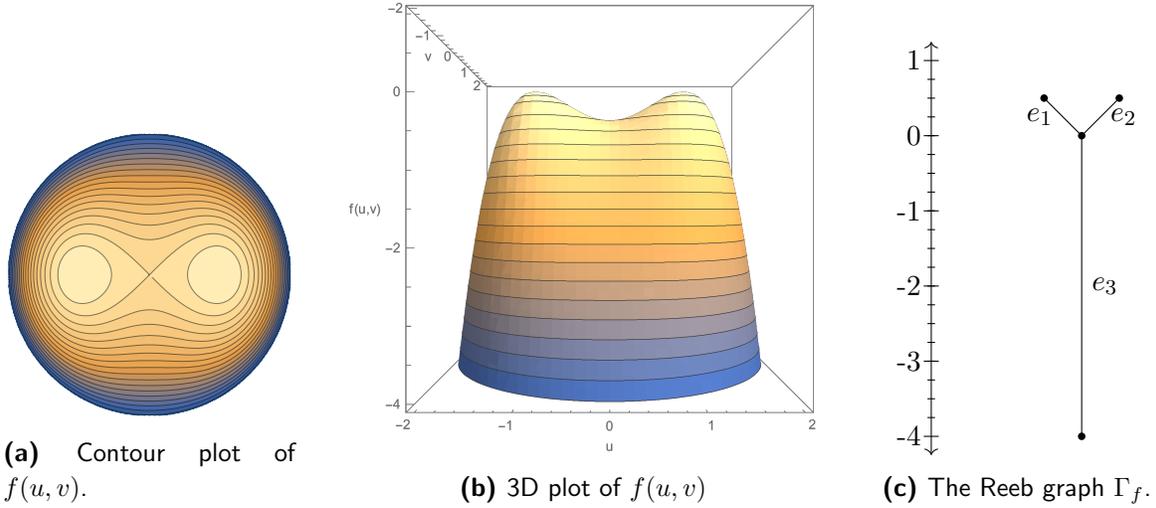
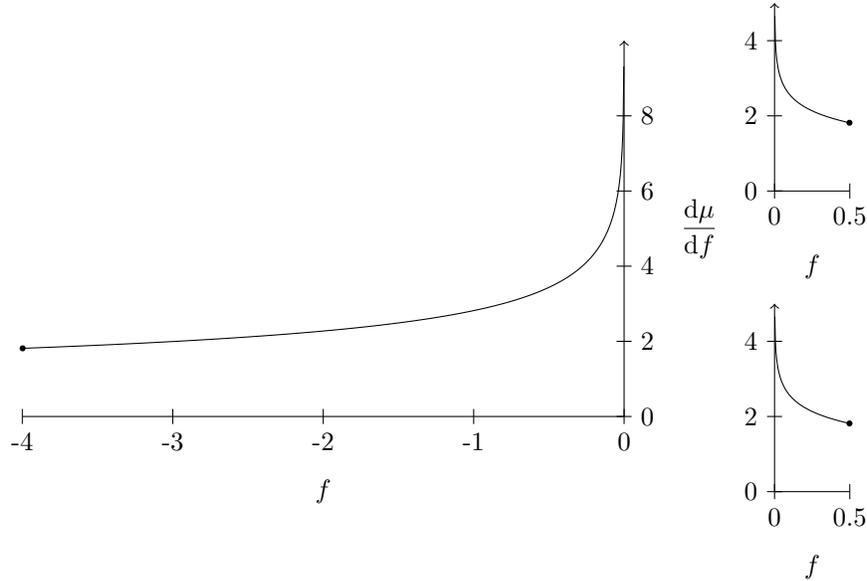

Vertices of the graph coincide with 
level sets that pass through critical points of $w$. 
When $w$ has a local maximum or minimum, the graph has a univalent vertex $v$ whose preimage 
under the projection $\pi$ is a single point.
When $w$ has a saddle point, $\Gamma_w$ has a trivalent vertex corresponding to the splitting or merging of the level sets of $w$.
The preimage of such a trivalent vertex takes the form of a figure-eight.
An illustrative example is shown in figure \ref{figure:reeb_example}, which will be explained further in an example below.

Since the construction of $\Gamma_w$ only refers to the structure of $S$ as a topological space, it is invariant under all diffeomorphisms of $S$.
When $S$ is a sphere, the Reeb graph of $w$ is a tree.\footnote{More generally, the number of vertices $v$ and number of edges $e$ of the Reeb graph are related to the genus of the surface $S$ by $v - e = 1 - g$.}
This illustrates why the generalized enstrophies are not sufficient to classify the coadjoint orbits.
By the Hausdorff moment problem \cite{Tagliani}, the generalized enstrophies allow for the reconstruction of the measure on the range of $w$, which is the sum of the measures associated to all edge of the Reeb graph.
As shown in Ref.~\cite{izosimov2017classification}, whenever there are two edges corresponding to the same range of values of $w$, the function $w$ can be modified so that the measures on the individual edges of the Reeb graph are changed but their sum remains the same.
This leads to the construction of a family of functions with different measured Reeb graphs, but the same generalized enstrophies.

The area form $\tn$ on $S$ defines a measure on $S$, which we can use to define a measure on the Reeb graph.
Given a subset $B$ of the Reeb graph, we define $\mu(B) = \int_{\pi^{-1}(B)} \tn$ by integrating over the preimage $\pi^{-1}(B) \subseteq S$ of the set $B$; this defines the pushforward of the measure $\tn d^2 \sigma$ under the map $\pi$.
This measure satisfies some additional properties: it is smooth along the edges of $\Gamma_f$ but has specified logarithmic singularities at trivalent vertices.\footnote{This feature is easily seen by noting that if we view $w$ as a Hamiltonian generating a flow via the Poisson bracket on the sphere, the points of $\Gamma_w$ are its orbits, and the measure on the Reeb graph is $T \rd w$, where $T$ is the period of the orbit. The logarithmic singularity in the measure corresponds to the way in which the flow slows down as the critical point is approached.}
A graph $\Gamma_f$ together with a measure satisfying these conditions is invariant under $\sdiff(S)$ and is called a \emph{measured Reeb graph}.
The central result of Ref.~\cite{izosimov2016coadjoint} is that the measured Reeb graph is a complete invariant of the coadjoint orbits of $\sdiff(S)$. 
As an alternative to the measure on the Reeb graph, one could equivalently specify the \emph{edge enstrophies}
\begin{equation}
C_k(e) =\int_{\pi^{-1}(e)} \tn \, w^k,
\end{equation}
where $e$ denotes a single edge of $\Gamma_w$. 
These carry exactly the same information as the measure on the Reeb graph, thanks to the Hausdorff moment problem.

\subsubsubsection{An example of a Reeb graph}
\label{sec:reeb_graph_example}
To see how the Reeb graph encodes invariant information of a function $f$, it is useful to introduce the Lambert azimuthal equal-area projection.
The coordinates $u,v$ are related to the embedding coordinates $(x,y,z) \in \R^3$ as:
\begin{equation}
x = \sqrt{1 - \tfrac14(u^2 + v^2)} \; u,\quad
y = \sqrt{1 - \tfrac14 (u^2 + v^2)} \; v, \quad 
z =  1 - \tfrac12(u^2 + v^2).
\end{equation}

\noindent The $u,v$ coordinates cover the disk $u^2 + v^2 < 4$, with the circle $u^2 + v^2 = 4$ mapping to the ``south pole'' $x = 0, y = 0, z = -1$.
This coordinate system has the feature that the natural volume form on the unit sphere is given by $\rd u \rd v$.

An illustrative example is the function\footnote{The curve $w=0$ belongs to a family of curves called lemniscates or hippopedes.}
 \begin{equation} \label{example-function}
     f = x^2 - y^2 - (1-z)^2 = u^2 - v^2 - \frac12 u^2(u^2+ v^2).
 \end{equation}
This function has maxima at $u = 0$, $v \pm 1$ where $f = -\tfrac12$, a minimum at the south pole where $f = -4$, and a saddle point at $u = 0, v = 0$ where $f = 0$.
While this function does not quite meet the criteria for a generic Morse function (the values of the function at the two maxima coincide) this poses no difficulties and can be resolved by adding a small perturbation $f \to f + \epsilon y$ at the expense of complicating the algebra.
The associated Reeb graph takes the shape of a letter ``Y'', with a single trivalent vertex corresponding to the figure-eight orbit passing through the saddle point.
The function $f$, its Reeb graph, and the measure on the Reeb graph are depicted in figure \ref{figure:reeb_example}.

\subsection{Construction of invariants}\label{subsec:construction of invariants}

Having classified the coadjoint orbits of the little group $\SDiff(S)$, we now show how this extends to a classification of coadjoint orbits of the hydrodynamical group $G_\R(S)$ in an invariant way, i.e. without resorting to fixing the density $\wt{n}$.

The action of the translation subgroup $C(S)$ on $\tilde{p}$ is controlled by 
 the \emph{moment map} 
\be  \label{fluid-moment-map}
\ell_{\wt{n}}(x): = \wt{n}\rd x.
\ee 
This is the analog of the orbital angular momentum in the Poincar\'e group.
The orbital angular momentum can be viewed as a map which, given an element $x$ of the translation group and an element $P$  of its Lie algebra, produces a Lorentz generator $P_{[\mu} x_{\nu]}$ as in \eqref{coadjoint-orbit-P}.
Similarly, \eqref{fluid-moment-map} can be viewed as a moment map $\ell: T^*  C(S) \to (\diff(S))^*$, $(\tilde n, x)\mapsto \ell_{\tilde n}(x)$, for the action of $\Diff(S)$ on the canonical phase space $T^*  C(S)$: given an element $x$ of the translation subgroup and a generator $\tilde n$, it produces a  generator $\tilde n \rd x$ in $\diff(S)^*$, as appears in \eqref{fluid-coadjoint}.

The moment map \eqref{fluid-moment-map} measures the transformation of the $\diff(S)^*$ element under translations.
To construct further invariants of the orbit, we need to identify the translation-invariant part of the $\diff$ generator.
This is analogous to the Pauli-Lubanski vector, which gives a translation-invariant part of the angular momentum and allows us to construct the spin.
To do this, we introduce the \emph{spin map}\footnote{In general the spin map is the dual of the little group inclusion map. Therefore its image is the dual of the little group \cite{Rawnsley197501,Baguis199705,Oblak201502}.}:
\be 
s_{\wt{n}}: (\diff(S))^* \to \Omega^2(S),\qquad 
\w := s_{\wt{n}} (\tilde{p}) = \rd \left(\frac{\tilde{p}}{\wt{n}}\right).
\ee 
In a fluid of density $\wt{n}$, the momentum density $\tilde{p}$ is related to the velocity $p$ by $\tilde{p} =\wt{n}p$. 
This means that $\w = s_{\wt{n}} (\tilde{p})$ is the fluid \emph{vorticity}
\be \label{vorticity}
\w =\rd p.
\ee
Since the sphere is simply connected, the moment map and the spin map satisfy the key property
\be 
\mathrm{Ker}(s_{\wt{n}})=\mathrm{Im}(\ell_{\wt{n}}).
\ee

We can now give an invariant description of the coadjoint orbit associated with a coadjoint vector $(\tilde p, \wt{n})$.
The first invariant we can construct is the total mass $M = \int_S \wt{n}$.
The vorticity  $\w = \rd p$ is covariant under diffeomorphisms, and transforms homogeneously under the coadjoint action as
\be
(g, x) \triangleright \w = g^* \w.
\ee
From this we can construct the scalar vorticity $w = \frac12 \nu^{AB} \w_{AB}$.
The pushforward of the measure $\tn$ to the Reeb graph of $w$ then defines a complete invariant of the orbit.
In particular, the measured Reeb graph encodes the Casimirs
\begin{equation}
C_k = \int_S  \tn \; w^k, \qquad k = 2, 3, \cdots,
\end{equation}
which can be interpreted as the generalized enstrophies in 
the fluid picture, and 
again $C_1$ vanishes since  
$w$ is the dual of an exact top form on $S$.
These data completely characterize the 
generic positive-mass coadjoint orbits of the group $G_\R(S)$.

\section{Coadjoint orbits of corner symmetry group}\label{sec:corner_symmetry_group}
In the previous section, we  studied the coadjoint orbits of the hydrodynamical group $G_{\mbb{R}}(S)$. 
This was a precursor for studying the coadjoint orbits of the corner symmetry group 
\be  \label{eqn:corner-sym}
G_{\SL(2,\R)}(S)= \Diff(S) \ltimes \SL(2,\R)^S,
\ee
where $\SL(2,\R)^S$ denotes the space of smooth functions $S \to \SL(2,\R)$ \cite{Donnelly:2016auv}. 
This is a semidirect product in which the normal subgroup $\SL(2,\mbb{R})^S$ is nonabelian.
As discussed in Appendix \ref{app:fiber}, this group 
appears naturally as the automorphism group of a trivial
$\SL(2,\R)$ bundle over $S$, and, in the case of general 
relativity, arises as  the symmetry group of the normal bundle 
of the $2$-surface $S$ embedded in spacetime. It turns out that the classification problem for coadjoint orbits of $\cog$ can be reduced to the one of the hydrodynamical group. Once this reduction is done, one needs to deal with the 
nonabelian nature of $\tenofo{SL}(2,\mbb{R})^S$.  However, the final result is similar to the one for the hydrodynamical group: the coadjoint orbits of $\cog$ are labelled by the total area, which plays the role analogous of mass for the Poincar\'e group, and invariants constructed from a modified vorticity, which
takes into account the additional noncommutativity of the 
normal subgroup. 
The correction to the vorticity associated with the $\SL(2,\R)^S$
factor can be interpreted physically as an internal component
that adds to the fluid vorticity. 

The adjoint representation of the corner symmetry algebra consists of pairs $(\xi,\alpha)$ where $\xi = \xi^A(\sigma) \partial_A$ is a vector field on $S$ and $\alpha = \alpha^a(\sigma) \tau_a$ is an $\slr^S$-valued function on $S$.
$\tau_a$ denotes a basis of the Lie algebra $\slr$.
For concreteness, we work with the explicit basis given by 
\begin{equation} \label{su11}
\tau_0= \frac{1}{2}
\begin{bmatrix}
0 & -1
\\
+1 & 0
\end{bmatrix}, 
\qquad
\tau_1 = \frac{1}{2} 
\begin{bmatrix}
+1 & 0
\\
0 & -1
\end{bmatrix}, 
\qquad
\tau_2 = \frac{1}{2} 
\begin{bmatrix}
0 & +1
\\
+1 & 0
\end{bmatrix}.
\end{equation}
These generators satisfy the defining relation 
\begin{equation}\label{eq:the product of taus}
    \tau_a \tau_b =  \frac14 \eta_{ab} \mathbbm{1} + \frac12 \varepsilon_{ab}{}^c \tau_c,
\end{equation}
where $\eta_{ab}=\mathrm{diag}(-1,1,1)$  is a Lorentzian metric 
and $\varepsilon_{abc}$ is the totally skew-symmetric tensor with $\varepsilon_{012}=1$.
$\slr$ indices $a,b,c,\ldots$ are raised and lowered with $\eta_{ab}$. The commutator is given by
\begin{equation} \label{commutator} 
[(\xi,\alpha), (\zeta,\beta)] = \left([\zeta,\eta]_{\mathrm{Lie}}, \Lie_\xi\beta - \Lie_\zeta \alpha + [\alpha,\beta]\right).
\end{equation}
Note that $[\xi,\zeta]_{\mathrm{Lie}}^B = \Lie_\xi \zeta^B = \xi^A \partial_A \zeta^B - \zeta^A \partial_A \xi^B$ denotes the Lie bracket of vector fields on $S$, while $[\alpha,\beta] = \varepsilon_{ab}{}^c \alpha^a \beta^b \tau_c$ denotes the pointwise Lie bracket of the $\slr$ algebra.
The Lie derivative acts on the scalar functions $\alpha$ and $\beta$ as $\mcal{L}_{\zeta}\alpha=\zeta^A\pa_A\alpha$ and $\Lie_\xi \beta = \xi^A \partial_A \beta$.

The coadjoint representation consists of pairs $(\wt p, \wt{n})$ where $\wt p = \wt p_A(\sigma) \rd \sigma^A$ is a covector density on $S$, and $\wt{n}= \wt{n}_a(\sigma) \tau^a$ is a density on $S$ valued in the coadjoint representation of $\slr^S$.
These have a natural pairing
\begin{equation} \label{pairing}
\big \langle (\wt p, \wt{n}), (\xi,\alpha) \big \rangle = \bigintsss_{S} (\wt p_A \xi^A + \wt{n}_a \alpha^a).
\end{equation}
In expressing $\tilde n$ as a matrix, we have adopted a pairing which is twice the trace: $\tilde n_a \alpha^a\equiv \tilde n \cdot \alpha = 2 \tr(\tilde n \alpha)$.
This pairing allows us to define the action of the corner symmetry group in the coadjoint representation. 
The action of the generator $(\xi,\alpha)$ on $(\wt p, \wt{n})$ is denoted $(\xi,\alpha) \vartriangleright (\wt p,\wt{n})$ and is defined by the relation
\begin{equation}
     \big \langle(\xi,\alpha) \vartriangleright (\wt p,\wt{n}), (\eta,\beta) \big \rangle = - \big \langle (\wt p, \wt{n}), [(\xi,\alpha), (\eta,\beta)] \big \rangle,
\end{equation}
for all adjoint vectors $(\eta,\beta)$.
Using the definitions \eqref{commutator} and \eqref{pairing} and integration by parts, we obtain the coadjoint action
\begin{equation} \label{coadjoint2}
(\xi, \alpha) \vartriangleright (\wt p, \wt{n}) = ( \Lie_\xi \wt p +  \rd \alpha \cdot \wt{n}, \Lie_\xi \wt{n}+ [\alpha,  \wt{n}]).
\end{equation}
In coordinates, the quantities appearing in \eqref{coadjoint2} are as follows: $\Lie_\xi \wt p_B = \xi^A \partial_A \wt p_B + (\partial_B \xi^A) \wt p_A + (\partial_A \xi^A) \wt p_B$ is the Lie derivative of a covector density, $(\rd \alpha \cdot \wt{n})_A = \partial_A \alpha^a \wt{n}_a$ is the contraction of the $\slr^S$-valued one-form $d \alpha$ with the $\slr^S$-valued density $\wt{n}$, $\Lie_\xi \wt{n}_a = \xi^A \partial_A \wt{n}_a + (\partial_A \xi^A) \wt{n}_a$ is the Lie derivative of a scalar density, and the last term $[\alpha, \wt{n}]_a = \varepsilon_{ab}{}^c \alpha^b \wt{n}_c$ is the pointwise coadjoint action of the $\slr^S$ Lie algebra. 
Note that $\wt p$ transforms nontrivially under a local $\slr^S$ transformation; this is analogous to the fact that in the Poincar\'e group the angular momentum transforms nontrivially under translations.
In this case $\wt p$ transforms like a component of a densitized connection; the geometric interpretation of this transformation rule will be expanded upon in Section \ref{sec:moment_map}.

For a complementary description of the corner symmetry group,
coadjoint representation, and pairing from the perspective
of the $\SL(2,\R)$ bundle on which it acts, see 
appendix \ref{app:fiber}.

\subsection{Orbit reduction}\label{subsec:orbit reduction}

In this section, we show by a symmetry breaking argument that
the classification of orbits for the corner symmetry group reduces to the classification of orbits for the hydrodynamical group, and then further to the classification of orbits for its little group, the group of area-preserving diffeomorphisms.
This shows that, similar to the hydrodynamical group, we can label the corner symmetry orbits by the total area $\mcal{A}$ and a vorticity form.
The vorticity form is related to the area-preserving diffeomorphism group.

The first step of the procedure is to restrict our analysis to representations in which the Casimir of $\slr^S$ algebra is positive:
\be\label{posdef}
\eta_{ab} \wt{n}^a\wt{n}^b > 0. 
\ee
These are analogous to the physical orbits of the hydrodynamical group in which the density is everywhere positive.\footnote{This assumption excludes singular or degenerate measures. 
While the continuity condition is absolutely natural at the classical level, it is possible that representations associated with a singular measures appear at the quantum level, as, for example, in loop quantum gravity, where surface areas are concentrated on a discrete set of punctures.
As discussed in \cite{Freidel:2020ayo}, the area quantization corresponds to a choice of representation associated with discrete measures.
    See also the discussion in section \ref{subsec:nonperturbative}.
    }
Geometrically, as shown in \cite{Donnelly:2016auv}, \eqref{posdef} corresponds to the condition that the measure  on the sphere $S$ is the measure associated with a nondegenerate Riemannian metric.
In that context, $|\wt n| = \sqrt{ \eta_{ab} \wt{n}^a \wt{n}^b}$ is the area form of $S$, so we refer to the representations satisfying \eqref{posdef} as \emph{positive-area} representations. For such representations, we  can  use the $\mathrm{SL}(2,\R)^S$ symmetry to fix the generator $\wt{n}_a$ at each point to be in a given direction. In particular we can fix $\wt{n}_a$ so that 
\begin{equation} \label{fixed-k1}
    \wt{n}_a =  (0, | \wt n| ,0).
\end{equation}
The subgroup of transformations that preserves this condition is then simply the hydrodynmical group \eqref{gh}, where the $\R^S$ factor is the diagonal subgroup of $\mathrm{SL}(2,\R)^S$ which preserves the condition \eqref{fixed-k1}.
The classification of the corner symmetry group therefore reduces to the classification of the hydrodynamical group done in Section \ref{subsec:hydrodynamical group}:
the orbits are classified by the total area $\mathcal{A} = \int_S |\wt{n}|$ and an orbit of the little group $\SDiff(S)$. However, there is one additional subtlety coming from the fact that the normal subgroup $\tenofo{SL}(2,\R)^S$ is nonabelian.
While the vorticity \eqref{vorticity} is invariant under the hydrodynamical group, we have to define a ``dressed'' vorticity to produce an invariant under the full $\SL(2,\R)^S$ group.

\subsection{Lifting invariants from the little group}
\label{lifting}

We have seen that the classification of coadjoint orbits for $G_{\SL(2,\R)}(S)$ reduces to the classification of orbits of its little group, the group of area preserving diffeomorphisms. We now have to show how to construct invariants of the full group from little group invariants.
Since the little group can be defined by a gauge fixing of the full symmetry group, we will solve this problem by constructing an invariant extension of the vorticity by unfixing the gauge.

Before proceeding, we need to find the coadjoint action of the group, which is obtained by exponentiating the infinitesimal action \eqref{coadjoint2}.
The diffeomorphism action can be trivially exponentiated; given diffeomorphism $F:S\to S$, we define
\be 
F \vartriangleright(\widetilde p, \wt{n})
= (F^*\widetilde p, F^*\wt{n}),
\ee
where $F^*$ is the pull-back action.
The $\slr$ coadjoint action can also be exponentiated: let $x \in \SL(2,\R)^S$ denote a normal group element. Then, its action is given by
\be  \label{finite-sl2r-transformation}
x\vartriangleright 
(\widetilde p, \wt{n}) =  (\widetilde p + x^{-1} \rd x \cdot \wt{n}, x \wt{n}x^{-1}).
\ee
Here, $x^{-1}\rd x:=\theta$ is the $\slr^S$-valued, left-invariant, Maurer-Cartan 1-form. The fact that these transformations compose correctly is the specified by the cocycle identity
\be
x_1\vartriangleright x_2\vartriangleright
(\widetilde p, \wt{n})
= 
(x_1 x_2)\vartriangleright (\wt p, \wt{n}),
\ee
The cocycle identity follows from
\be 
x_2^{-1} \rd x_2  \cdot \wt{n}+ 
x_1^{-1} \rd x_1  \cdot (x_2 \wt{n}x_2^{-1})
=  (x_1x_2)^{-1} \rd (x_1 x_2) \cdot \wt{n},
\ee 
which means that $\beta(x)= x^{-1} \rd x  \cdot n$ is a one-cocycle. For convenience, we can use the scalar density $|\wt{n}|$ to define de-densitized versions of the generators
\beq\label{de-densitized}
\wt p = |\wt{n}| p, \qquad \wt{n}^a = |\wt{n}| n^a, \quad a=0,1,2
\eeq
in terms of a one-form $p$ and $\slr$-valued scalar $n$ satisfying $n^a n_a= -n_0^2+n_1^2+n_2^2=1$.
Geometrically, $\Sl(2,\R)$ with the Killing metric $\eta_{ab}$ is isometric to Minkowski spacetime $\R^{1,2}$.
The $\slr$ elements satisfying $n^a n_a = 1$ form a 2-dimensional hyperboloid $H$ inside $\R^{1,2}$ of Lorentzian signature isometric to two-dimensional de Sitter space dS${}_2$. Under an infinitesimal $\slr^S$ transformation with parameter $\alpha$, the generators transform as
\be \label{variations I}
\delta_\alpha p = \rd \alpha \cdot n,\qquad
\delta_\alpha n =[\alpha,n].
\ee
These can be seen easily from \eqref{finite-sl2r-transformation} by writing $x=\exp(\alpha)\simeq 1+\alpha=1+\alpha^a(\si)\tau_a$. We can now choose a ``standard boost'', which is a choice of group element $n \to x_n$ such that
\be\label{boost}
x_n n x_n^{-1} = \tau_1 = \frac{1}{2} \begin{bmatrix} +1 & 0 \\ 0 & -1 \end{bmatrix}. 
\ee
One natural choice of standard boost is given by
\begin{equation}
x_n = \frac{1}{\sqrt{2 (1 + n_1)}} \begin{bmatrix} 1 + n_1 & -n_0 + n_2 \\ -n_0 - n_2 & 1 + n_1 \end{bmatrix}.
\end{equation}
This choice is not unique: given a standard boost $x_n$ which solves \eqref{boost}, we can construct another one by $x_n' = e^{f(\sigma) \tau_1}  x_n $  where $f(\sigma)$ is any real function on $S$.
In the process of lifting the little group invariants to the full orbit, 
we must demonstrate that 
the end result does not depend on the choice of standard boost.

Under an infinitesimal $\slr^S$ transformation with parameter $\alpha$, the standard boost transforms as:
\be \label{variations II} 
x_n^{-1} \delta_\alpha x_n = -\alpha + L_\alpha(n) n.
\ee
The variation of $x_n$ is determined by demanding that the condition \eqref{boost} is preserved under variation, which implies that $[x_n^{-1} \delta_\alpha x_n + \alpha, n] = 0$. $L_\alpha(n)$ is a function of $\alpha$ and $n$ depending on the choice of standard boost $n \to x_n$.
Taking a second variation of $L_\alpha(n)$ we see that is satisfies the cocycle condition:
\be\label{cocycle2}
L_{[\alpha,\beta]} = \delta_\beta L_\alpha - \delta_\alpha L_\beta.
\ee
A derivation of this identity is given in Appendix \ref{app:cocycle-identity}.
A change of standard boost $x_n \to x_n' = e^{f(\sigma) \tau_1} x_n  $ implies a transformation\footnote{This can be seen easily by first using \eqref{variations II} for $e^{f(\sigma)\tau_1}x_n$ and then using \eqref{boost} to replace $\tau_1x_n$ with $x_nn$.}
\be
L_\alpha(n)\to L'_\alpha(n) = L_\alpha(n) +\delta_\alpha f.
\ee
The additional term is a trivial cocycle transformation which also solves \eqref{cocycle2}.

Given a choice of standard boost, we can use it to relate the momentum $p$ in an arbitrary $\slr^S$ frame to the momentum $\bar p$ in the frame \eqref{fixed-k1} in which $n_a$ is fixed.
We call the momentum in this frame the \emph{dressed momentum} and denote it with a bar.
From the transformation law for the momentum under a finite $\slr^S$ transformation \eqref{finite-sl2r-transformation}, we find the dressed momentum is given by:
\be\label{eq:the definition of dressed momentum}
 \bar{p} = p + x_n^{-1} \rd x_n  \cdot n
\ee
This depends on $n_a$ as well as the choice of standard boost $x_n$.
Under an infinitesimal $\slr^S$ transformation, $\bar p$ transforms as:
\bea
\delta_\alpha \bar{p} &=& 
\delta_\alpha p + \delta_\alpha(x_n^{-1}\rd x_n) \cdot n + 
x_n^{-1}\rd x_n \cdot \delta_\alpha n,\cr
&=& \rd \alpha \cdot n +  \rd (\delta_\alpha x_n x_n^{-1})  \cdot (x_n n x_n^{-1})
+ (x_n^{-1}\rd x_n) \cdot [\alpha, n],\cr
&=& \rd \alpha \cdot n -  \rd ( x_n \alpha x_n^{-1} )  \cdot (x_n n x_n^{-1})
+ [x_n^{-1}\rd x_n, \alpha] \cdot n + \rd L_\alpha( n),\cr
&=& \rd L_\alpha( n). \label{deltapbar}
\eea
We can now define the dressed vorticity\footnote{We use that 
$\rd n = [n, x_n^{-1}\rd x_n]$ and 
$\rd (x_n^{-1}\rd x_n) = -\frac12 [ x_n^{-1}\rd x_n,x_n^{-1}\rd x_n]$.
More details are given in Appendix \ref{app:vorticity}.}
\be \label{vorticity1}
\bar \w = \rd \bar p = \rd p + \frac12 [x_n^{-1} \rd x_n, x_n^{-1} \rd x_n]\cdot n.
\ee
which by \eqref{deltapbar} is invariant under local $\slr^S$ transformations. 
Since $\bar{p}$ represents the momenta associated with the hydrodynamical group $G_\R(S)$, the dressed vorticity is simply the associated fluid's vorticity. 

While the construction of $\w$ involves a choice of standard boost, it is in fact independent of such a choice.
Under a change of standard boost $x_n\to x'_n=e^{f(\si)\tau_1}x_n$, we have $x_n'^{-1}\rd x'_n = x_n^{-1} \rd x_n  + (\rd f) n$, under which \eqref{vorticity1} is invariant.
While the expression \eqref{vorticity1} is valid for an arbitrary group equipped with an invariant pairing, in the case of $\slr^S$, we can write $\bar{\w}$ directly in terms of $n$ making its independence of the standard boost manifest. 
As shown in Appendix \ref{app:vorticity}, the dressed vorticity can be expressed as 
\begin{equation}\label{mainomega}
\bar \w = \rd p - \frac12 \varepsilon_{abc}n^a \rd n^b \wedge \rd n^c,
\end{equation}
When $n_a$ is fixed to a constant as in \eqref{fixed-k1}, $\bar \w$ reduces to the vorticity in the hydrodynamical subgroup.
Moreover, $\bar \w$ transforms as a $2$-form under diffeomorphisms.
This means we can define a complete invariant of the orbit by defining the scalar dressed vorticity 
\begin{equation}\label{star} 
\bar w = \tfrac12 \nu^{AB} \bar w_{AB}.
\end{equation}
The measured Reeb graph associated with the function $\bar w$, together with the total area $\mcal{A}$, defines a complete invariant of the coadjoint orbit of the group $G_{\SL(2,\R)}(S)$.
In particular the generalized enstrophies:
\beq\label{eq:the definition of generalized enstropy in terms of dressed vorticity}
\bar{C}_k = \bigintsss_S |\wt{n}| \,  \bar w^k, \qquad k=2,3,\cdots,
\eeq
are Casimirs of $G_{\SL(2,\R)}(S)$.

Just as in the case of the hydrodynamical group, the first 
generalized enstrophy $\bar C_1$ vanishes for the 
corner symmetry group orbits, since $\bar w$ is 
realized as the dual of an exact form $\rd\bar p$ according
to equation (\ref{vorticity}).   Note, however, the 
expression for the dressed vorticity given by (\ref{mainomega})
is not obviously exact, and it is only through the expression
(\ref{boost})
of $n$ in terms of a standard boost from a constant configuration
that we can determine that the second term in (\ref{mainomega})
is also exact.  The existence of such a constant configuration
for $n$ depends crucially on the fact that the corner 
symmetry group (\ref{eqn:corner-sym}) is a semidirect product,
as opposed to a nontrivial group extension of $\Diff(S)$.  
From the fiber bundle perspective developed in appendix 
\ref{app:fiber}, such a constant configuration for $n$ defines
a section of the principal $\SL(2,\R)$ bundle associated with 
the symmetry group, and hence implies that the bundle is trivial.  
If we had instead allowed for nontrivial bundles, the surface
symmetry group would be modified to a nontrivial extension,
and $\bar C_1$ would no longer vanish; rather, it would describe a 
topological invariant of the bundle.  

\section{Corner symmetries in general relativity}\label{sec:algebra}

Having studied the coadjoint orbits of the corner symmetry group, we turn now to a concrete realization of a phase space $\mathcal{P}$ possessing this symmetry.  
This phase space appears when partitioning a Cauchy slice in general relativity into spatial 
subregions, and the surface symmetries act on the edge modes that appear due to broken diffeomorphism invariance in the presence of a boundary.  We explicitly describe the generators of the corner symmetry
group on this phase space.  Furthermore, we construct the moment map
$\mu:\mathcal{P}\rightarrow \mathfrak{g}^*$, which is
simply the map that sends the phase space $\mathcal{P}$ to the 
space of coadjoint orbits \cite{kirillov2004lectures}, and hence provides the link between
our corner symmetry phase space and the orbits constructed in section \ref{sec:corner_symmetry_group}.
We further relate the Casimirs
to geometric characteristics of the surface, such as its area
and moments of the outer curvature of the normal bundle.  
The relation between this outer curvature and  
area-preserving diffeomorphisms is later discussed in section \ref{sec:vorticity}.  

\subsection{Description of phase space} 

The basic idea of \cite{Donnelly:2016auv} is to construct a gauge-invariant presymplectic form  for gravity in metric variables.\footnote{Different choices of variables, such as vielbeins, can lead to different surface symmetry algebras.  Note also that the  the choice of corner terms and boundary conditions can also affect the surface symmetries, see e.g. \cite{Freidel:2020xyx}.  In this work, we employ the covariant Iyer-Wald symplectic form \cite{Iyer1994a}.  }
Here,  gauge invariance means the invariance under all diffeomorphisms, including those 
which act at the boundary. 
This basic idea forces us to introduce new degrees of freedom on the codimension-$2$ corner $S$ bounding a Cauchy surface $\Sigma$
for the subregion. These new degrees of freedom are invariant under a new group $G_S$ of physical, as opposed to gauge, symmetries. 
They arise from the would-be gauge degrees of freedom, and necessarily appear once a boundary is introduced, which partially breaks the gauge symmetry. 
To characterize them locally, we introduce new fields $X^\mu$, $\mu = 0,\ldots, 3$, which can be thought of as coordinates describing the location of $S$ and its local neighborhood in the spacetime manifold $M$.\footnote{The construction of Ref.~\cite{Donnelly:2016auv} works for arbitrary spacetime dimension, but here we specialized to $D = 4$.} 
Abstractly, the fields $X^\mu$ together define a map $X:\mathbb{R}^4 \to M$.
We let $s$ denote a fixed $2$-dimensional surface in $\mbb{R}^{4}$, defining the location of $S$ by  $S=X(s)$. 
It was shown in \cite{Donnelly:2016auv} that the symplectic form for gravity can be modified in a canonical way by a boundary term that restores invariance under all diffeomorphisms.  It then takes the following form:
\beq \label{eqn:Om}
\Omega = \Omega_\Sigma + \Omega_S,
\eeq
where $\Omega_\Sigma$ is the covariant symplectic form for general relativity \cite{Burnett1990, Wald2000b}, depending on variations of the metric $\delta g_{\alpha\beta}$, and $\Omega_S$ is the boundary
modification, which contains variations both of the metric and of the fields $X^\mu$.

Diffeomorphisms act on both the metric and $X^\mu$ fields,
and, because they are pure gauge, have identically vanishing 
Hamiltonians.  The surface symmetries arise from transformations
acting only on $X^\mu$.  They can be parameterized by a spacetime
vector field $\xi^\mu$, and their action is given
by
\beq
\delta_\xi g_{\alpha\beta} = 0,\qquad \delta_\xi X^\mu = \xi^\mu.
\eeq
Only vectors $\xi^\mu$ that are nonzero in a neighborhood
of $S$ have nonvanishing Hamiltonians, as is clear from (\ref{eqn:Om}),
since the bulk term $\Omega_\Sigma$ only involves 
$\delta g_{\alpha\beta}$.  
More precisely, we can
decompose $\xi^\mu$ into its transverse and parallel components
near $S$
via
\beq
\xi = \xi^i_\top \partial_i + \xi^A_\parallel \partial_A,
\eeq
where $x^i$ denote transverse coordinates and $\sigma^A$ parallel coordinates (see equation (\ref{eqn:2+2})).
The canonical analysis shows that vector fields for which
$\xi^A_{\parallel}\Seq \xi^i_{\top}\Seq\partial_{i}\xi^j_\top\Seq0 $ are pure gauge and do not correspond to physical symmetries. It also shows that  dilatations of the normal bundle, for which
\be \label{eqn:dil}
\pa_{i} \xi_\top^j \Seq f(\sigma)\delta_i^j,
\ee
for a function $f(\sigma)$, are also pure gauge.\footnote{Such a $\xi$ characterizes the trace part of $\mfk{gl}(2,\mbb{R})^S$, which denotes the set of all maps $S\to\mfk{gl}(2,\mbb{R})$.} The group of physical symmetries we are left with consists of\footnote{There is another class of transformations
\begin{equation}
	\xi^A_\parallel\Seq 0,\qquad\qquad \xi^i_\top\Seq v^i, \qquad\qquad \partial_{i}\xi_\top^j\Seq 0.
\end{equation}
These transformations move $S$ itself and as such are called {\it surface deformations} (called \emph{surface translations} in Ref.~\cite{Donnelly:2016auv}). 
For these transformations to be integrable, one generically needs
to impose boundary conditions on the field variations \cite{Donnelly:2016auv, Speranza:2017gxd}, or work 
with nonintegrable, quasilocal charges using the 
Wald-Zoupas construction \cite{Wald2000b, Chandrasekaran2020}. 
Their treatment is more subtle and will not be pursued in the present work, but 
they are nevertheless quite interesting due to the generic appearance 
of central extensions in their symmetry algebra
\cite{Speranza:2017gxd, Chandrasekaran2020}.}

\begin{itemize}
	\item {\bf Surface Boosts}:  they are generated by vector fields with the following properties
	\begin{equation}\label{eqn:xitop}
	\xi^A_\parallel\Seq 0,\qquad\qquad \xi^i_\top\Seq 0, \qquad\qquad \partial_{i}\xi_\top^j\Seq 
	\alpha_i{}^j\equiv \alpha^c\left(\tau_c\right)\indices{_i^j},
	\end{equation}
	where $\alpha_i{}^j$ is a traceless matrix, which can be expanded 
	in the $\tau_c$ basis  defined in (\ref{su11})
	using the coefficients $\alpha^c$. They correspond to position-dependent transformations of the normal plane of $S$ that leave the unit binormal invariant.
	
	\item {\bf Surface Diffeomorphisms}:  they correspond to vector fields tangent to $S$ 
	\begin{equation} \label{eqn:xipar}
	\xi_\parallel^A \Seq \xi^A\neq 0,\qquad\qquad \xi^i_\top\Seq 0, \qquad\qquad \partial_{i}\xi^i_\top\Seq 0.
	\end{equation}
	They define diffeomorphisms mapping $S$ to itself. 
	\end{itemize}
These transformations preserve $S$ and are hence {\it corner-preserving transformations}. 
Their algebra can be computed from the Lie brackets of the vector fields, and coincides with the algebra
\begin{equation}
    \diff(S) \oplus_{\Lie} \slr^S,
\end{equation}
whose generators $(\xi^A,\alpha^c)$ are the quantities appearing in \eqref{eqn:xitop} and \eqref{eqn:xipar}.
The Lie algebra generated by these vector fields is given explicitly in \eqref{commutator}.

We now have a symplectic manifold with a group action, and we would like to understand how the phase space is divided into representations of this group.
To do this, we first map the phase space into the coadjoint representation of the group via the \emph{moment map}:  this simply amounts to identifying the generators of the Lie algebra in terms of the quantities in the phase space.
To do this we first have to define some geometric structures associated with the surface $S$ and its normal bundle.

\subsection{Normal bundle geometry}
\label{subsec:outer-curvature}

The generators and Casimirs of the corner symmetry algebra are closely tied to the geometry of the normal bundle of $S$, so we begin with a description of this geometry. The discussion largely follows Ref.~\cite{Speranza:2019hkr}.
Near the surface $S$, we can perform a $2+2$ decomposition of the metric, in which we foliate the region near $S$ by 2-dimensional surfaces.  
We introduce
coordinates adapted to the foliation $(x^i, \sigma^A)$, 
where $i = 0,1$, $A= 2,3$, 
in which the leaves of the foliation coincide with surfaces of 
constant $x^i$, and $\sigma^A$ serve as intrinsic coordinates
on these surfaces.  
In these coordinates, we can parameterize the metric as 
\begin{equation} \label{eqn:2+2}
\rd s^{2} = h_{ij}\rd x^{i}\rd x^{j} + q_{AB}\left(\rd \sigma^A - V^A_{i}\rd x^{i}\right)
\left(\rd \sigma^B - V^B_{j}\rd x^{j}\right).
\end{equation}
Here, $q_{AB}$ is the induced metric on the leaves, $h_{ij}$ 
is the normal metric, which acts as a generalized lapse, 
and $V_i^A$ is a generalized shift, which
measures the failure of the coordinate vectors $\partial_i$
to be orthogonal to the constant $x^i$ surfaces. 
The unit binormal to the leaves is a $2$-form given by
\beq
N = -\frac12\sqrt{|h|}\, dx^0\wedge dx^1,
\eeq
where the prefactor of $\frac12$ is nonstandard, and is employed to
simplify some later expressions.  
The mixed-index binormal obtained by raising one of the indices 
of $N$ is used to construct the generators of the normal boosts,
so we write out its components,
\beq
N\indices{_i^j} = -\frac12 \sqrt{|h|} \varepsilon_{ik}h^{kj}
\eeq
where $h^{jk}$ is the inverse of $h_{ij}$, and $\varepsilon_{ik}$
is the antisymmetric symbol with $\varepsilon_{01} = 1$.

The normal bundle of $S$ consists of all vectors 
of the form $Y = Y^i(\partial_i + V_i^A\partial_A)$ that are 
orthogonal to the surface.  The spacetime covariant derivative 
$\nabla_\mu$ defines a canonical connection on the normal bundle,
obtained by projecting $\nabla_\mu Y^\alpha$ tangentially
on the $\mu$ index and normally on the $\alpha$ index.  This 
results in a derivative operator $D_A$ on normal vectors
which acts as 
\beq
D_A Y^i = \partial_A Y^i +M_A{}^i{}_jY^j.
\eeq
The connection coefficients $M\indices{_A^i_j}$ have a direct
interpretation in terms of the geometry of the foliation \cite{Speranza:2019hkr}.\footnote{In \cite{Speranza:2019hkr},
the normal connection was called the normal extrinsic curvature
tensor, and was denoted $L\indices{_A_j^i}$.  The convention
we employ in the present work is such that $M\indices{_A^i_j} = L\indices{_A_j^i}$, which affects some of 
the signs in the expression for 
$M_{Aij}$.} 
By lowering an index using $h_{ij}$, it can be decomposed into its 
symmetric and antisymmetric parts on the normal components,
\beq\label{eqn:LAij}
M_{Aij}  = A_{Aij} - P_A N_{ij}.
 \eeq
The symmetric part $A_{Aij}$ is called the acceleration
tensor, since it measures the accelerations of the vectors
normal to the foliation.  In particular, if $y^\alpha$ is 
everywhere normal to the leaves of the foliation, the
tangential component of its acceleration is 
\beq
q_{\alpha\beta} Y^\mu\nabla_\mu Y^\beta = - A_{\alpha ij}Y^i Y^j .
\eeq
The coordinate expression for this tensor 
in terms of the decomposition \eqref{eqn:2+2} 
is derived in Appendix B.3 of 
\cite{Speranza:2019hkr},
and  is given simply by the gradient of the 
normal metric
\beq
A_{Aij} = \frac12\partial_A h_{ij}, 
\eeq
and we can also express the mixed index form appearing in 
$M\indices{_A^i_j}$ in terms of the gradient of $N\indices{_i^j}$
using that $h_{ij} = 4 N\indices{_i^k}N\indices{_k_j}$
and $N_{ij} = -\frac12 \sqrt{|h|}\varepsilon_{ij}$,
\begin{align} 
A\indices{_A^j_i} &= 
2\partial_A(N\indices{_i^k}) N\indices{_k^j}
+2N\indices{_i^k} h^{jl} \partial_A N_{kl}\nonumber\\
&=2\partial_A(N\indices{_i^k}) N\indices{_k^j} +\frac12 \partial_A\log\sqrt{|h|}  \delta^j_i .\label{eqn:AAij}
\end{align}
It is often convenient to work in a gauge where $\sqrt{|h|} = 1$,
since this condition is invariant under $\SL(2,\mathbb{R})$ 
transformations in the normal plane.  
Doing so reduces $A\indices{_A^j_i}$ to the first term in (\ref{eqn:AAij}),
and makes it a traceless tensor on the $i,j$ indices. 

The antisymmetric part in (\ref{eqn:LAij}) must be proportional
to the binormal $N_{ij}$, where the coefficient $P_A$
is the twist one-form.  It measures the failure
of the normal directions 
of the foliation to be integrable, since if $X^i$ and $Y^j$ are 
normal vectors, it can be shown that \cite{Speranza:2019hkr}
\beq \label{eqn:PAcommutator}
q_{AB} [X,Y]^B = -2P_A N_{ij}X^i Y^j.
\eeq
Its coordinate
expression is given in terms of the shift via
\beq \label{PA}
P_A = \frac{1}{\sqrt{|h|} } q_{AB} \left(\partial\indices{_0} V_1^B - \partial\indices{_1} V_0^B
+V_0^C\partial\indices{_C} V_1^B - V_1^C\partial\indices{_C} V_0^B\right).
\eeq

As a connection, $M\indices{_A^j_i}$ is not itself an invariant
tensor defined on $S$, since it shifts under a change in foliation
away from the surface.  On the other hand, its curvature
is tensorial, and is given by 
\begin{align}
W\indices{^i_j_A_B}  &=
\partial\indices{_A} M\indices{_B^i_j} 
-\partial\indices{_B} M\indices{_A^i_j} 
+ M\indices{_A^i_k}M\indices{_B^k_j} -
M\indices{_B^i_k}M\indices{_A^k_j}.
\label{eqn:WijAB}
\end{align}
Lowering a normal 
index with $h_{ij}$, the tensor
$W_{ijAB}$ is antisymmetric in each pair of lower indices, and, 
as such, it can be written in terms of a 2-form $W_{AB}$ on the surface
via
$W_{ijAB} = N_{ij}W_{AB}$. As shown in Section 5.3 of \cite{Speranza:2019hkr} and also Appendix \ref{app:outerident} 
of the present work, this 2-form can be 
expressed in terms of $P_A$ and $A_{Aij}$ as 
\begin{align} 
W_{AB} &= 2\partial_{[A} P_{B]} + 
4N\indices{_j^i}A\indices{_A_i^k}A\indices{_B_k^j}\label{eqn:dPNAA}
\\
&=
2\partial_{[A} P_{B]}
+2 \partial\indices{_A}(N\indices{_i^j}) N\indices{_j^k} 
\partial\indices{_B}(N\indices{_k^i}) \label{eqn:OmAB} 
 \end{align}
where the second line follows from (\ref{eqn:AAij}) using the gauge
$\sqrt{|h|}=1$, which can be imposed because $W\indices{^i_j_A_B}$ is 
independent of the trace $C_A \equiv A\indices{_A_i^i}
= \partial_A\log\sqrt{|h|}$, as is 
apparent from the definition (\ref{eqn:WijAB}).  
Note that the second term in (\ref{eqn:OmAB}) is antisymmetric in 
$A$ and $B$.

The fact that $W_{AB}$ is an invariant tensor independent of the choice of foliation makes it useful in constructing invariant quantities for the corner symmetry algebra, and we will see it is directly analogous to the vorticity \eqref{vorticity} used to construct invariants for the coadjoint orbits.  
One way to see the independence of the 
foliation extension is by noting that it is related to the spacetime
Riemann tensor $R_{\alpha\beta\gamma \delta}$ and the extrinsic curvature tensor $K\indices{^i_A_B}$
of the surface via the Ricci equation,
\beq \label{eqn:Ricci}
W_{AB} = -2 N^{ij} R_{ijAB} -4
N^{ij}K\indices{_i_A^C}K\indices{_j_C_B}.
\eeq
Since neither the Riemann tensor nor the extrinsic curvature depends on the choice of foliation away from $S$, it is clear that $W_{AB}$ also has no such dependence.  
It is also interesting to note that the tensor $W_{AB}$ is Weyl invariant, as has been emphasized by Carter \cite{Carter1992}.  
This follows from the fact that (\ref{eqn:Ricci}) only depends on the curvature through the Weyl tensor, and on the traceless part of $K_{iAC}$.

\subsection{Moment map}
\label{sec:moment_map}
With the normal bundle geometry in hand, we can proceed to the 
construction of the moment map.  This is done by identifying the 
geometric objects comprising the corner phase space with objects
defined in the coadjoint orbit.  The easiest way to identify
this map is to construct the Hamiltonians generating the symmetry.
It was shown in Ref.~\cite{Donnelly:2016auv} that the Hamiltonian is 
constructed from the generators according to  
\beq \label{eqn:Hxipair}
H_\xi = \frac{1}{16\pi G}\int_{S} \rd^2\sigma
\sqrt{q}\left(2N\indices{_j^i}\partial\indices{_i}\xi^j_\top
+ P_A \xi^A_\parallel \right).
\eeq
We find that the Hamiltonian is constructed as a linear pairing between the the vector fields and the geometric fields 
on the surface.  The densitized twist $\tilde P_A = \sqrt{q} P_A$
pairs with the $\diff(S)$ generator $\xi^A_\parallel$, and 
the densitized binormal $\tilde N\indices{_j^i} =\sqrt{q} N\indices{_j^i}$ pairs with the $\slr$ generator $\partial_i\xi^j_\top= \alpha\indices{_i^j} = \alpha^a(\tau_a)\indices{_i^j} $.
This can be put in an even more suggestive form by expanding $\tilde N$ in the Lie algebra basis as\footnote{In components, this relation
reads
\beq
\tilde N_0 = \tilde N\indices{_0^1} - \tilde N\indices{_1^0}, \quad 
\tilde N_1 = \tilde N\indices{_0^0} - \tilde N\indices{_1^1},\quad
\tilde N_2 = \tilde N\indices{_0^1} - \tilde N\indices{_1^0}.
\eeq}
\begin{equation}
\tilde N_a = 2\tilde N\indices{_j^i} \tau\indices{_a_i^j}.
\end{equation}
Comparing with the pairing (\ref{pairing}) for the coadjoint
orbit, we immediately 
find the correspondence 
\beq\label{momentmap}
\frac{\tilde N_a}{16 \pi G} \rightarrow \tilde n_a, \qquad
\frac{\tilde P_A}{16\pi G} \rightarrow \tilde p_A,
\eeq  
which gives the images of 
of $\tilde{P}_A$ and $\tilde{N}\indices{_i^j}$ under the 
moment map.  This also implies that $\tilde N\indices{_i^j}$ maps to
\beq \label{eqn:tNijmap}
\frac{\tilde N\indices{_i^j}}{16\pi G} 
\rightarrow \tilde n_a(\tau^a)\indices{_i^j}.
\eeq
Going forward, we will set $16\pi G = 1$.  

From the moment map \eqref{momentmap} we can continue to map the remaining quantities associated with the normal bundle geometry to the coadjoint 
orbits.  The first quantity to extract is the quadratic Casimir for the 
$\SL(2,\mathbb{R})$ group.  This is none other than the 
local area form on the surface, which can be defined in terms of 
$\tilde N\indices{_j^i}$ by 
\beq
\sqrt{q} = \sqrt{2 \tilde N\indices{_i^j}\tilde N\indices{_j^i}} = \sqrt{-4 \det \tilde N\indices{_i^j}}, 
\eeq
which, using  (\ref{eqn:tNijmap}) and $\tr\tau_a\tau_b = \frac12\eta_{ab}$, maps to 
\beq
\sqrt{q}\rightarrow \sqrt{\eta^{ab}\tilde n_a \tilde n_b} = |\tilde n|.
\eeq
This is a restatement of the result of \cite{Donnelly:2016auv}
that the area density $\sqrt{q}$ coincides with the local 
$\tenofo{SL}(2,\mathbb{R})$ Casimir. 

We can further identify the quantities associated with the 
undensitized generators $N\indices{_j^i}$ and $P_A$ by dividing 
$\tilde n_a$ and $\tilde p_A$ by the local density $|\tilde n|$,
as was done in equation (\ref{de-densitized}) to give 
\beq \label{eqn:undensitizedmoment}
N\indices{_j^i}\rightarrow  n_a
\,(\tau^a)\indices{_j^i}, 
\qquad P_A\rightarrow p_A.
\eeq
Since the dilatation \eqref{eqn:dil} is pure gauge, we can act
with dilatations to set the normal metric determinant to unity,
$\sqrt{|h|} = 1$, and this condition is $\SL(2,\mathbb{R})$-invariant.  
Using equation (\ref{eqn:AAij}), we then can identify the image of the 
acceleration tensor $A\indices{_A^j_i}$ under the moment map with 
\beq
A\indices{_A^j_i}\rightarrow \partial_A(n_a)n_b \,
\varepsilon\indices{^a^b_c}(\tau^c)\indices{_i^j}
\eeq
using (\ref{eq:the product of taus}) and the fact that 
$\eta^{ab}\partial_A(n_a) n_b = 0$ since $n_a$ is normalized. 
The moment map therefore sends the 
entire connection on the normal bundle $M\indices{_A^j_i}$ to 
\beq \label{momentLAij}
M\indices{_A^j_i}\rightarrow m_A^c (\tau_c)_i{}^j,\qquad
m_A^c := \partial_A(n_a) n_b \,\varepsilon\indices{^a^b^c} 
+ p_A n^c .
\eeq
$M_A^c$ is an 
$\slr$-valued one-form on the sphere, and hence can be viewed as  an $\slr$
 connection naturally constructed from the orbit data.  This also
gives a geometric interpretation for the transformation law
for $p_A$ in \eqref{coadjoint2}, where it was remarked that $p_A$  
transforms like the component of a connection.  Equation
\eqref{momentLAij} gives the full connection of which $p_A$ is a 
component, which is seen to simply be the image under the moment
map of the normal bundle connection $M\indices{_A^j_i}$.  
In fact such a connection is similar to the connection that appears in the Georgi-Glashow model of symmetry breaking with an adjoint Higgs field \cite{Corrigan, Shnir2005}.
Here, the analog of the Higgs field is the dual-Lie-algebra element $n^a$,
which arises from the mixed-index binormal of $S$. The Higgs mechanism is at play because the metricity of the connection implies that the normal connection $M$ preserves $N\indices{_i^j}$:
\bea
D_A N\indices{_i^j} =0.
\eea
This is equivalent to preserving a metric on the normal bundle, and 
so we see that he symmetry breaking pattern  is the reduction $\SL(2,\R) \to \SO(1,1)$, and the ``Higgs field'' $n^a$ is valued into the de Sitter hyperboloid $ \SL(2,\R)/\SO(1,1)$.
This also implies, as we have seen, that the $\SL(2,\R)$ curvature has to commute with $N\indices{_i^j}$ and is therefore given by $W N\indices{_i^j}$. 

The curvature of $M\indices{_A^j_i}$ is an important object
for constructing geometric invariants, and using equation
\eqref{eqn:OmAB}, we can determine the image of $W_{AB}$ under 
the moment map 
\beq\label{Vorticity}
W_{AB}\rightarrow  2\partial_{[A} p_{B]} -  \varepsilon_{abc}
n^a \partial_{A} n^b  \partial_{B} n^c =  \bar{w}_{AB}.
\eeq
We see that it is proportional to the vorticity constructed 
for the orbit in equation \eqref{mainomega}.  
This relation is one of the main results of this work, 
providing a geometric interpretation of the $\SL(2,\mathbb{R})$ 
invariant constructed for the coadjoint orbits in Section
\ref{lifting}.

\section{Poisson algebra}\label{sec:poisson}

In this section, we consider the Poisson brackets, the algebra of area-preserving diffeomorphisms, and the Casimirs of $\gsl$. Furthermore, we show that the smeared version of the outer curvature discussed in the previous section satisfies the algebra of area-preserving diffeomorphisms and commutes with $\slr^S$ algebra. 

As established in \cite{Donnelly:2016auv}, the Poisson brackets  of the momentum generators $\widetilde{P}_A(\sigma)$
and $\slr^S$ generators $\widetilde{N}_a(\sigma)$ are given by 
\bea
\left\{\widetilde{\N}_a(\sigma),\widetilde{\N}_b(\sigma')\right\}&=&\kappa\varepsilon\indices{_{ab}^c} \widetilde{\N}_c(\sigma)\delta^{(2)}(\sigma-\sigma'),\label{lBNN}\\
\left\{\tP_A(\si),\tP_B(\si')\right\} 
&=& 
 \kappa \left( \tP_A(\si') \pa_B -
\tP_B(\si) \pa_{A}' \right)\delta^{(2)}(\si-\si'),\label{lBPP}\\
\left\{\widetilde{P}_A(\sigma),\widetilde{\N}_a(\sigma')\right\}
&=&-\kappa\widetilde{\N}_a(\sigma)\partial_A'\delta^{(2)}(\sigma-\sigma'),\label{lBPN}
\eea
where $\kappa := 16\pi G$ is the gravitational coupling constant.
One can recognize \eqref{lBNN}-\eqref{lBPN} as the Kirillov-Kostant bracket associated with the dual of the corner symmetry algebra.
To do so, we  introduce the smeared  generators of $\diff(S)$ and 
$\mathfrak{sl}(2,\mbb{R})^S$  
\begin{equation}\label{eq:the smeared generator for g_S}
   P[\xi] := \bigintsss_S \xi^A \widetilde{P}_A,\qquad \N[\alpha]:= \bigintsss_S \alpha^a \widetilde{\N}_a.
\end{equation}
The smearing parameters for $\diff(S)$ generators are smooth vector fields $\xi:= \xi^A\pa_A $ on $S$ while the smearing parameters for $\slr^S$ generators are defined using the smooth $\mfk{sl}(2,\mbb{R})$-valued function $\alpha:=\alpha^a(\si)\tau_a$. 
This smearing reflects that $(\widetilde{P}_A,\widetilde{\N}_a)$ can be viewed as elements of $\gsl^*$, the dual Lie algebra of $\gsl$. The Poisson algebra can be conveniently written in terms of the smeared generators as follows
\begin{equation}
    \begin{aligned}
    \label{eq:full-algebra}
\{\N[\alpha_1],\N[\alpha_2]\}&=\kappa \N\left[ [\alpha_1, \alpha_2] \right],\cr
\{P[\xi_1],P[\xi_2]\}&=\kappa P\left[[\xi_1,\xi_2]_{\mathrm{Lie}}\right], \cr
\qquad
\qquad\{ P[\xi], \N[{\alpha}] \}&= \kappa \N\left[{\Lie_{\xi}\alpha}\right],
    \end{aligned}
\end{equation}
where we have used the $\slr$ commutator $[\alpha_1,\alpha_2] ^a= \epsilon^a{}_{bc} \alpha_1^b\alpha_2^c$. In the following, we take the normalization $ \kappa =1$.

\subsection{Casimir and the area element}
As we have seen in Section \ref{sec:moment_map}, the area element can be expressed in terms of Casimir invariant of $\slr^S$. 
\begin{equation} \label{casimir-area}
\sqrt{{C}[{\mfk{sl}(2,\mbb{R})}^S]} = \sqrt{\eta^{ab}\widetilde{\N}_a\widetilde{\N}_b} = \sqrt{q}, 
\end{equation}
where  $\eta:=\text{diag}(-1,+1,+1)$ is the metric on $\slr$.
From  this relationship, we get the nondegeneracy condition ${C}[{\mfk{sl}(2,\mbb{R})^S}]>0$. This property provides us an hint for quantization as this  implies that the $\mathfrak{sl}(2,\mbb{R})^S$ algebra should be quantized in terms of its \emph{continuous series} \cite{Kitaev201711}. 

This also means that the area element $\sqrt{q}$ is invariant under $\slr^S$. From the bracket \eqref{lBPN}  we can establish that it transforms as a scalar density under diffeomorphisms.
This means that we have 
\be\label{density}
\{P[\xi],\sqrt{q}(\si')\}=-\pa_A'(\xi^A(\si')\sqrt{q}(\si')), \qquad  \{\N[\alpha],\sqrt{q}(\sigma) \}=0.
\ee
Given this element, we can construct the total area
\be
\AA := \bigintsss_S \sqrt{q}. 
\ee
After integrating the first equation of \eqref{density} over $S$, it is clear that the total area is a Casimir of of $\gsl$ since it commutes with both generators $P[\xi]$ and $\N[\alpha]$ of $\gsl$. 

It is also useful for the analysis to consider the Poisson brackets of the de-densitized
momentum $P_A(\sigma) = \tilde P_A(\sigma)/{\sqrt{q}(\sigma)}$, which are given by
\bea
\sqrt{q(\si)} \left\{{P}_A(\si),{P}_B(\si')\right\} 
&=& -
\nu_{AB} J(\si)\,
\delta^{(2)}(\si-\si')\label{PPJ}
\\
\sqrt{q(\si)}\left\{{P}_A(\sigma),{\N}_a(\sigma')\right\}&=& - (\pa_A N_a )(\si)\,\delta^{(2)}(\si-\si')
\label{PNN}\\
\sqrt{q(\si)} \left\{{\N}_a(\sigma),{\N}_b(\sigma')\right\}&=&\varepsilon\indices{_{ab}^c} {\N}_c(\sigma)\delta^{(2)}(\sigma-\sigma')\label{NNN}
\eea  
as verified in Appendix \ref{app:Commutators}. 
In the fluid analogy,  $\tilde{P}_A$ is the fluid momentum density, and $P_A$ is the fluid velocity. The fluid vorticity  $J =\nu^{AB}\pa_A P_B$ 
 encodes  the local  angular momentum of the fluid. 
 Hence, we refer to $J$ as the angular momentum observable,
\be\label{ang-momenta}
\widetilde{J} :=  \varepsilon^{AB} \pa_A P_B, \qquad J= \frac{\widetilde{J}}{\sqrt{q}}.
\ee
The derivation of these commutators is given in Appendix \ref{app:Commutators}.
Equation \eqref{PPJ} has an interesting interpretation: it shows that the angular momenta can be understood as a curvature tensor controlling the noncommutativity of the de-densitized momenta.

The analysis performed in Section \ref{sec:corner_symmetry_group} has established that the generalized enstrophies \eqref{eq:the definition of generalized enstropy in terms of dressed vorticity} are coadjoint orbit invariants. The analysis performed in Section \ref{sec:moment_map} has established that the generalised fluid  vorticity is represented by the outer curvature on the gravitational phase space (see Equation \eqref{Vorticity}). This means that we expect 
 two results:
 First the smeared vorticity 
 (or smeared outer curvature) 
 \be\label{eq:the smeared version of outer curvature}
W[\phi] :=  \bigintsss_S \left(\rd P - \frac12 [\rd \N,\rd \N ] \cdot \N \right) \phi, 
\ee
commutes with $\slr^S$. It also generates the subalgebra $\mathfrak{sdiff}(S)$ of area-preserving diffeomorphisms.
From this, we conclude that the Casimirs of $\gsl$ can be constructed as \emph{gravitational enstrophies}
\be
\bar{C}_k := \bigintsss_S \sqrt{q}\,W^k,\qquad k=2,3\cdots. 
\ee
The first generalized enstrophy $\bar{C}_1$, or total vorticity, 
vanishes when working with a trivial normal bundle, as we have 
assumed throughout this work.  In Section \ref{sec:discussion}, we comment on a generalization that allows $\bar{C}_1$ to be nonzero,
and its relation to NUT charges.  The remaining sections provides proofs of the above expectations.

\subsection{Area-preserving diffeomorphisms revisited}
In this section, and for the readers convenience 
we describe the area-preserving diffeomorphism subalgebra $\sdiff(S)$ repeating some of the formulas discussed earlier. This prepares the ground for the proof of the previous statement in the following sections.

As has been explained in Section \ref{subsec:area-preserving diffeomorphisms}, this is the subalgebra of $\diff(S)$ that preserves a given area element. Let us recall that $\nu_{AB} := \sqrt{q} \varepsilon_{AB}$ defines\footnote{$\varepsilon_{AB}= \varepsilon^{AB}$ denotes the Levi-Civita symbol which is  a skew symmetric tensor density and normalized by $\varepsilon^{12}=\varepsilon_{12}=1$.} the volume form on $S$ 
\begin{equation}
    \upi:=\frac12 \upi_{AB} \rd \sigma^A \wedge \rd \sigma^B=\sqrt{q} \rd^2\sigma.
\end{equation} 
We denote the inverse volume form\footnote{ $\nu^{AB}$ is such that $\nu^{A C}\nu_{BC}= \delta\indices{^A_B}$} by 
$\nu^{AB} $. It is given by 
\begin{equation}
    \nu^{AB}= \frac{\varepsilon^{AB}}{\sqrt{q}}.
\end{equation}
$\nu^{AB}$ defines a natural Poisson structure on the sphere as follows:
given $\phi,\psi \in C^{\infty}(S)$, where $C^{\infty}(S)$ denotes the space of smooth function on $S$, we define their Poisson bracket to be  
\be\label{eq:the Poisson bracket of functions on sphere}
\pb{\phi}{\psi}:=\nu^{AB} \pa_A \phi \pa_B\psi.
\ee
The fact that $\rd \nu=0$ implies that $\pb{\cdot}{\cdot}$ satisfies the Jacobi identity.
This bracket enters the commutators of area-preserving vector fields. 
These vector fields are the ones which preserve the volume form $\upi$ and are  characterized by
\be 
\Lie_{\xi} \upi= 0,\quad  \iff  \quad \rd \imath_\xi \upi=0,
\ee
where we have used The Cartan Formula $\mathcal{L}_\xi=\rd\imath_\xi+\imath_\xi \rd$, $\imath_\xi$ denotes the interior product along $\xi$, and we have used 
that the volume form is closed, $d\nu=0$. Since $S$ is simply connected, this means that $\imath_\xi \rd \upi$ is exact. Such vector fields form an algebra which we denote by $\diff_\nu(S)$.

An area-preserving  vector field is entirely determined by a function on sphere called the \emph{stream function}. There is a map
$C^\infty(S) \to \mathfrak{X}(S)$ from differentiable functions to vector fields,  $\phi\to \xi_\phi$, given by
\be\label{Ham}
 \iota_{\xi_\phi}\nu = -\rd \phi, \quad \mathrm{or}\qquad \xi_\phi^B=\nu^{AB}\partial_A\phi.
\ee
This map is such that 
\be \label{Poisson2}
\nu(\xi_\phi,\xi_\psi) =\pb{\phi}{ \psi} ,\qquad\mathrm{and} \qquad [\xi_\phi,\xi_{\psi}]_{\mathrm{Lie}}= \xi_{\pb{\phi}{ \psi} },
\ee
which shows that  $\sdiff_\nu(S)$  forms a subalgebra  of $\diff(S)$.

\subsection{Vorticity decomposition and centralizer algebra} \label{sec:vorticity}
We are now in a position to show that the smeared outer curvature $W(\phi)$, defined in \eqref{eq:the smeared version of outer curvature},  satisfies the algebra of area-preserving diffeomorphisms and  commutes with the $\slr$ generators:
\be\label{sdiffalg}
\{W[\phi], W[\psi]\} = -W[\pb{\phi}{ \psi}], \qquad \{W[\phi], \N[\alpha]\}= 0.
\ee
To see this, we use \eqref{eq:the smeared version of outer curvature} to express the area-preserving diffeomorphism generator 
\be\label{eq:the curvature of P_A}
J[\phi]:=\bigintsss_S \phi \, \widetilde{J} = \int_S \phi \, \rd P,
\ee
as a sum of the smeared outer curvature and a term $S[\phi]$
\be\label{eq:the decomposition of outer curvature}
J[\phi] = W[\phi] + S[\phi].
\ee
The second component in the decomposition \eqref{eq:the decomposition of outer curvature} is an element of the enveloping algebra $U(\slr^S)$ given by
\be
S[\phi]:= \frac12 \int_S \phi(\si) \,  [\rd \N, \rd \N]\cdot \N =
\frac12 \bigintsss_S \rd^2\si\,\,\phi(\si)  \,  \varepsilon^{AB} \pa_A \N^b \pa_B \N^c  \N^a \epsilon_{abc},
\ee
Remarkably, both $J[\phi]$ and $S[\phi]$ form a representation of 
$\sdiff_\nu(S)$
\be\label{JS}
\{J[\phi],J[\psi]\} =- J[\pb{\phi}{\psi}],
\qquad
\{S[\phi],S[\psi]\} = - S[\pb{\phi}{\psi}],
\ee
while $J[\phi]$ acts on $S[\phi]$ by the adjoint action
\be\label{Saction}
 \{J[\phi], S[\psi]\}=- S[\pb{\phi}{ \psi}].
\ee
Writing $W = J -S$, it then follows that $W$ forms its own $\sdiff_\nu$ algebra and commutes with $S$:
\begin{equation}
\{W[\phi], W[\psi]\} = -W[\pb{\phi}{ \psi}], \qquad \{W[\phi], S[\psi]\}= 0,
\end{equation}
which establishes the first part of \eqref{sdiffalg}.
Moreover, we see that $W[\phi]$ and $S[\phi]$ generate the algebra $\sdiff_\nu(S) \oplus \sdiff_\nu(S)$.
Note that this result resembles a consequence of the first Whitehead lemma 
\cite{Jacobson},
which implies that the semidirect product of a semisimple Lie algebra 
with itself via the adjoint action is in fact a direct product.

To evaluate the Poisson bracket between the outer curvature $W$ and the $\slr$ generator $N$, we note that both $J$ and $S$ act covariantly on the 
$\slr^S$ generator
\be\label{JSN}
\{J[\phi],\N[\alpha]\} = -\N[\pb{\phi}{\alpha}],
\qquad
\{S[\phi],\N[\alpha]\} =  -\N[\pb{\phi}{\alpha}].
\ee
From this, we see that
\begin{equation}
\{ W[\phi], N[\alpha] \} = \{J[\phi], N[\alpha] \} - \{ S[\phi], N[\alpha] \}= 0,
\end{equation}
which establishes the second part of \eqref{sdiffalg} and reflects the $\slr$ invariance of the generator $W$.
The derivation of these commutators is given in section \ref{subsec:main proofs}.

Using these results, we see that $J[\phi]$ together with the Lie algebra generators $\N[\lambda]$ 
form a Poisson-Lie subalgebra $\mfk{c}_{\mfk{sl}(2,\mbb{R})}(S)\subset \mfk{g}_{\mfk{sl}(2,\mbb{R})}(S)$
\bea\label{eq:algebra of sdiff times SL(2,R)}
\{\N[\lambda_1],\N[\lambda_2]\}&=& \N\left[[\lambda_1,\lambda_2]\right],\cr
\{J[\phi],\N[\lambda]\} &=& -\N[\pb{\phi}{\lambda}], \cr
\{J[\phi],J[\psi]\} &=& - J[\pb{\phi}{\psi}].
\eea
This subalgebra is the semidirect product of area-preserving diffeomorphisms with $\slr^S$, where $\sdiff_\nu(S)$ acts on $\slr^S$ by the Lie derivative
\be\label{eq:the stabilzer algebra}
 \sg_{\mfk{sl}(2,\mbb{R})}(S)=\mathfrak{sdiff}(S) \oplus_{\Lie} \slr^S.
\ee
It can be characterised as the \emph{centraliser algebra}, i.e., the algebra that commutes with the area form $\sqrt{q}$:
\be\label{centralizer}
\{J[\phi], \sqrt{q}(\sigma) \} =  0,
\qquad \{N[\lambda],\sqrt{q}(\sigma)\}=0.
\ee
The proof of the first equality is given in Appendix \ref{app:Commutators}. 
The second equality follows from the fact that the area element $\sqrt{q}$ is the Casimir of $\slr^S$ according to equation \eqref{casimir-area}. 

\subsection{Main proofs}\label{subsec:main proofs}
The fact that the generators $J[\phi]$ for various smooth functions $\phi$ form a $\diff_\nu(S)$  subalgebra 
follows directly from the identity
\begin{equation}
    \begin{aligned}
    \P[\xi_\phi]  &=\bigintsss_S  \;    \nu^{AB}  \pa_A\phi   \widetilde{\P}_B =
    \int_S  \;     {\varepsilon^{AB }} \pa_A\phi   {\P}_B
\\
&= - \bigintsss_S \phi \; \varepsilon^{AB} \pa_A {\P}_B 
=  -\bigintsss_S \phi \widetilde{J} = -J[\phi].
    \end{aligned}
\end{equation}
This suggests the following proof of the commutator \eqref{JS} 
\be\label{JJ}
\{J[\phi],J[\psi]\}= \{P[\xi_\phi],P[\xi_\psi]\}\stackrel{*}{=} 
P[[\xi_\phi,\xi_\psi]_{\mathrm{Lie}}] =
P[\xi_{ \pb{\phi}{\psi} }] =-J[\pb{\phi}{\psi}].
\ee
This is, however, not valid as such. The  equality indicated  $\stackrel{*}{=}$  
assumes that the vector $\xi_\phi$ is field-independent and therefore commutes with the momentum.
Since $\xi_\phi$ depends on the measure, one cannot make this assumption.
The commutator $\{J[\phi]),J[\psi]\}$ therefore involves in addition the following sum of commutators $\int_S (\{P[\xi_\phi],\xi_\psi^A(\si)\} - \{P[\xi_\psi],\xi_\phi^A(\si)\})P_A(\si)$. While it is not obvious, it turns out that this sum vanishes. A more direct way to prove \eqref{JS} is  to remember that $\sqrt{q}\xi_\phi$ is field-independent and use the de-densitized commutator \eqref{PPJ}. 
This gives 
\bea 
\{J[\phi],J[\psi]\}&=&
\bigintsss_S \bigintsss_{S'} (\sqrt{q}\xi_\phi^A)(\si)
(\sqrt{q} \xi_\psi^B)(\si') \{P_A(\si),P_B(\si')\} \cr
&=& -\bigintsss_S \nu(\xi_\phi,\xi_\psi) \widetilde{J}
= -J[\pb{\phi}{\psi}].
\eea
The first commutator in \eqref{JSN} of
$J[\phi]$  with $\N[\lambda]$ can be evaluated more directly. Although $\xi_\phi$ is field-dependent, it commutes with the $\slr^S$ generators: $\{\xi_\phi(\si), N[\alpha]\}=0 $. 
This means that 
\be\label{JN1}
\{J[\phi],N[\alpha]\}= -\{P[\xi_\phi],N[\alpha]\}=
-N[{\cal L}_{\xi_\phi}\alpha] =
-N[\pb{\phi}{\alpha}].
\ee
One then establishes \eqref{Saction} by using the fact that 
$S[\phi]$ is the integral of a density that transforms covariantly under diffeomorphisms:
\be
\{P[\xi],S[\phi]\}= 
S[{\cal L}_\xi \phi].
\label{PS}
\ee
The identity $J[\phi]= P[\xi_\phi]$ then implies \eqref{Saction} by an argument similar to \eqref{JN1}. 

We focus next on the second identity of \eqref{JSN}.
We first recall  that  \be
 \{ \N^a(\sigma), \N[\alpha]\}  = [\alpha(\sigma) , N(\sigma)]^a.
 \ee
Using the Leibniz identity we end up with
\bea
 \{ S[\phi], \N[\alpha]\}  &=&
 \bigintsss_S \phi  [[  \rd \alpha, \N], \rd \N] \cdot \N
 + \bigintsss_S \phi  [[ \alpha, \rd \N ], \rd \N] \cdot \N
 + \frac12 \bigintsss_S \phi  [ \rd\N, \rd \N] \cdot [\alpha,\N]\cr
&=& \bigintsss_S \phi  [[ \rd \alpha, \N ], \rd \N] \cdot \N,
 \eea
 where we have used the pairing's invariance   $[ \rd\N, \rd \N] \cdot [\alpha,\N] 
 = [[ \rd\N, \rd \N], \alpha] \cdot\N$ and the Jacobi identity for the second equality.
 To continue this evaluation, one uses that 
 \be
N^2(\sigma)=1, \qquad [N,[A,N]]= (A\cdot N) N - A. 
\ee
when $N$ and $A$ are elements of the $\slr^S$ algebra.
 We then get that
 \bea
 \{ S(\phi), \N[\alpha]\}  &=&\bigintsss_S \phi  [[  \rd \alpha, \N], \rd \N] \cdot \N 
 = \bigintsss_S \phi \, \rd \alpha\cdot [ \N, [\rd \N, \N]]\cr
 &=& \bigintsss_S \phi \, \rd \alpha\cdot ((\N\cdot \rd \N) \N  -\rd \N)
 =- \bigintsss_S \phi \, (\rd \alpha\cdot \rd \N )\cr
 &=& - \bigintsss_S \rd \phi \wedge  \rd \alpha \cdot \N
 = -\bigintsss_S \pb{\phi}{\alpha }\cdot\widetilde{\N}
 = -\N[\pb{\phi}{\alpha }].
 \eea
where in the second line we have used $2N\cdot dN=d(N\cdot N)=d(1)=0$. This establishes the second equality in \eqref{JSN}.
 To complete the proof, we have to establish the second equality of  \eqref{JS}. 
 We start with the identity just established,
 \be 
 \{ S[\phi], N_a(\si) \}= \pb{\phi}{N_a}(\si),\qquad
 \{ S[\phi], \pa_A N_a(\si) \}= \pa_A \pb{\phi}{N_a}(\si).
 \ee 
 This means that 
 \begin{alignat}{2}
    \{S(\phi),S(\psi)\}&=
    \frac{1}{2}\bigintsss_S\psi\{S(\phi),[\rd\N,\rd\N]\cdot N\}\cr&=
    \frac{1}{2}\bigintsss_S
    \psi \left(
    [\rd\pb{\phi}{\N},\rd\N]\cdot N +
     [\rd\N,\rd \pb{\phi}{\N}]\cdot N  
    + [\rd\N,\rd\N]\cdot \pb{\phi}{N} 
    \right)\cr
    &= \frac{1}{2}\varepsilon^{abc}
    \bigintsss_S
    \sqrt{q} \psi \left(
    \pb{\pb{\phi}{\N_a}}{\N_b} N_c +
     \pb{\N_a}{\pb{\phi}{\N_b}} N_c 
    + \pb{\N_a}{\N_b}  \pb{\phi}{N_c} 
    \right).\nonumber
\end{alignat}
where we used the definition of the bracket
\be 
\int_S \phi( \rd a \wedge \rd b )= \int_S \sqrt{q} \phi \pb{a}{b}.
\ee 
Using the Jacobi identity for the sphere bracket and integrating by parts, we get 
\begin{alignat}{2}
    \{S[\phi],S[\psi]\}&=
    \frac{1}{2}\varepsilon^{abc}
    \bigintsss_S
    \sqrt{q} \psi \left(
    \pb{\phi}{\pb{ \N_a}{\N_b}} N_c 
    + \pb{\N_a}{\N_b}  \pb{\phi}{N_c}\right)\cr 
    &= 
    \frac{1}{2}\varepsilon^{abc}
    \bigintsss_S
    \sqrt{q} \psi \,
    \pb{\phi}{\pb{ \N_a}{\N_b} N_c }\cr 
    &= 
    - \frac{1}{2}\varepsilon^{abc}
    \bigintsss_S
    \sqrt{q} \pb{\phi}{\psi}  \left(
    \pb{ \N_a}{\N_b} N_c\right) = - S[\pb{\phi}{\psi}].
 \end{alignat}
 This establishes \eqref{JS}.
 
\section{Discussion and future directions} \label{sec:discussion}

In this work, we have studied the coadjoint orbits of the corner symmetry group $G_{\tenofo{SL}(2,\mbb{R})}$. 
Starting from $G_{\tenofo{SL}(2,\mbb{R})}$, and a symmetry breaking explained in Section \ref{subsec:orbit reduction}, we have seen that the problem of classification of coadjoint orbits of $G_{\tenofo{SL}(2,\mbb{R})}$ reduces to the problem of classification of coadjoint orbits of $G_{\mbb{R}}$, the hydrodynamical group. Since this is a semidirect product group with infinite-dimensional abelian normal factor $\mbb{R}^S$, the space of functions on $S$, there is a general formalism to study 
its coadjoint orbits \cite{Rawnsley197501, Baguis199705}. As explained in Sections \ref{subsec:hydrodynamical group} and \ref{subsec:area-preserving diffeomorphisms}, the coadjoint orbits of $G_{\mbb{R}}$ are classified by the total mass of the fluid and a coadjoint orbit of the subgroup 
$\tenofo{SDiff}_{|\widetilde{n}|}$
of area-preserving diffeomorphisms. 
The measured Reeb graph associated to the scalar vorticity function $w=\frac{1}{2}\nu^{AB}{w}_{AB}$, where $w_{AB}$ are the components of the 2-form vorticity $\mbs{w}$ defined in \eqref{vorticity}, encodes the invariants $C_k$ of the coadjoint orbits of $G_{\mbb{R}}$. 
However, since the normal factor of $G_{\tenofo{SL}(2,\mbb{R})}$, i.e.\ $\tenofo{SL}(2,\mbb{R})^S$, is nonabelian and the vorticity $\mbs{w}$ is not invariant under this normal factor, we need to replace $\mbs{w}$ by its dressed version $\bar{\mbs{w}}$, defined in \eqref{vorticity1}, which is invariant under $\tenofo{SL}(2,\mbb{R})^S$. Similar to $G_{\mbb{R}}$, the invariants of the coadjoint orbits are given by  the total area $\mcal{A}$ and  the measured Reeb graph associated to the function $\bar w$. Furthermore, we have realized the generators of corner symmetries on a phase space $\mcal{P}$ and described the moment map which sends the quantities on phase space to the corresponding quantities on the coadjoint orbits. In particular, we have shown that $W_{AB}$, the outer curvature of the normal bundle 
of $S$, is mapped to the dressed vorticity $\bar{w}_{AB}$. We also have seen that this construction involves a noncompact analog of the Higgs mechanism which reduces  $\SL(2,\R)$  to $\SO(1,1)$. 
The generator $n^a$ of $\slr$ plays the role of the Georgi-Glashow Higgs field.
Finally, we considered the smeared version of the outer curvature and we concluded that it satisfies the algebra of area-preserving diffeomorphisms. 
 
The overarching motivation for studying the  group $G_{\SL(2,\R)}$ is its role as a 
fundamental symmetry group of general relativity restricted to a subregion.  Associated to these symmetries is a collection of charges that comprise an algebra of observables for the subregion.  
Understanding this algebra therefore is relevant both for characterizing observables in the classical theory, and for making progress toward quantizing gravity.  
In the following subsections, we will discuss a number of avenues for generalizing the results of this paper.  
We also opine on various applications of this work, including the connections between gravity and hydrodynamics, the role of NUT charges in the construction, the question of dynamics for the Casimir invariants identified in this work, the implications for defining entanglement entropy in a gravitational theory, the construction of unitary representations of this algebra via geometric quantization of the coadjoint orbits, and the implications for quantum gravity coming from a nonperturbative quantization of this algebra.

\subsection{Immediate generalizations}
\label{subsec:generalizations}

This work focused on the case where the surface $S$ is a sphere, 
but the generalization to higher-genus surfaces is straightforward.
The reduction to the little group proceeds in the same way as in section \ref{sec:corner_symmetry_group}, and one can still construct the measured Reeb graph associated with the vorticity function $w$.
However, this is not a complete invariant in the higher genus case, because the vorticity $w$ determines the momentum $p$ only up to the addition of a closed form. To fully specify the orbit for a surface of genus $g$, we have to specify the flux of the momentum around the $2g$ nontrivial cycles.
The algebra of area-preserving diffeomorphisms of surfaces of genus $g$ admits $2g$ central charges \cite{Bars:1988uj}, and it would be interesting to understand the implications of these central extensions for representations of the full corner symmetry group.
While at the classical level we can restrict to a sector in which the topology of the surface $S$ is fixed, we can expect the quantum theory to include all possible topologies for the surface $S$.

The generalization to higher dimensions introduces additional considerations when carrying out a complete classification.  
Specifically, the complete classification of invariants for 
the area-preserving diffeomorphisms of surfaces of dimension $\geq 3$ is expected 
to be more involved than the case $d=2$.
Nevertheless, we can still construct invariants analogous to the Casimirs $C_k$.
In $D$ spacetime dimensions, these invariants must be constructed from the vorticity 2-form $\w$ and the density $\sqrt{q}$. 
When $D$ is even, we can construct a $(D-2)$-form $\w \wedge \w \wedge \cdots \wedge \w$, which can be turned into a scalar by dividing by the density $\sqrt{q}$. The moments of the resulting scalar are invariants, analogous to the Casimirs $C_k$.
When $D$ is odd, we can construct the Chern-Simons invariant $\int \w \wedge  \w \wedge \cdots \wedge \w \wedge p$ (where recall that $\w = \rd p$).
In three dimensions the Chern-Simons form is called the helicity and its integral is the only known integral invariant \cite{Morrison}.
Unlike in even dimensions, there does not appear to be an infinite family of invariants that can 
be constructed from this.
It remains an interesting open question to classify coadjoint orbits of area-preserving diffeomorphism groups in higher dimensions and to construct a complete list of invariants.

Another generalization would be to consider other diffeomorphism-invariant 
theories, including higher curvature theories and dilaton gravity.  The charges for these theories
were worked out quite generally in Ref.\ \cite{Speranza:2017gxd}, but it would be  interesting to determine the analogues of the Casimirs $C_k$ in these more 
general contexts.

\subsection{Nontrivial bundles and NUT charges}
\label{subsec:nut}

In the classification of the orbit invariants in terms of generalized
enstrophies, we found in sections \ref{subsec:construction of invariants}
and \ref{lifting} that the first generalized enstrophy $C_1=\int_S \w$ vanishes 
both for the hydrodynamical group and the corner symmetry group.  
The vanishing of this invariant is directly tied to working with 
symmetry groups that are semidirect products.  As explained in 
appendix \ref{app:fiber}, these groups arise as automorphism groups of 
trivial principal bundles over $S$.  Nontrivial bundles are associated
with nontrivial extensions of $\text{Diff}(S)$ by the group 
of gauge transformations, and the corresponding $2$-cocycle 
characterizing the extension is constructed from the curvature of 
a connection on the bundle.  This curvature can be used to construct
a topological invariant of the associated vector bundle: the Chern class of the bundle. And this 
invariant coincides with the first generalized enstrophy $C_1$.  
Hence, we can view $C_1$ as a charge measuring the nontriviality of 
the group extension, and is associated with working with a symmetry group of a nontrivial bundle.  

In the case of the hydrodynamical group discussed in section 
\ref{sec:coadjoint orbits of hydrodynamical group}, $C_1$ always
vanishes, because there are no nontrivial principal $\mathbb{R}$ 
bundles over $S$ \cite[ch.\ Vbis., sec.\ B.1]{Choquet2004}.  The 
vanishing of $C_1$ is also apparent from the fact that it is the 
integral of an exact form $\w = \rd p$, and the form 
$p_A$ is well-defined, independent of a choice of section of the 
bundle (see appendix \ref{app:fiber}). By contrast, there exist 
nontrivial $\SL(2,\R)$ bundles over $S$, and for these, 
$C_1$ is nonvanishing.   Equation (\ref{mainomega}) still gives 
a valid equation for the dressed vorticity $\bar\w$, but the 
second term in this expression involves $n^a$, which, according 
to appendix \ref{app:fiber}, depends on a choice of section of the 
bundle.  Only when the bundle admits a global section is this second
term exact, and hence for nontrivial bundles the integral of the 
second term will produce a nonzero result for $C_1$.  

On the gravitational phase space, the generalized vorticity 
$\bar\w$ was shown in section \ref{sec:moment_map} 
to coincide with the outer curvature tensor 
of the normal bundle of $S$ under the moment map.
The outer curvature can be viewed as a representative of the Euler 
class of the normal bundle, and its integral over $S$ therefore determines
a topological invariant \cite{Carter1992}.  
In standard setups, this integral must vanish since the null normals provide
a trivialization of the normal bundle.  However, notable exceptions are given by 
configurations possessing nontrivial NUT charges, where $C_1$ is seen to 
be proportional to the NUT charge (see, for example, the 
analysis near null infinity in Bondi coordinates of \cite{Godazgar2019}). 
In order for the surface $S$ to be a regular embedded 
spacelike surface when the NUT charge is nonvanishing, there
must be a physical Misner string penetrating the surface, which provides a 
compensating contribution to the outer curvature, to enforce that the normal
bundle remains globally trivial
\cite{Bonnor1969, Kalamakis2020, Hennigar2019, Clement2015}.  The NUT charge is then the contribution
to the integral excluding the Misner string singularity.  It may be 
interesting to investigate how to appropriately handle such
defects in the corner symmetry algebra of general relativity, and understand
their relations to nontrivial extensions in the corner symmetry algebra.

An alternative possibility is to consider spacetimes
in which the Misner string merely represents a coordinate singularity,
as occurs in the Taub-NUT solution and its generalizations \cite{Taub1951, Newman1963, Misner1963, Gibbons1979}.  These spacetimes can be interpreted as 
gravitational magnetic monopoles, and it is natural to conjecture that 
they would be associated with corner symmetry groups involving
nontrivial extensions of $\Diff(S)$ by the $\SL(2,\R)$ gauge group.  
The interpretation in this case is less straightforward, since regularity of 
the spacetime requires closed timelike curves.  Furthermore, it is generally
not possible to find regular embeddings of spacelike 2-spheres in such
spacetimes \cite{Misner1963}, necessitating a re-examination of the
construction of charges and corner symmetries in such cases, which are 
generally defined on spacelike codimension-2 surfaces.\footnote{A simpler
setup would be to examine the corner symmetry algebra for Einstein-Yang-Mills theory (generalizing the analysis of \cite{Setare2020}), 
where monopoles for the gauge field would also be associated 
with nontrivial extensions of $\Diff(S)$ by the Yang-Mills
gauge group  in the full corner symmetry algebra.
In this case, one can examine subtleties in constructing the covariant 
phase space on bundles that do not admit sections, using the 
fiber bundle formalism developed in \cite{Prabhu2017},
without worrying about causality issues that occur in Taub-NUT spacetimes.}
Some ideas in this direction have already been explored in the context of QED in \cite{Freidel:2018fsk}.  
Although classically these properties pose interpretational issues, the inclusion
of nontrivial spacetime topologies may have important consequences in the 
quantum theory.  In gauge theories, the inclusion of monopole configurations
in the sum over bundle topologies leads to quantization conditions
on the charges \cite{Shnir2005}, and we would expect similar quantization conditions
to appear for gravitational charges when summing over nontrivial corner symmetry bundles \cite{Dowker1974, Mazur1986}.  Additionally, sums over nontrivial Euclidean topologies
in gravity have recently been argued to be important for understanding 
the black hole information problem
\cite{Almheiri2020, Penington2019a, Almheiri2019, Penington2019, Almheiri2020a}, 
and it would be interesting to 
investigate the consequences of including sums over NUT charge sectors 
in the gravitational path integral.

\subsection{Hydrodynamics}

The hydrodynamical group appeared in our analysis as a 
reduction of the full corner symmetry group to the boost
subgroup that preserves the normal metric. Although presented here as
a warmup to the analysis of the full corner symmetry group, 
the connection between gravity and hydrodynamics is 
intriguing in its own right.  Such a relationship has been noted
before in the context of the membrane paradigm \cite{Thorne:1986iy, Damour}. The 
equations governing the evolution of a black hole's stretched 
horizon are recast as hydrodynamical equations for a 2+1 dimensional membrane stretching outside a black-hole horizons. One cornerstone of this connection is the interpretation of the Raychaudhuri and Damour equation as thermodynamical fluid conservation laws for the fluid's energy and momenta.  
More recently, this correspondence has been upgraded by the realization that the near-horizon limit is ultra-relativistic. This new understanding implies that one can describe the horizon's fluid as an ultra-relativistic fluid subject to Carrollian symmetry \cite{Campoleoni:2018ltl,Donnay:2019jiz}.

 The work of Arnold, Marsden,  and followers \cite{Arnold,HolmMarsdenRatiu1998, khesin2020geometric} showed that perfect fluid can be understood as a generalised Euler-Poincar\'e top for a hydrodynamical symmetry group.
 Our analysis therefore suggests that one could  extend further the membrane paradigm  connection into the canonical framework.
It is natural to wonder, in  particular,  whether the Euler top equation for the corner symmetry group can be derived from the Einstein's equation of motion.

The membrane paradigm  connection also features prominently
in the fluid/gravity correspondence \cite{bhattacharyya2007}, which utilizes the power
of AdS/CFT to relate long wavelength hydrodynamics of a dual 
conformal field theory to gravitational dynamics in black hole 
backgrounds \cite{Rangamani2009}.  In the present analysis, we found
that the Casimirs of the corner symmetry group were constructed
using the outer curvature tensor, which, in the fluid picture, was 
interpreted as the vorticity.  If we choose the surface $S$ to be 
a cut near infinity of $\text{AdS}$, the analogy becomes 
precise, with the outer curvature mapping to the vorticity of the 
CFT dual.  Rotating fluids have recently been investigated 
in the context of the fluid/gravity correspondence \cite{Leigh2011, Kalamakis2020}, particularly with emphasis
on fluids with total net vorticity.  These fluids were related to 
spacetimes with nonvanishing NUT charge, which, by the discussion
of section \ref{subsec:nut}, would relate to corner symmetry algebras
in which the first generalized enstrophy $C_1$ is nonvanishing.
It would be interesting to further explore the connection
between the corner symmetries of the present work and the 
connection to hydrodynamics and holography.

\subsection{The role of Casimirs}
The Casimirs $C_k$ associated with the corner symmetry algebra play a central role in our analysis, and they will play an important role in the quantization of this algebra.
It is therefore natural to wonder what role they play in gravitational dynamics. 
While the Casimirs are invariant under motions that preserve the surface $S$, one would like to understand how they evolve when the corner $S$ is translated in time and space.

This points to one limitation of our approach: we have focused so far on the corner symmetries, which only include transformations that fix the surface $S$.
A necessary step for understanding the dynamics of the Casimirs would be to extend the algebraic picture described here to include transformations that move the surface.
Such transformations are not Hamiltonian: they involve  nontrivial fluxes, since the system under consideration is no longer closed \cite{Barnich:2001jy, Compere:2018aar}.
An intriguing feature of these transformations involving so-called non-integrable charges is  that it is still possible to introduce a nontrivial bracket \cite{Barnich:2011mi} and an associated notion of boundary symmetry algebra \cite{Adami:2020amw, Chandrasekaran2020}.
Moreover it is interesting to notice that fluxes not only imply that corner charges are not conserved but also can lead to the non-conservation of the Casimirs.
In the fluid picture, the nonconservation of the generalized enstrophies is associated with dissipative effects.
In the gravitational phase space, the loss of charges is associated with radiation, which may give insight into the definition of radiation at the nonperturbative level. 

\subsection{Geometric quantization}

Our main motivation for classifying the coadjoint orbits of $\cog$ is that it provides us with an organizing principle that can be used to understand the quantum Hilbert space that general relativity associates to a region bounded by $S$.
Specifically, it allows us to foliate the phase space of a finite region by coadjoint orbits whose quantization is well-studied.
As we pointed out in Section \ref{subsec:the canonical symplectic form on coadjoint orbits}, the coadjoint orbits carry a natural symplectic structure defined by the so-called Kirillov-Kostant-Souriau (KKS) symplectic form $\omega_{\tenofo{KKS}}$ \cite{Kostant1965,Souriau1970,Kirillov1976}. Choosing a specific coadjoint orbit $(\mcal{O}^{G_{\tenofo{SL}(2,\mbb{R})}},\omega_{\tenofo{KKS}})$ of $\cog$, the quantization problem thus reduces to the quantization of $(\mcal{O}^{G_{\tenofo{SL}(2,\mbb{R})}},\omega_{\tenofo{KKS}})$ as a symplectic manifold. One often uses geometric quantization for that purpose\footnote{There are other methods for quantizing a symplectic manifolds such as Fedosov deformation quantization \cite {Fedosov1994} and A-model quantization \cite{GukovWitten200809}. Deformation quantization is a perturbative quantization that does not produce a Hilbert space. In contrast, the A-model quantization provides a Hilbert space as the state space of open strings stretched between two types of specified branes, which could be supported on Lagrangian or coisotropic submanifolds of the symplectic manifold in question.} \cite{Souriau1970, Kostant1970,GuilleminSternberg1977,GuilleminSternberg1980,Woodhouse1997,BatesWeinstein,Blau}. 

Geometric quantization can be used to construct all irreducible unitary representations of compact or solvable Lie groups. 
We are, however, dealing with infinite-dimensional coadjoint orbits of $\cog$ and there may be complications such as the existence of a well-defined measure on the orbit, the existence of polarization, and also the question of whether this method can produce all irreducible representations. It has nevertheless been used to construct the representations of $\tenofo{BMS}_3$ \cite{BarnichOblak201502}. One could hope that the same should be true for $\cog$. {\it A priori}, this procedure should give us the same Hilbert space as the one obtained by inducing representations of $G_{\tenofo{SL}(2,\mbb{R})}$ from its subgroups, just like the case of $\tenofo{BMS}_3$ \cite{BarnichOblak201502}. However, this relation is a tricky correspondence with various delicate aspects \cite{DuvalElhadadGotaySniatyckiTuynman199102}. For example, in the situation that the semidirect product group has an abelian normal factor, its coadjoint orbits can be constructed from the coadjoint orbtis of its little group by symplectic induction \cite{DuvalElhadadTuynman199202}. One can expect that the same result holds in our case where the normal factor is nonabelian. In that situation, the representations of $\cog$ can be constructed by inducing  the representations from its little group, and this would coincide with the representations obtained via the geometric quantization of the little group and full group orbits. For more details on 
geometric quantization and the relation to induced representations, we refer to \cite[Ch.\ 5]{Li199306}. 

Eventually, one should be able to compute the characters associated with the representation of $\cog$ and understand the decomposition of the partition function in terms of these characters. 
Finally, one need to understand the the fusion and intertwining property of representations of the corner symmetry group.
It  was argued in \cite{Freidel:2016bxd, Freidel:2019ees}, that the intertwining property of the corner symmetry representations encodes the gluing of subregions.
This fact could provide us with a new way to understand, at the quantum level,  important aspects of the quantum gravitational dynamics.

\subsection{Entanglement} 

If the quantization of the coadjoint orbits of the corner symmetry group can be carried out, this will have profound implications for entanglement between gravitational subsystems.
For a classical system with surface symmetry group $G(S)$,
we expect each region of space in the quantum theory with boundary $S$ to be associated with a Hilbert space containing a representation of $G(S)$.
When combining two regions $A$ and $B$ along a common boundary $S$, we then have a Hilbert space $\mathcal{H}_A \otimes \mathcal{H}_B$ with a representation of $G(S) \times G(S)$.
To construct the phase space for the region $A \cup B$, we have to quotient by the diagonal action of $G(S)$, an operation called the \emph{entangling product} in Ref.~\cite{Donnelly:2016auv}, which we denote as $\mathcal{H}_A \otimes_S \mathcal{H}_B$.
This quotient amounts to a matching condition on the generators of the Lie algebra across the surface $S$.

It can be shown that each state of $\mathcal{H}_A \otimes \mathcal{H}_B$ results in a reduced density matrix of system $A$ which is invariant under the corner symmetry group $G(S)$.
When $G(S)$ is compact, we can decompose any reduced density matrix $\rho_A$ into irreducible representations as $\rho = \bigoplus_R p_R \rho_R$, and the entanglement entropy takes the form \cite{Donnelly:2011hn,Donnelly:2014gva} 
\begin{equation} \label{entropy}
    S(\rho) = - \tr (\rho \log \rho) = \sum_R \left[ p_R \log \dim R - p_R \log p_R - p_R \tr (\rho_R \log \rho_R) \right].
\end{equation}
The similarity between \eqref{entropy} and the generalized entropy formula 
\begin{equation} \label{generalized-entropy}
    S = \frac{\langle \mathcal{A} \rangle}{4G} + S_\text{out}
\end{equation}
has been noted before \cite{Donnelly:2016auv,Harlow:2016vwg,Lin:2017uzr}.
The first term in \eqref{entropy} resembles \eqref{generalized-entropy} in that both are expectation values of operators localized on $S$.
Moreover, when the support of the distribution $p(R)$ is concentrated on high-dimensional representations, this first term of \eqref{entropy} may dominate over the others.
This can occur, for example, for 2D Yang-Mills theory with surface symmetry group $\SU{N}$ in the large $N$ limit \cite{Donnelly:2019zde}.

We expect similar results to hold for the corner symmetry group $\cog$ of general relativity.
In this case, the entangling product involves matching the twist and mixed-index binormal across the surface $S$, which further implies matching of the area element $\sqrt{q}$ and the outer curvature scalar.
This will require a generalization of \eqref{entropy} to infinite-dimensional noncompact groups.
Carrying out this calculation will require a more detailed study of the representation theory of $\cog$ and will be the subject of future work.
Without doing this calculation, we can already see that $\log \dim R$ is an invariant in the group and so can depend only on the area and invariants constructed from the outer curvature such as the Casimirs $\bar C_k$.
Because the moment map \eqref{momentmap} contains a factor of $G$, these invariants naturally appear in Planck units.
In the semiclassical regime, the area is large in Planck units, while for a smooth surface, the outer curvature invariants are small.
So in this regime, we expect the entropy to be a function of $\mathcal{A}/G$, with subleading corrections coming from the outer curvature.
While more work is needed to establish the functional form of the entropy, it is intriguing possibility that a universal density of entanglement entropy could arise as a consequence of hydrodynamical symmetries.

\subsection{Nonperturbative quantization}
\label{subsec:nonperturbative}

In this work, we have studied the classification of coadjoint orbits of the corner symmetry group - this can be viewed as a semiclassical version of the representation theory of the corner symmetry group and is a natural prelude to quantization. 
We expect to explore the non-perturbative quantization of this algebra in a future publication.
The quantization of sphere groups $G_{\mathfrak{h}}(S)$ first relies on the choice of a measure $\nu$ on the sphere.
Here we have studied the case where the measure $\nu$ is absolutely continuous with respect to the Lebesgue measure \emph{and} strictly positive.
This choice was justified by the semiclassical demand of non-degeneracy of the induced metric, but at the quantum level such conditions could be relaxed.

It is also possible that discrete measures play a role at the quantum level: the work  \cite{Freidel:2020svx, Freidel:2020ayo}  showed that this happens for gravity in the first-order formalism and in the presence of the Immirzi parameter $\gamma$.
In this case, the measure $\nu/\gamma $ appears as the Casimir of an internal $\SU{2}^S$ group. Accordingly, this means that the spectrum of $\nu$ is quantized. The discrete representations are labelled by an integer $N$ that characterizes the number of points on $S$ in the support of the measure, and a collection of spins $j_i$ attached to each sphere point. We expect the quantum measure entering the label of corner sphere representation to be given by 
\be
\nu =\gamma\left( \sum_{i=1}^N  \sqrt{j_i(j_i+1)} \delta^{(2)}(\sigma-\sigma_i) \right) \rd^2\sigma,
\ee 
where $\sigma_i\in S$ labels the support of the discrete measure. Only the diffeomorphism class of this measure is relevant, and therefore the precise location of the points does not matter.
Such a quantum measure corresponds to a quantization where the area admits a discrete spectrum.
The second-order metric formalism is recovered in the limit where $\gamma\to 0$ and the spacing of the discrete spectrum tends to $0$.

Even if one the measure on the sphere is taken to be continuous, one still has 
to quantize  the area-preserving diffeomorphisms, which form the little group preserving the measure.
There too, we have the option of continuous or discrete representations --- the discrete representations correspond to the quantization of vortices in quantum hydrodynamics.
Such quantization arises in the study of superfluids \cite{RasettiRegge}, Bose condensates \cite{Goldin} and quantum Hall fluids \cite{Wiegmann_2013}. We clearly could learn a lot from the study of quantum hydrodynamics \cite{tsubota2010quantized,Tsubota_2013, foskett2020holonomy}.
This discussion implies that there is the possibility to have discrete spectra at the quantum level, not only for the area element but also for the vortices.
That is, there is the possibility to propose a fully discrete representation theory for quantum gravity. The fact that this possibility arises from the study of continuous sphere algebra is quite exciting and deserves a more in-depth analysis.

Let us finally mention that since the group of area-preserving diffeomorphisms plays a central role in our construction, it leads to an exciting prospect:
we know that  area-preserving diffeomorphism can be deformed at the quantum level into the group $\SU{N}$, with $N$ large \cite{Hoppe},  which is the symmetry group of matrix models.
Understanding how such deformation can arise from the study of the corner symmetry algebra and whether it is related to the area and vortex quantization just discussed is a fascinating quest we expect to come back to soon.

\acknowledgments
We would like to thank Marc Geiller,  Hal Haggard, Ted Jacobson, Djordje Minic, Daniele Pranzetti, and Lee Smolin for helpful discussions. 
Research at Perimeter Institute is supported in part by the Government of Canada through the Department of Innovation, Science and Industry Canada and by the Province of Ontario through the Ministry of Colleges and Universities. The work of SFM is also funded by the National Science and Engineering Council of Canada and the Fonds de Recherche du Qu\'ebec.

\appendix

\section{Fiber bundle description of the corner symmetry group}
\label{app:fiber}
The corner symmetry group $\text{Diff}(S)\ltimes \tenofo{SL}(2,\mathbb{R})^S$ 
arises as an automorphism group of the normal bundle of the 
surface $S$ embedded in spacetime.  It can equivalently be described 
as the automorphism group $\text{Aut}(P)$ of the associated principal $\text{SL}(2,\mathbb{R})$
bundle, which we call $P$.   The description of this group in terms 
of diffeomorphisms of $P$ helps clarify some of the properties 
of this symmetry group, and in this appendix we develop this principal 
bundle description.  For a review of principal bundles and connections
defined on them, see, e.g.\ \cite{Choquet2004}.  

We recall that a principal bundle $P$ is a manifold equipped with a 
global right action by a group $G$ that acts \emph{freely}, i.e.\ without
fixed points, and \emph{properly}, which ensures that the quotient space 
$P/G$ is a smooth manifold. 
This quotient is called the base space, which we will denote $S$, and there is 
a canonical projection $\pi:P\rightarrow S$ from the total space $P$ 
to the base.  The fibers for this projection are the orbits of the $G$ action,
and they are all diffeomorphic to $G$.  

The automorphism group $\text{Aut}(P)$ is simply the collection of 
diffeomorphisms of $P$ that commute with the global right $G$-action
\cite{Abbati1989, Neeb2008, Gay-Balmaz2010}.  This group contains a normal subgroup of 
\emph{gauge transformations}, which consist of automorphisms that map the 
fibers into themselves.  The gauge group is isomorphic to the space $G^S$ 
of smooth maps from the base $S$ into $G$.  The quotient group $\text{Aut}(P)/G^S$ is a subgroup of $\text{Diff}(S)$, the group of 
diffeomorphisms of the base.  Hence, there is an exact sequence 
of groups
\beq
1\rightarrow G^S \rightarrow \text{Aut}(P) \rightarrow \text{Diff}(S).
\eeq
For nontrivial bundles, the final map is not generally surjective, 
since large diffeomorphisms
of $S$ can fail to be automorphisms of $P$.  In such a situation,
the large diffeomorphisms
produce inequivalent  bundles under pullbacks \cite{Neeb2008}.
However, the connected components of the identity of each 
of these groups do form a short exact sequence,
\beq\label{eqn:ses}
1\rightarrow G_0^S \rightarrow\text{Aut}_0(P)\rightarrow 
\text{Diff}_0(S)\rightarrow 1.
\eeq

In the case we are considering, $G = \text{SL}(2,\mathbb{R})$, and the
base is a 2-sphere, $S = S^2$.  Furthermore, we are assuming that the normal
bundle for $S$ is trivial, which means that the associated 
principal bundle is also trivial, implying that $P$ admits a section
$s:S\rightarrow P$, a smooth map from the base into $P$.  This section
allows the bundle to be realized as a direct product, $P = S\times \SL(2,\mathbb{R})$.  Diffeomorphisms of this section can be extended 
to fiber-preserving automorphisms of the bundle, which then shows that 
$\text{Diff}(S)$ is in fact a subgroup of $\text{Aut}(P)$.  This means that 
the sequence (\ref{eqn:ses}) splits, and hence the automorphism
group is  simply a semidirect product, $\text{Aut}(P)=
\text{Diff}(S)\ltimes\text{SL}(2,\mathbb{R})^S$.  If we were instead
dealing with a nontrivial principal bundle which did not 
admit global sections, the sequence (\ref{eqn:ses})
would not split, and $\text{Aut}(P)$ would be a nontrivial extension of 
(a subgroup of)
$\text{Diff}(S)$ by $\text{SL}(2,\mathbb{R})^S$.  

The Lie algebra of $\text{Aut}(P)$ also has a simple description in
terms of the bundle: it is given by the set of vector fields on $P$ that 
are invariant under the right action.  The Lie algebra for the gauge 
group $G^S$ 
consists of all right-invariant vertical vectors, i.e.\ vectors that 
are tangent to the fibers.  To identify the remaining generators
corresponding to
$\text{Diff}(S)$, we must choose a connection on the bundle.  One way
to specify a 
 connection is simply to make a right-invariant 
choice of horizontal subspaces on the bundle \cite[ch.\ Vbis., sec.\ A.1]{Choquet2004}.  
Each choice of section is canonically associated with a connection,
 with the horizontal directions 
simply being the directions tangent to the section.  More generally, the 
horizontal spaces can be defined by a Lie-algebra-valued one-form,
$\omega_\alpha^a$, with the horizontal vectors $\xi^\alpha_H$ 
coinciding with those annihilate it, 
$\xi_H^\alpha\omega_\alpha^a = 0$.\footnote{Additionally, on vertical 
vectors generating the gauge group, $\omega_\alpha^a$ is required 
to implement the canonical isomorphism between the vertical tangent
space of the fibers and the Lie algebra of $G$
\cite[ch.\ Vbis., sec.\ A.1]{Choquet2004}.  }
Connections arising from a choice of section have vanishing curvature,
and flatness 
is the property that ensures that the algebra of horizontal vectors
closes on itself.  This is because the Lie bracket of two horizontal vectors
when contracted into $\omega_\alpha^a$ satisfies
\beq
[\xi_H, \zeta_H]^\alpha\omega_\alpha^a = 
\Lie_{\xi_H}(\zeta_H^\alpha\omega_\alpha^a) - \zeta_H^\alpha\Lie_{\xi_H}
\omega_\alpha^a
=-\zeta_H^\alpha \partial_\alpha(\xi_H^\beta\omega_\beta^a) 
-\zeta_H^\alpha \xi_H^\beta (d\omega^a)_{\alpha\beta} 
= \xi_H^\alpha\zeta_H^\beta \Omega^a_{\alpha\beta}
\eeq
where $\Omega^a_{\alpha\beta}$ is the curvature of the connection 
$\omega_\alpha^a$.  We recall that the curvature is given by the
covariant exterior derivative of $\omega_\alpha^a$, which is defined 
by $D\omega^a(\xi, \zeta) = d\omega^a(\xi_H, \zeta_H)$, where $\xi_H^\alpha, \zeta_H^\alpha$ are the horizontal projections of the vectors \cite[ch.\ Vbis., sec.\ A.3]{Choquet2004}.  This shows that $[\xi_H,\zeta_H]^\alpha$ is itself horizontal for all choices of $\xi_H^\alpha$, $\zeta_H^\beta$ only when the curvature vanishes.  

Such a globally flat connection exists for the trivial $\text{SL}(2,\mathbb{R})$
bundles considered in this work, although there is no canonical choice 
of such a connection.  This is simply the statement that in the semidirect
product $\text{Diff}(S)\ltimes \text{SL}(2,\mathbb{R})^S$, there are many different
choices for the $\text{Diff}(S)$ subgroup, each coinciding with a different
choice of flat connection on the bundle, or, equivalently, a different 
choice of section.  

The connection $\omega_\alpha^a$ allows us to relate the bundle
description to the presentation
of the Lie algebra of $\Diff(S)\ltimes \SL(2,\R)^S$ in section 
\ref{sec:corner_symmetry_group}.
Given an arbitrary vector $\xi^\alpha$ 
on $P$ generating a bundle automorphism, its 
pushforward under the projection $(\pi_* \xi)^A \equiv \xi^A$ 
defines a vector on $S$ that can be viewed as a generator of $\Diff(S)$,
and this vector is independent of the choice of connection and section.  The 
contraction of $\xi^\mu$ with the connection produces an $\Sl(2,\R)$-valued
function  on $P$, and pulling this function
back to $S$ using the section produces an $\Sl(2,\R)$-valued function on 
$S$, $\alpha^a = s^*(\xi^\mu\omega_\mu^a)$.  
The pair $(\xi^A, \alpha^a)$ then coincide with the description of the 
Lie algebra given in section \ref{sec:corner_symmetry_group}.  Hence, we see that 
 $\alpha^a$ depends on choices of connection and section, while 
$\xi^A$ depends on neither.  

The dual Lie algebra is parameterized on the bundle by the space of 
one-form-valued horizontal top forms $u_\alpha\otimes \nu$ that are invariant
under the global right action.  By horizontal top form, we mean that $\nu$
annihilates all vertical vectors, and its degree is equal to the dimension
of the base.  The pairing between the Lie algebra and its dual 
is given by first forming the contraction $\xi^\alpha u_\alpha \nu$, which 
is a horizontal top form invariant under the right action, and as such, 
it descends to a well-defined top form $\pi_*(\xi^\alpha u_\alpha \nu)$ on $S$.  Integrating $\pi_*(\xi^\alpha u_\alpha \nu)$ over $S$ 
then defines the pairing.  Note that this pairing is manifestly independent
of the choice of connection or section, and hence is valid even for 
nontrivial bundles which do not admit sections.  

To connect to the description of section \ref{sec:corner_symmetry_group}, we first note that 
$u_\alpha$ can be pulled back to the fibers, in which case it defines 
a form acting on vertical vectors.  Using the canonical isomorphism between
vertical vectors and $\Sl(2,\R)$ Lie algebra elements, this allows us to 
extract an element $\hat u_a$ of $\Sl(2,\R)^*$ by requiring that $\hat u_a \hat\xi^a = 
u_\mu \xi_V^\mu$, where $\hat\xi^a$ is the Lie-algebra element corresponding 
to the vertical vector $\xi^\mu_V$.  We can then get a function on $S$ 
using a choice of section to pull back the function $\hat u_a$, producing
$n_a = s^* \hat u_a$.  Here, we see that $n_a$ depends on the choice of 
section, but not on the choice of connection.  Finally, we can extract
the horizontal piece of $u_\alpha$ using the connection by the equation
$u_\alpha^H = u_\alpha - \hat u_a\omega^a_\alpha$.  This form is horizontal
and invariant under the right action, and hence descends to a well-defined 
form $p_A = (\pi_* u^H)$ on the base, independent of a choice of section.  
In this case, we find that $p_A$ depends on the choice of connection, but 
not on the choice of section.  In this argument, we have dropped the density
factor $\nu$, but it similarly descends to a density on the base, which can
be used to densitize $n_a$ and $p_A$.  The resulting pair $(\tilde p_A, \tilde n_a)$ then coincides with the data for the coadjoint representation
presented in section \ref{sec:corner_symmetry_group}.  

As a final note, we comment on how the Lie algebra is parameterized in the 
case of a nontrivial bundle.  One again uses a connection to 
identify a class of horizontal vectors.  This vector space is equivalent
to the space of vectors on the base, but because the connection must
necessarily not be flat, the Lie bracket of these vectors does not close 
on itself.  Instead, the curvature $\Omega_{\alpha\beta}^a$ defines a 
nontrivial 2-cocycle representing the obstruction to the splitting 
of the sequence (\ref{eqn:ses}).  We can continue to use the variables 
$(\xi^A, \alpha^a)$ on the base to parameterize 
the Lie algebra,\footnote{There are subtleties at this point associated 
with the nonexistence of global sections, causing the connection one-form to 
not be well-defined on the basis.  This can lead to subtleties in
interpreting the function $\alpha^a$, and one must either work in patches
and relate $\alpha^a$ in different patches using transition functions, 
or give a description with specified singularities associated with 
Dirac strings. }
and the Lie bracket is a modification of (\ref{commutator}), given by
\beq
[(\xi,\alpha), (\zeta,\beta)] = ([\xi,\zeta]_\text{Lie}, 
\Lie_\xi \beta - \Lie_\zeta \alpha + [\alpha,\beta] + \Omega(\xi,\zeta) ).
\eeq
It would be interesting to further carry out the analysis of the orbits
for these automorphism groups of nontrivial bundles.  We note that 
the classification of the orbit invariants will parallel the discussion
of section \ref{lifting}, the only difference being that the first 
generalized enstrophy $C_1 =\int_S \bar{\boldsymbol{w}}$, also known as the 
total vorticity, will not vanish.

 \section{Frame fields for the normal bundle}\label{app:frame fields for the normal bundle}

The normal bundle geometry in section \ref{subsec:outer-curvature}
was described in 
terms of a coordinate basis adapted to the local foliation by 
codimension-2 surfaces.  Another description is obtained by working with 
a null basis for the normal bundle, which has the advantage of producing
a relatively simpler expression for the outer curvature.  In this 
appendix, we describe the normal bundle geometry in terms of then null
basis, and explain the relation to the coordinate basis description 
of section \ref{subsec:outer-curvature}.

We  choose null basis vectors $(\lpup i , \lmup i )$ for
the normal bundle, normalized by $\lp\cdot\lp = \lm \cdot\lm = 0$,
$\lp \cdot\lm = 1$.  These conditions determine
the null basis up to an overall rescaling by a positive function, 
\beq \label{eqn:boost}
(\lp, \lm)\rightarrow (e^\phi \lp, e^{-\phi}\lm),
\eeq
which reflects the local boost redundancy in the frame field 
description.  A useful characterization of the covectors 
$(\lpdn i, \lmdn i)$ is that they
are the eigenvectors of the mixed-index binormal
$2N\indices{_i^j}$, with eigenvalues $(+1, -1)$. This allows $2N\indices{_i^j}$
to be expressed as 
\beq
2N\indices{_i^j} =\lpdn i \lmup j - \lmdn i \lpup j
\eeq
An object $\eta$ that transforms under the rescaling (\ref{eqn:boost})
as $\eta\rightarrow e^{w\phi} \eta$ is said to have boost weight $w$.  
For example, if we express a normal vector $U^i$ in terms of its components 
in the null basis $U^i  = \bar{U} \lpup i + U \lmup i$, we see that $\bar U$ has 
boost weight $-1$ and $U$ has boost weight $1$.  We can then define the 
covariant derivative of a scalar object of definite boost weight by
\beq
D_A \eta = \partial_A \eta -w \Xi_A \eta,
\eeq
where the connection $\Xi_A$ is known as the \Hajicek one-form
\cite{Hajicek1974, Hajicek197503},
and is canonically determined from the spacetime covariant derivative  
via 
\beq
\Xi_A = q\indices{^B_A} \lmup i \nabla_B \lpdn i.
\eeq
Since this connection only involves derivatives tangential to $S$, it
is manifestly independent of the choice of foliation away from $S$;
however, under local boosts of the normal frame, it transforms 
by a shift, $\Xi_A\rightarrow \Xi_A + \partial_A \phi$.  As an abelian
connection, its curvature is simply given by the tangential
exterior derivative $d\Xi$, which is boost and foliation independent,
and hence should be related to the outer curvature tensor 
(\ref{eqn:OmAB}). 

The precise relation can be derived by expressing $\Xi_A$ in
terms of the tensors characterizing the foliation.  Using the relation
(\ref{eqn:PAcommutator}), we find that 
\beq
P_A = q_{AB}[\lp, \lm]^B,
\eeq
from which we can then derive
\beq \label{eqn:XiF}
\Xi_A = -\frac12\left(P_A + \Bp_A - \Bm_A\right),
\eeq
where
\beq\label{eqn:boosts}
\Bp_A = q\indices{^B_A} \lmup i (d\lp)\indices{_i_B},\qquad
\Bm_A = q\indices{^B_A} \lpup i (d \lm)\indices{_i_B}.
\eeq
and $d\lp$, $d\lm$ are the exterior derivatives of the null covectors 
$\lp_\alpha$, $\lm_\alpha$, defined in a neighborhood of $S$ using the local
foliation.  
It is also useful to write the decomposition of the acceleration tensor
in the null basis.  We have 
\beq \label{eqn:Adecomp}
A_{Aij} = -\Ap_A \lmdn i \lmdn j - \Am_A \lpdn i \lpdn j -\frac12(\Bp_A + \Bm_A) h_{ij}
\eeq
where $\Ap_A = q\indices{_A_B} \lpup i \nabla_i \lpup A$ is the tangential
acceleration of $\lpup i$, and similarly for $\Am_A$.

The quantities $\Bp_A, \Bm_A$ depend on the boost frame, and hence
do not correspond to any invariant tensors associated with the foliation,
although (\ref{eqn:Adecomp}) shows that their sum is boost-invariant
and coincides with the trace $ A\indices{_A_i^i} = -(\Bp_A + \Bm_A)$.
Additional relations can be derived by working out the consequences of 
integrability of the foliation in a neighborhood of $S$.  
The Frobenius integrability conditions imply that the null covectors
$\lp, \lm$ satisfy
\begin{align}
d\lp &= \lm\wedge \Ap + \lp\wedge \Bp - \kappa\, \lp\wedge \lm\\
d\lm &= \lp \wedge \Am + \lm \wedge \Bm + \km \, \lp\wedge \lm 
\end{align}
where $\kappa, \km$ are the inaffinities of $\lp, \lm$.  The relation
we need comes from computing $d(d \lp\wedge \lm)$ in two different ways.  
First we have 
\beq
d(d\lp\wedge \lm) = -d(\Bp\wedge \lp\wedge \lm) = -d\Bp\wedge \lp\wedge\lm
-\Bp\wedge \Bm \wedge \lp\wedge \lm,
\eeq
while on the other hand, 
\beq
d(d\lp \wedge \lm) = d\lp \wedge d\lm = \Ap \wedge \Am \wedge \lp \wedge \lm
-\Bp \wedge \Bm \wedge \lp\wedge \lm.
\eeq
This gives the relation
\beq \label{eqn:dBp}
d\Bp_\parallel = - \Ap\wedge \Am,
\eeq
where the subscript $\parallel$ denotes the parallel projection of $d\Bp$.  
We can derive in a similar manner
\beq \label{eqn:dBm} 
d \Bm_\parallel = \Ap\wedge \Am
\eeq

Now by taking an exterior derivative of (\ref{eqn:XiF}) pulled back to 
$S$ and applying
(\ref{eqn:dBp}), we derive the relation
\beq\label{eqn:dXi}
-d\Xi = \frac12 dP -\Ap \wedge \Am
\eeq
Comparing to equation (\ref{eqn:dPNAA}), we see that it agrees with 
(\ref{eqn:dXi}), upon substituting in the decomposition (\ref{eqn:Adecomp})
for $A\indices{_A_i^j}$.  

Finally, we can work out the coadjoint orbit quantity to which $\Xi$ corresponds 
under the moment map.  According to 
(\ref{eqn:undensitizedmoment}), $N\indices{_i^j}$ simply
maps to the $\Sl(2,\R)$ coadjoint element $n\indices{_i^j}$, and $(\lp_i, \lm_j)$ will map to eigenvectors of $n\indices{_i^j}$, which we 
will denote by the same letters as the spacetime covectors.  
The boost parameters (\ref{eqn:boosts}) then map to 
\beq
\Bp_A\rightarrow b_A = - \lm^i \partial_A \lp_i,\qquad
\Bm_A\rightarrow \bar b_A = -\lp^i\partial_A \lm_i 
\eeq
From here, we express the eigenvectors in terms of the 
standard boosts introduced in section \ref{lifting}.  Examining
equation (\ref{boost}), we see that the eigenvectors 
can be expressed in terms of the constant basis vectors $\lp^0 = \begin{bmatrix} 1\\0\end{bmatrix}$, $\lm^0 = \begin{bmatrix} 0\\1\end{bmatrix}$ and the standard boost $x_n$ via
\beq
\lp = x_n^{-1} \lp^0, \qquad \lm = x_n^{-1} \lm^0.
\eeq
The boost parameter can then be computed to be
\begin{align}
b_A = -\lm^i \partial_A\left((x_n^{-1})\indices{_i^j}\lp^0_j\right)
=\lm^i(x_n^{-1})\indices{_i^j}\partial_A(x_n)\indices{_j^k}
(x_n^{-1})\indices{_k^l}\lp_l^0
= \left(x_n^{-1} \partial_A x_n\right)\indices{_i^j} \lp_k\lm^i,
\end{align}
and similarly 
\beq
\bar b_A = (x_n^{-1}\partial_A x_n)\indices{_i^j}\lm_j \lp^i.
\eeq
Recalling that $2n\indices{_i^j} = \lp_i \lm^j-\lm_i\lp^j$, 
we see that 
\beq
b-\bar b = 2\tr\left(x_n^{-1} d x_n n\right) = x_n^{-1}dx_n \cdot n.
\eeq
With this expression, along with (\ref{eqn:XiF}), we find that the 
\Hajicek one-form is related to the dressed momentum $\bar p$ 
defined in (\ref{eq:the definition of dressed momentum})
under the moment map according to
\beq
\Xi\rightarrow -\frac12(p + x_n^{-1}dx_n\cdot n) 
= -\frac12\bar p.
\eeq

\section{Additional calculations}\label{app:additional identities}
\subsection{The cocycle identity}\label{app:cocycle-identity}
In this appendix, we give the derivation of the cocycle identity \eqref{cocycle2}. To find the variation of $L_\alpha$ we have to vary the identity \eqref{variations II}, which requires second-order variations. Using \eqref{variations I},
\begin{equation}
    \delta_\alpha n=[\alpha,n],
\end{equation} 
the second variation of $n$ is given by
\begin{equation}\label{eq:the second variation of x_n}
    (\delta_\alpha \delta_\beta - \delta_\beta \delta_\alpha) n 
    = [\alpha,[\beta,n]]-[\beta,[\alpha,n]]=  [[\alpha,\beta], n] =  \delta_{[\alpha,\beta]} n.
\end{equation}
Since $x_n$ is a function of $n$, this implies
\begin{equation}
    (\delta_\alpha \delta_\beta - \delta_\beta \delta_\alpha) x_n = \delta_{[\alpha,\beta]} x_n.
\end{equation}
The variation of \eqref{variations II} gives
\begin{alignat}{2}\label{eq:derivation of cocycle identity I}
\delta_\alpha(x_n^{-1}\delta_\beta x_n)-\delta_{\beta}\left(x_n^{-1}\delta_\alpha x_n\right)&=\delta_\alpha L_\beta(n)n+L_\beta(n)\delta_\alpha n-\delta_\beta L_\alpha(n)n-L_\alpha(n)\delta_\beta n\nonumber
\\
&=\delta_\alpha L_\beta(n)n-\delta_\beta L_\alpha(n)n+L_\beta(n)[\alpha,n]-L_\alpha(n)[\beta,n].
\end{alignat}
On the other hand, using \eqref{eq:the second variation of x_n}, we have
\begin{alignat}{2}\label{eq:derivation of cocycle identity II}
\delta_\alpha(x_n^{-1}\delta_\beta x_n)-\delta_{\beta}\left(x_n^{-1}\delta_\alpha x_n\right)&=-[x_n^{-1} \delta_\alpha x_n, x_n^{-1} \delta_\beta x_n] +x_n^{-1} \delta_{[\alpha,\beta]} x_n
\end{alignat}
Using \eqref{variations II} and the fact that $L_{\alpha}(n)$ is a real-valued function, we get
\begin{alignat}{2}
[x_n^{-1} \delta_\alpha x_n, x_n^{-1} \delta_\beta x_n]&=
[-\alpha+L_\alpha(n)n,-\beta+L_\beta(n)n]\nonumber
\\
&=[\alpha,\beta]-[\alpha,L_{\beta}(n)n]-[L_\alpha(n)n,\beta]+[L_{\alpha}(n)n,L_{\beta}(n)n]\nonumber
\\
&=[\alpha,\beta]-L_\beta(n)[\alpha,n]+L_\alpha(n)[\beta,n].
\end{alignat}
We thus have
\begin{alignat}{2}\label{eq:derivation of cocycle identity III}
-[x_n^{-1} \delta_\alpha x_n, x_n^{-1} \delta_\beta x_n]+x_n^{-1}\delta_{[\alpha,\beta]}x_n=L_\beta(n)[\alpha,n]-L_\alpha(n)[\beta,n]+L_{[\alpha,\beta]}(n)n.
\end{alignat}
Taking into account \eqref{eq:derivation of cocycle identity II} and equating \eqref{eq:derivation of cocycle identity I} and \eqref{eq:derivation of cocycle identity III} yields
\begin{equation}
L_{[\alpha,\beta]} = \delta_\beta L_\alpha - \delta_\alpha L_\beta,
\end{equation}
which is the cocyle identity \eqref{cocycle2} that we sought to prove.

\subsection{Computing the vorticity}\label{app:vorticity}
In this appendix, we establish the validity of \eqref{vorticity1} and the correspondence between \eqref{vorticity1} and \eqref{mainomega}.
First, we need two preliminary results which involve the $\slr$-valued 1-form
\be 
\theta:= x_n^{-1} \rd x_n,
 \ee
which is the pullback of the Maurer-Cartan form under the map $x_n:S\rightarrow \SL(2,\mathbb{R})$.
Using the relation $n= x_n^{-1} \tau_1 x_n$ \eqref{boost} and the definition of $\theta$, we have
\begin{alignat}{2}\label{rdn}
    \rd n &= \rd x^{-1}_n\tau_1x_n+x^{-1}\tau_1\rd x_n=-x_n^{-1} \rd x_n x_n^{-1}\tau_1 x_n + x_n^{-1}\tau_1x_nx_n^{-1}\rd x_n\nonumber=[n,\theta].
    \\
    \rd\theta &= \rd (x_n^{-1}) \rd x_n =-x_n^{-1} \rd x_n x_n^{-1} \rd x_n=-\frac{1}{2}[\theta,\theta].
\end{alignat}
Also we need the cross product identity
\be\label{cross-prod}
[n,[A,n]]= (A\cdot n) n - n^2 A. 
\ee
valid for any element $n,A\in \slr$.
This identity is equivalent in coordinates to
\begin{equation}\label{eq:the property of Levi-Civita symbol}
    \varepsilon_{abc} \varepsilon_{fg}{}^c = \sigma (\eta_{af} \eta_{bg} - \eta_{ag} \eta_{bf}),
\end{equation}
for the 3-dimensional Levi-Civita symbol, where $\sigma$ denotes the product of the eigenvalues of metric $\eta$, which is $\sigma = -1$ in our case. Using  \eqref{rdn} we evaluate 
\bea
\rd(\theta \cdot n) &=& - 
\frac12 [\theta,\theta] \cdot n -
\theta \cdot \rd n \cr
&=&
-  \frac12 [\theta,\theta] \cdot n
-  \theta
   \cdot   [n , \theta]
=
+ \frac12 [\theta,\theta] \cdot n.
  \eea
We have used the ad-invariance of the pairing $[A,B]\cdot C= A\cdot [B,C]$ in the last equality. This establishes \eqref{vorticity1}.
A more direct way to establish \eqref{vorticity1} is to  use that 
\be 
\theta \cdot n =
x^{-1}_n\rd x_n \cdot (x^{-1}_n\tau_1 x_n)=
\rd x_n x_n^{-1}\cdot \tau_1.
\ee
So we have 
\bea 
\rd(\theta \cdot n) &=&\rd( x_n^{-1}\rd x_n\cdot n)
= \rd( \rd x_n x_n^{-1}\cdot \tau_1)
=- \rd x_n\wedge \rd x_n^{-1} \cdot \tau_1
= \frac12 [ \rd x_n x_n^{-1},\rd x_n x_n^{-1}]\cdot \tau_1\cr
&=& \frac12 [ x_n^{-1}\rd x_n,x_n^{-1}\rd x_n]\cdot (x_n^{-1}\tau_1 x_n)=
\frac12 [\theta,\theta]\cdot n.
\eea
For the equivalence between \eqref{vorticity1} and \eqref{mainomega}, one  establishes that 
\bea
[\rd n, \rd n ]&=&[[ n,\theta], [ n,\theta]] = 
[n,[[\theta, [ n,\theta]]]- [\theta, [n, [ n,\theta]]]
\cr
&=&- [\theta,\theta] + (\theta\cdot n) [ n, \theta]
- \frac12 
[n,[[\theta, \theta], n]],
\eea
where we used Jacobi identity in the first equality and \eqref{cross-prod} in the second. 
The last two terms vanish when contracted with $n$. This gives 
\be
n\cdot [\rd n, \rd n ] =- n\cdot [\theta ,\theta ],
\ee
which is the desired result. Finally, we provide a slightly simpler derivation of this result using index notation:
\begin{align}
n \cdot [\rd n, \rd n] &= 2 \varepsilon_{abc} n^a (\varepsilon_{de}{}^b n^d \theta^e) \wedge (\varepsilon_{fg}{}^c n^f \theta^g) \cr
&= 2 (\eta_{af} \eta_{bg} - \eta_{ag} \eta_{bf}) \varepsilon_{de}{}^b n^a n^d n^f \theta^e \wedge \theta^g \cr
&= 2 \varepsilon_{deg} n^d \theta^e \wedge \theta^g \cr
&= - [\theta,\theta]\cdot n.
\end{align}
Here we used $n^a n_a = 1$ and the property \eqref{eq:the property of Levi-Civita symbol} of the 3-dimensional Levi-Civita symbol with $\si=-1$.

\subsection{Outer curvature identity} \label{app:outerident}
Here we directly compute the contraction of the outer curvature
with the binormal to produce the expression for $W_{AB}$ in 
(\ref{eqn:OmAB}).  For a more elegant derivation following from
the integrability conditions of the foliation, see \cite{Speranza:2019hkr}. We work in the gauge where $\sqrt{|h|} = 1$, so that 
\beq
A\indices{_A_i^j} = 
2\partial_A (N\indices{_i^k}) N\indices{_k^j},
\eeq
in which case $C_A = A\indices{_A_i^i} = 0$; there is no
loss of generality in making this choice since the outer curvature 
is independent of $C_A$.  
Beginning with (\ref{eqn:WijAB}), we have 
\begin{align}
-2N\indices{_i^j}W\indices{^i_j_A_B} 
&= 
2N\indices{_i^j}\partial_A M\indices{_B^i_j} +2N\indices{_i^j}
M\indices{_A^i_k}M\indices{_B^k_j} - (B\leftrightarrow A)
\nonumber \\
&=
2\partial_A\left(N\indices{_i^j}M\indices{_B^i_j}\right) 
-2\partial_A N\indices{_i^j} M\indices{_B^i_j} 
+2\left(A\indices{_A_k^i}+ P_A N\indices{_k^i}\right)
N\indices{_i^j}
\left(A\indices{_B_j^k}+ P_B N\indices{_j^k} \right)
- (B\leftrightarrow A) \nonumber \\
&=
\partial_A P_B-4A\indices{_A_i^j}N\indices{_j^k} M\indices{_B^i_k}
+2A\indices{_A_k^i} N\indices{_i^j} A\indices{_B_j^k} 
- (B\leftrightarrow A) \nonumber \\
&=\partial_A P_B - 4 A\indices{_A_i^j}N\indices{_j^k} A\indices{_B_k^i}
+2A\indices{_A_i^j}N\indices{_j^k} A\indices{_B_k^i}
- (B\leftrightarrow A) \nonumber \\
&= 2 \partial_{[A} P_{B]} +4
N\indices{_k^j}A\indices{_A_j^i} A\indices{_B_i^k}
\end{align}
Note that the second term is antisymmetric in $A$ and $B$, due to 
the antisymmetry of $N_{ij}$ and the symmetry of $A\indices{_A_i_j}$ on
$i$ and $j$.

\subsection{Commutators}\label{app:Commutators}
In this section, we give the proofs of the Poisson brackets used in section \ref{sec:poisson}.

\newl{\bf\underline{the first identity of \eqref{density}}:}
This can be established using that $q= \wt{n}_a \wt{n}^a$ and $N= \wt{n}/\sqrt{q}$. We have
\bea
\left\{\wt{P}_A(\si), \sqrt{q}(\sigma')\right\}&=&
 N_a(\si') \{\wt{P}_A(\si), \wt{N}^a(\sigma')\}\cr
&=& -  N_a(\si') \wt{N}^a(\sigma) \pa_A'\delta^{(2)}(\si-\si')\cr&=& 
-  N_a(\si')  \pa_A'(\wt{N}^a(\sigma') \delta^{(2)}(\si-\si') )\cr
&=& -  \pa_A'(\sqrt{q} (\sigma') \delta^{(2)}(\si-\si') )\cr 
&=& -  \sqrt{q}(\sigma) \pa_A'(  \delta^{(2)}(\si-\si') ) .\label{Pq}
\eea
From this, the first bracket of \eqref{density} follows easily.

\newl{\bf\underline{the identity \eqref{PPJ}}:} We now evaluate \eqref{PPJ} using that \be
\{\wt P_A(\si), \tP_B(\sigma')\}=  \left( \wt{P}_A(\si') \pa_B -
\wt{P}_B(\si) \pa_{A}' \right)\delta^{(2)}(\si-\si'),
\ee
we get 
\begin{alignat}{2}
    &\sqrt{q(\si) q(\si')}\left\{\wt{P}_A(\si),{P}_B(\si')\right\}\nonumber
\\
=&
\left\{\wt{P}_A(\si),\wt{P}_B(\si')\right\}
- 
\left\{\wt{P}_A(\si),\sqrt{q(\si')} \right\} {P}_B(\si')
-
\wt{P}_A(\si) \left\{\sqrt{q(\si)},\wt{P}_B(\si')\right\}\nonumber
\\
=& 
  \left(\wt{P}_A(\si') \pa_B -
\wt{P}_B(\si) \pa_{A}' \right)\delta^{(2)}(\si-\si')+
\partial'_A(\sqrt{q}(\si')\delta^{(2)}(\si-\si')) {P}_B(\si')
-
\partial_B(\sqrt{q}(\si)\delta^{(2)}(\si-\si')) {P}_A(\si)\nonumber
\\
=&   \pa_B\left[ \wt{P}_A(\si)\delta^{(2)}(\si-\si')\right]   -
 \pa_{A}' \left[ \wt{P}_B(\si')\delta^{(2)}(\si-\si')\right]\nonumber
\\
+&
\partial'_A\left[\sqrt{q}(\si')\delta^{(2)}(\si-\si')\right] {P}_B(\si')
-
\partial_B\left[\sqrt{q}(\si)\delta^{(2)}(\si-\si')\right] {P}_A(\si).
\end{alignat}
In the last equality we have used that 
$\wt P_A(\si') \pa_B\left[ \delta^{(2)}(\si-\si')\right]=
\pa_B\left[\wt{P}_A(\si')\delta^{(2)}(\si-\si')\right]
=\pa_B\left[\wt{P}_A(\si)\delta^{(2)}(\si-\si')\right]$. One can now use that $\tP_A= \sqrt{q} P_A$ to simplify the expression
\bea
\sqrt{q(\si) q(\si')}\left\{{P}_A(\si),{P}_B(\si')\right\}&=&
\sqrt{q(\si)} \left[\pa_B {P}_A(\si')- \pa_{A}' {P}_B(\si')\right]\delta^{(2)}(\si-\si')\cr
 &=&- \sqrt{q(\si)}\varepsilon_{AB}  \wt{J}(\si) \delta^{(2)}(\si-\si')  .
\eea
where we used the definition of $\wt{J}= \varepsilon^{AB} \pa_A P_B$. Dividing out by ${q}(\si)$, one finally gets  that 
\be
\left\{{P}_A(\si),{P}_B(\si')\right\}  = -\varepsilon_{AB}  \wt{J}(\si) 
\frac{\delta^{(2)}(\si-\si') }{\sqrt{q}(\sigma)} =-  \varepsilon_{AB} J(\si)
\delta^{(2)}(\si-\si').
\ee
where we use the undensitized expression
$J = \wt{J}/\sqrt{q}$.

\newl{\bf\underline{the identity \eqref{PNN}}:} this can be established in a similar manner
 \bea
\sqrt{q(\si) q(\si')}\left\{{P}_A(\si),N_a(\si')\right\} &=&
\left\{\tP_A(\si),\widetilde{N}_a(\si')\right\}
- 
\left\{\tP_A(\si),\sqrt{q(\si')} \right\} N_a(\si')
\cr
&=& 
-\widetilde{N}_a(\si) \pa_{A}'\delta^{(2)}(\si-\si') +
N_a(\si')\partial'_A(\sqrt{q}(\si')\delta^{(2)}(\si-\si'))\cr
&=&  - \pa_{A}'\left[ \widetilde{N}_a(\si)\delta^{(2)}(\si-\si')\right] +
N_a(\si')\partial'_A(\sqrt{q}(\si')\delta^{(2)}(\si-\si'))\cr
&=& - \sqrt{q(\si)}(\partial'_AN_a(\si'))  \delta^{(2)}(\si-\si').
\eea
{\bf \underline{The first equality of \eqref{centralizer}}:}  
Using $J[\phi] = \int \xi^A_\phi \tilde P_A$ and \eqref{Pq}, we have 
\bea
\{ J[\phi], \sqrt{q(\sigma')} \} &=& -\int_S d^2 \sigma \; \xi_\phi^B(\sigma) \sqrt{q(\sigma)}  \partial'_B ( \delta^{(2)}(\sigma - \sigma'))\cr &=& 
-\int_S \rd^2 \sigma \epsilon^{AB}\pa_A \phi(\sigma) \partial'_B ( \delta^{(2)}(\sigma - \sigma'))\cr 
&=&- 
\int_S \rd^2 \sigma \epsilon^{AB} \phi(\sigma) \pa_A \partial_B ( \delta^{(2)}(\sigma - \sigma'))
=0.
\eea
The last equality follows from integration by parts and the fact that 
$\partial'_B ( \delta^{(2)}(\sigma - \sigma')) =- \partial_B ( \delta^{(2)}(\sigma - \sigma'))$

\newl {\bf \underline{the first identity of \eqref{PS}}:}
Using \eqref{PNN} one gets that
\bea
\left\{\tP_A(\sigma),\widetilde{S}(\si')\right\}
=\sqrt{q(\si)}\left\{{P}_A(\sigma),\widetilde{S}(\si')\right\}
&=& - \pa_A'(\widetilde{S}(\si'))\delta^{(2)}(\si-\si'),
\eea 
where 
\be
\widetilde{S}(\si') = (\varepsilon^{AB} \pa_A \N^b \pa_B \N^c  \N^a \epsilon_{abc})
\ee
Therefore we have that 
\be 
\{ P[\xi],\widetilde{S}(\si')\} =
- \xi^A \pa_A \widetilde{S}(\si'), \qquad
\{ P[\xi],{S}[\psi]\}
=S[{\cal L}_\xi\psi].
\ee
Also since $\{\xi_\phi^A,\widetilde{S}(\si')\}=0$, we have 
\be 
\{ P[\xi_\phi],{S}[\psi]\}
=S[{\cal L}_{\xi_\phi}\psi]
= S[ \pb{\phi}{\psi}].
\ee

\bibliographystyle{utphys} 
\bibliography{reps}

\end{document}